\documentclass[prb,reprint,superscriptaddress,amsmath,amssymb,aps]{revtex4-2}

\usepackage{graphicx}
\usepackage{dcolumn}
\usepackage{bm}
\usepackage[section]{placeins}

\usepackage{amsmath,amssymb}
\usepackage{graphicx}
\usepackage{hyperref}
\hypersetup{colorlinks=true,urlcolor=blue,citecolor=blue,linkcolor=blue,breaklinks=true}
\usepackage[noabbrev]{cleveref}
\usepackage{adjustbox}
\usepackage{footnote}
\usepackage{grffile}
\usepackage{color}
\usepackage{xcolor}
\usepackage[normalem]{ulem}
\usepackage{dsfont}
\usepackage{mathtools}
\usepackage{verbatim}
\usepackage{bbold}
\usepackage{hhline}
\usepackage{textcomp}
\usepackage{comment}
\usepackage{tikz}

\usetikzlibrary{arrows}
\usetikzlibrary{decorations.pathmorphing}
\usetikzlibrary{decorations.pathreplacing}
\usetikzlibrary{fit,backgrounds}
\usetikzlibrary{patterns}
\usetikzlibrary{shapes.callouts}
\usetikzlibrary{shapes,snakes}
\usetikzlibrary{plotmarks}

\usepackage{xcolor}
\usepackage{physics}
\definecolor{unirot}{RGB}{165,30,55}
\definecolor{unigelb}{RGB}{168,148,99}
\definecolor{pythongreen}{RGB}{0,128,0}
\definecolor{unigrau}{RGB}{50,65,75}

\allowdisplaybreaks

\begin{document}

\newcommand{\bfk}{\mathbf{k}}
\newcommand{\bfq}{\mathbf{q}}
\newcommand{\bfR}{\mathbf{R}}
\newcommand{\bfe}{\mathbf{e}}
\newcommand{\bfu}{\mathbf{u}}

\newcommand{\bfX}{\mathbf{X}}
\newcommand{\X}{\mathrm{X}}
\newcommand{\SC}{\mathrm{SC}}
\newcommand{\dSC}{\mathrm{dSC}}
\newcommand{\dM}{\mathrm{dM}}
\newcommand{\dD}{\mathrm{dD}}
\newcommand{\dX}{\mathrm{dX}}
\newcommand{\D}{\mathrm{D}}
\newcommand{\M}{\mathrm{M}}
\newcommand{\pp}{\mathrm{pp}}
\newcommand{\ph}{\mathrm{ph}}
\newcommand{\xph}{{\overline{\mathrm{ph}}}}

\newcommand{\dwave}{\mathrm{dw}}
\newcommand{\swave}{\mathrm{sw}}

\newcommand{\SA}[1]{{\color{orange}{[SA: #1]}}}
\newcommand{\MS}[1]{{\color{cyan}{\textsf{[MS: #1]}}}}
\newcommand{\ms}[1]{{\color{cyan}{#1}}}
\newcommand{\AAE}[1]{{\color{blue}{[AA: #1]}}}

%------------
\title{Intertwined fluctuations and isotope effects in the  Hubbard-Holstein model\\
on the square lattice from functional renormalization}
%------------

\author{Aiman Al-Eryani}
\email{aiman.al-eryani@rub.de}
\affiliation{Theoretical Physics III, Ruhr-University Bochum, 44801 Bochum, Germany}

\author{Sabine Andergassen}
\email{sabine.andergassen@tuwien.ac.at}
\affiliation{Institute of Information Systems Engineering, Vienna University of Technology, 1040 Vienna, Austria}
\affiliation{Institute for Solid State Physics, Vienna University of Technology, 1040 Vienna, Austria}

\author{Michael M. Scherer}
\email{scherer@tp3.rub.de}
\affiliation{Theoretical Physics III, Ruhr-University Bochum, 44801 Bochum, Germany}

%------------
\begin{abstract}
Electron-electron and electron-phonon interactions are responsible for the formation of spin, charge, and superconducting correlations in layered quantum materials.
A paradigmatic model for such materials that captures both kinds of interactions is the two-dimensional Hubbard-Holstein model with a  dispersionless Einstein phonon.
In this work, we provide a detailed analysis of the magnetic, density, and superconducting fluctuations at and away from half-filling.
To that end, we employ the functional renormalization group using the recently introduced extension of the single-boson exchange formulation. 
More precisely, we go beyond previous approaches to the model by resolving the full frequency dependence of the two-particle vertex and taking into account the feedback from the electronic self-energy.
We perform broad parameter scans in the space of Hubbard repulsion, electron-phonon coupling strength, and phonon frequency to explore the leading magnetic, density, and superconducting susceptibilities from the adiabatic to the anti-adiabatic regime.
Our numerical data reveal that self-energy effects lead to an enhancement of the $d$-wave superconducting susceptibility towards larger phonon frequencies, in contrast to earlier isotope-effect studies.
At small phonon frequencies, large density contributions to the $s$-wave superconducting susceptibility change sign and eventually lead to a reduction of $s$-wave superconductivity with increasing electron-phonon coupling, signaling the breakdown of Migdal-Eliashberg theory. 
We analyze our findings systematically, employing detailed diagnostics of the intertwined fluctuations and pinning down the various positive and negative isotope effects of the physical susceptibilities.
Our results pave the way for a more comprehensive understanding of the interplay of electron-electron and electron-phonon interactions in correlated quantum materials.
\end{abstract}
%------------

\maketitle

%-----------
\section{Introduction}
%-----------

Electron-electron and electron-phonon interactions are key players in controlling correlation effects across different material classes. 
For example, in normal metals, the electron-phonon coupling mediates a retarded attractive interaction between the electrons, leading to Cooper pairing and superconductivity at low temperatures.
According to Bardeen-Cooper-Schrieﬀer~(BCS) theory~\cite{PhysRev.108.1175}, the superconducting transition temperature is proportional to the Debye frequency, which in turn is inversely proportional to the square root of the mass of  ion cores constituting the crystal lattice, i.e., $T_c\sim 1/\sqrt{M_\mathrm{ion}}$. 
This phenomenon is typically referred to as the isotope effect and was known before BCS theory~\cite{PhysRev.78.477,PhysRev.78.487}.
However, in the presence of nesting or Van Hove points, the electron-phonon coupling can also induce different types of charge or bond order, depending on which phonon excitations dominate the coupling to the electronic degrees of freedom; see, e.g., Refs.~\cite{PhysRevB.40.197,PhysRevLett.66.778,PhysRevB.42.2416,PhysRevLett.91.056401,PhysRevLett.126.017601,PhysRevLett.127.247203} for the two-dimensional (2D) case.

Naturally, the electron-phonon-mediated attraction competes with the repulsive Coulomb interaction between electrons.
While the latter is well screened in ordinary metals, it is the game changer in strongly correlated materials. 
There, it can give rise to competing spin- and charge-density wave instabilities, or even unconventional superconductivity as, e.g., found in the paradigmatic Hubbard model~\cite{Hubbard1963,Arovas_2022,Qin_2022}.
The need for a better understanding of the interplay of electron-phonon and electron-electron interactions in strongly correlated materials has been discussed early on for the cuprate superconductors~\cite{PhysRevLett.58.2333,PhysRevLett.71.283,zech1994,muller2014} and later also for the iron-based superconductors~\cite{Liu_2009,PhysRevLett.103.257003}.
Recent breakthroughs in the synthesis and analysis of two-dimensional correlated materials, such as graphene multilayers~\cite{Andrei_2020,Pantale_n_2023} or moir\'e semiconductors~\cite{MakShan_2022}, have renewed the interest in advancing the theoretical description of this interplay.

A paradigmatic model for layered quantum materials that includes both, electron-phonon coupling  and Coulomb interactions, is the single-band Hubbard-Holstein model~\cite{HOLSTEIN1959325} on the square lattice~\cite{freericks_competition_1995,capone_phase_2004,koller_dynamic_2004,paci_polaronic_2005,sangiovanni_electron-phonon_2005,perroni_effects_2005,piegari_signatures_2005,sangiovanni_electron-phonon_2006,paci_isotope_2006,barone_effective_2006,sangiovanni_relevance_2006,APREA2006277,werner_efficient_2007,tezuka_phase_2007,barone_gutzwiller_2008,di_ciolo_charge_2009,
bauer_competition_2010,macridin_suppression_2012,nowadnick_competition_2012,hohenadler_excitation_2013,johnston_determinant_2013,murakami_ordered_2013,nowadnick_renormalization_2015,WWW2025,ohgoe_competition_2017,karakuzu_superconductivity_2017,weber_two-dimensional_2018,ghosh_study_2018,han_strong_2020,wang_zero-temperature_2020,costa_magnetism_2021,Costa2020phasediagram_hubbardholstein, issa2025learning}. 
More precisely, in this model, the electrons couple to a dispersionless Einstein phonon of frequency $\omega_0$ and the Coulomb interaction is represented by an on-site Hubbard repulsion $U$.
Effectively, the phonons give rise to a frequency-dependent local attraction that leads to a phonon-mediated retarded electron-electron interaction competing with the instantaneous Hubbard term.
Variations of the phonon frequency $\omega_0\sim 1/\sqrt{M_\mathrm{ion}}$ will generally have an impact on the phase diagram of the model, similar to (but potentially also different from) the case of BCS superconductivity; i.e., there will be an isotope effect, too.
The Hubbard-Holstein model on the square lattice represents a prime challenge for quantum many-body methods.
Therefore, recent numerical studies, such as quantum Monte Carlo~(QMC) simulations, have focused specific parameter regimes, e.g., to the case of pure nearest-neighbor hopping and half filling~\cite{Costa2020phasediagram_hubbardholstein} to avoid the occurrence of a fermion sign problem. 
The case of finite doping, instead, is less explored \cite{johnston_determinant_2013, macridin_suppression_2012} and in some cases has been limited to $U = 0$ where the sign problem is absent~\cite{issa2025learning}, see also Ref.~\cite{Cohen_Stead_2024_julia_DQMC_library} for a DQMC Julia library for electron-phonon systems.

A suitable approach to study the 2D Hubbard-Holstein model away from half filling is the functional renormalization group (fRG), which includes all
types of correlations on equal footing; cf.~Refs.~\cite{Kopietz2010,Metzner2012,Dupuis2021} for reviews.
The fRG for interacting fermion systems has developed into a versatile tool to identify the leading Fermi-surface instabilities in strongly correlated quantum materials in
the presence of competing interactions, e.g., cuprates, iron superconductors, graphene and its stacks, kagome metals, nodal-line semimetals, and moir\'e heterostructures; see the reviews~\cite{Metzner2012,Platt_2013,Dupuis2021,honerkamp2022recent}, Ref.~\cite{Profe_2024} for a  C/C++/Python library, and references therein.
First successful applications of the fRG to Hubbard models appended with different types of electron-phonon interactions have also been put forward, relying on the exclusive consideration of the renormalization of the two-particle interaction vertex without explicit  (or reduced) frequency dependence~\cite{Fu2006,PhysRevB.90.035122,WWW2025,HHH_paper,Cichutek2022,anomalous_isotope_effect,Reckling_2018} or 
for a finite size lattice~\cite{Yirga_2023}.

Recent methodological developments and algorithmic improvements of the fRG approach to correlated fermion systems have shown that the fRG can be pushed toward quantitative accuracy.
For example, employing the multiloop extension~\cite{Kugler2018a,Kugler2018b,Tagliavini_2019}, a comparison of fRG data for the pure Hubbard model with exact QMC data at half filling has shown very good agreement even for intermediate interaction strengths~\cite{Hille2020}.
Moreover, the recently introduced single-boson exchange~(SBE) formulation~\cite{Krien_SBE_original}  provides a physically transparent decomposition of the two-particle vertex that leads to computationally efficient approximations in which the flow of multiboson contributions can be neglected while maintaining quantitative accuracy~\cite{fraboulet2023singlebosonexchangefunctionalrenormalization,Bonetti_2022,aleryani2024screeningeffectiverpalikecharge}.
An extension to frequency-dependent interactions can be introduced by a generalization of the SBE to nonlocal 
interactions~\cite{aleryani2024screeningeffectiverpalikecharge} that can be readily applied to extended interactions in the time direction.

Here, we present an fRG analysis of the Hubbard-Holstein model that goes substantially beyond previous studies by employing various of the above developments.
Specifically, we consider the full frequency dependence of the two-particle vertex, the flow of the self-energy, and we utilize the Katanin substitution~\cite{Katanin_substitution_2004} in the flow equations that ensures Migdal-Eliashberg (ME) theory~\cite{Migdal1958,Eliashberg1960,MARSIGLIOreview2020168102}.
In addition, we supplement our investigation by a systematic fluctuation diagnostics~\cite{Gunnarsson2015,Schaefer2021b,heinzelmann2023entangledmagneticchargesuperconducting} that allows us to identify the dominant fluctuations underlying the physical behavior. 
We perform broad scans in the parameter space of the Hubbard-Holstein model, tuning the Hubbard repulsion, the electron-phonon coupling strength, and the phonon frequency to explore the leading magnetic, density, and superconducting susceptibilities.
We consider both, the half-filled case with Fermi-surface nesting where QMC data are available and the case of finite doping, where QMC simulations have a fermion sign problem.
Below we highlight our main results.

The {remainder of the} paper is structured as follows: in Secs.~\ref{sec:model} and~\ref{sec:method} we introduce the model and present the SBE extension for the treatment of retarded interactions, respectively.
We briefly review the SBE formulation of the fRG and describe its application to frequency-dependent interactions. Readers not interested in these method development aspects may skip this section.
In Sec.~\ref{sec:results}, we present the results: the phase diagrams determined by the leading susceptibilities, both at half filling and finite doping. 
We discuss how the enhanced density fluctuations affect the $s$-wave superconductivity leading to signatures beyond Migdal-Eliashberg~(ME) theory, as well as their impact on the phonon renormalization and softening.
We also determine the role of the electron self-energy in reversing the isotope effect on the $d$-wave superconducting susceptibility at finite doping. 
In Sec.~\ref{sec:fluct}, we perform a fluctuation diagnostics to investigate the effect of Holstein phonons and retardation on the electronic ordering tendencies.
We conclude with a summary and an outlook in Sec.~\ref{sec:concl}.

%-----------
\section{Summary of the main results}
\label{sec:summary}

\subsection{Methodological aspects}

\textit{Extension of the SBE formalism to retarded interactions.}
We show that the newly developed and computationally efficient %
SBE formulation
for extended interactions \cite{aleryani2024screeningeffectiverpalikecharge} can be  applied also to retarded interactions; the details for the fRG implementation are given in Sec.~\ref{sec:flow_of_vertex}. The SBE approximation of neglecting the flow of the rest functions $M^\X$ is found to provide quantitatively accurate results, with appreciable corrections only for small $\omega_0$ as we approach the adiabatic limit (see Appendix~\ref{app:sbe_approximation} for a detailed discussion).

\textit{Role of the electronic self-energy $\Sigma$ on $\chi^{\dSC}$}. Taking into account the frequency dependence of the vertex, we were able to include in our fRG scheme also the flow of the electronic self-energy. It leads to a sign change in the dependence of the $d$-wave superconducting susceptibility on the ion mass affecting the isotope effect in $\dd \chi^\dSC / \dd M_{\textrm{ion}}$, see Fig.~\ref{fig:effect_of_selfenergy_on_isotope_effect}). By performing a fluctuation diagnostics, we can trace back this behavior to the self-energy entering the $d$-wave bubble (see Fig.~\ref{fig:finitedoping_dwave_susceptibility_fluctdiag}). Further diagnostics on the bubble reveals its origin in 
the dampening of coherent quasiparticles by phonons (see Fig.~\ref{fig:bubble_fluct_diag}). 

\subsection{Aspects 
of the Hubbard-Holstein model}

\textit{Effect of phonon-induced charge fluctuations on $\chi^\SC$.}
Including the self-energy together with the full frequency dependence of the vertex within the so-called Katanin substitution, constitutes a scheme that encapsulates and goes beyond 
ME theory.
In particular, it accounts for the inter-channel feedback in an unbiased fashion. For small phonon frequencies $\omega_0$, we observe a maximum in $\chi^\SC$ as the dimensionless electron-phonon coupling $V_H$ is increased, see Fig.~\ref{fig:sc_d_suscs_vs_Vh_doped}. This is in contrast to the expectation from ME theory which predicts a monotonic growth with $V_H$. This behavior occurs concomitantly with an enhancement of charge-density wave fluctuations (see also Fig.~\ref{fig:sc_susc_vs_filling}). Through a fluctuation diagnostics analysis both at half filling and finite doping, we show the susceptibilities to be responsible for the above suppression, see Figs.~\ref{fig:halffilling_susceptibility_fluctdiag} and~\ref{fig:finitedoping_susceptibility_fluctdiag}. 

\textit{Phonon softening and absence of lattice instabilities.} 
We investigate the renormalization of the phonon propagator by deriving an exact formula for the phononic self-energy $\Xi$ in terms of the charge density susceptibility, see Eq.~\eqref{eq:phonon_self_energy_from_chi_D}. As a result, the previously predicted lattice instability occurring for large $V_H$ and finite $\chi^\D$ within the widely used approximation of Eq.~\eqref{eq:RPA_dispersion} vanishes. 
We find instead a phonon softening as $V_H$ is increased. The phonon dispersion never truly becomes imaginary but tends to zero as $\chi^\D$ diverges, see Fig.~\ref{fig:phonon_softening}. As such, there is no lattice instability in the model without a charge-density wave instability. 

\textit{Role of Holstein phonons in the interplay of electronic fluctuations.} 
The fluctuation diagnostics provides a detailed analysis of the influence of the effective retarded interaction resulting from the phonons on the interplay of electronic fluctuations, see  Figs.~\ref{fig:halffilling_susceptibility_fluctdiag} and~\ref{fig:finitedoping_susceptibility_fluctdiag}. We supplement the numerical results by deriving analytic expressions for the diagnostic matrices. Primarily, the role of low-frequency phonons is found to invert the influence of charge density fluctuations on the spin and $s$-wave pairing with respect to the pure Hubbard model, such that they suppress $s$-wave superconductivity and encourage spin fluctuations, see Fig.~\ref{fig:interplay_purehubbard}. 
%-----------

%-----------
\section{Hubbard-Holstein model}
\label{sec:model}
%-----------

%---------
\begin{figure}[t]
    \centering  \includegraphics[width=0.9\linewidth]{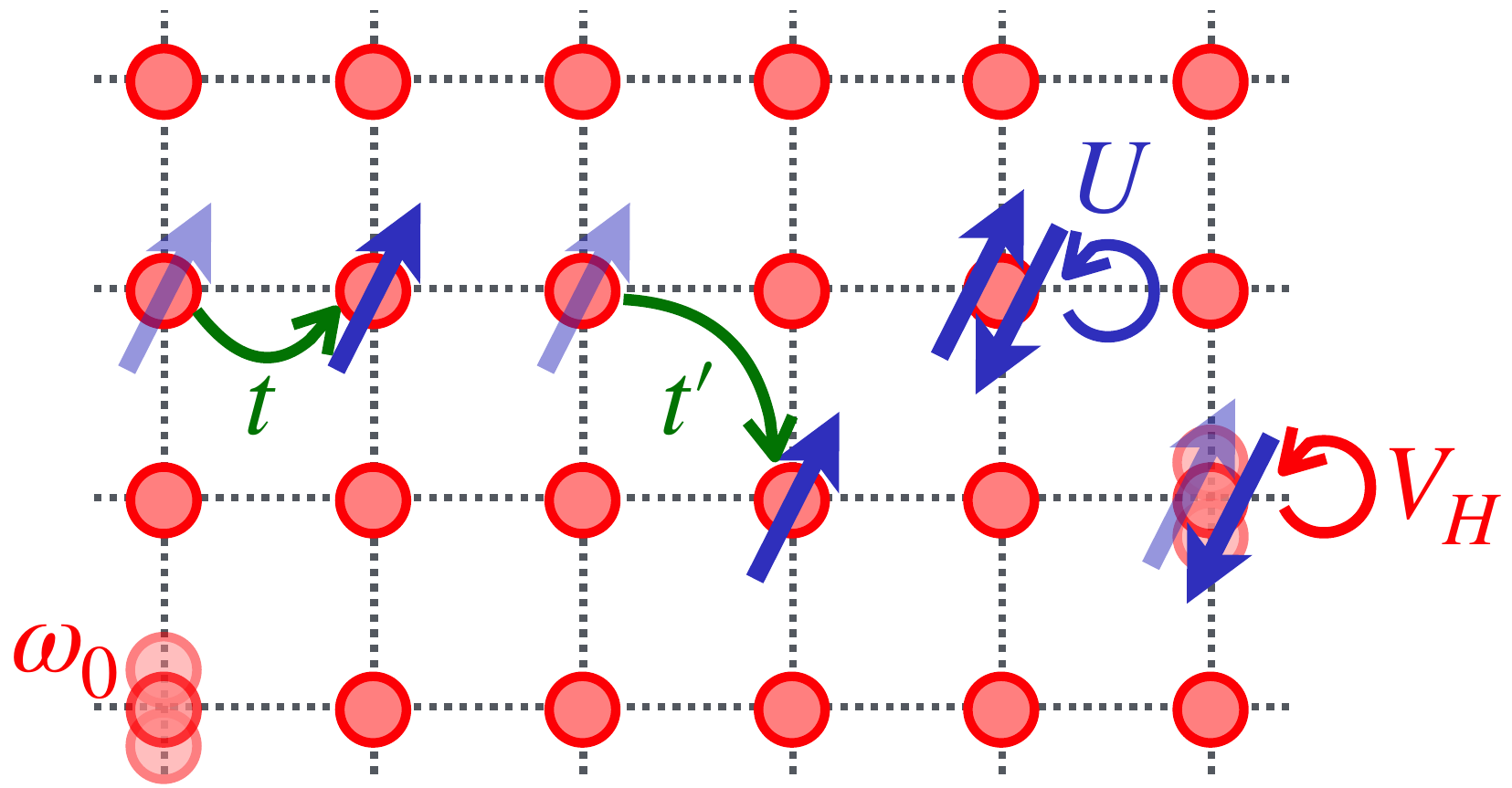}
    \caption{Representation of Hubbard-Holstein model. Electrons hop with amplitude~$t$ to their nearest neighbors and with amplitude $t'$ to their second-nearest neighbors. They interact instantaneously via the Hubbard repulsion~$U$.  Lattice sites vibrate with frequency~$\omega_0$ and the effect of an electron visiting a vibrating site leaves an imprint felt by an electron passing by that site at a later time. This induces an effective retarded interaction of strength~$V_H$ between the electrons.}
    \label{fig:hubbard-holstein_graphic}
\end{figure}
%---------

The Coulomb interaction couples electrons in a solid with each other and to the ion cores that constitute the crystal lattice.
To first order in the displacement from their equilibrium positions, the ions oscillate in different vibrational modes which -- when quantized -- give rise to phonon branches. 
A general model that describes a system of interacting electrons coupled to phonons is given by the Hamiltonian
\begin{align}
H =& \sum_{\mathclap{\mathbf{k} \sigma}} (\epsilon(\mathbf{k}) - \mu_0) c^\dagger_{\mathbf{k}  \sigma} c_{\mathbf{k} \sigma} + \sum_{\mathclap{\mathbf{q}j}}  \omega_j(\bfq)  \left(b^\dagger_{\bfq j} b_{\bfq j} + \frac{1}{2} \right) \nonumber \\ &+ \sum_{\mathbf{k}\sigma} \sum_{\mathbf{q}j} g^j_{\mathbf{k}\bfq} c^\dagger_{\bfk + \bfq \sigma} c_{\bfk \sigma} \left(b_{\bfq j} + b^\dagger_{-\bfq j}\right) \nonumber \\
&+ U \sum_{\bfq \bfk \bfk'} c^\dagger_{\bfk+\bfq\uparrow} c^\dagger_{\bfk'-\bfq\downarrow} c_{\bfk'\downarrow} c_{\bfk\uparrow}\, . \label{eq:the_hamiltonian}
\end{align}
where $c^{(\dagger)}_{\bfk, \sigma}$ and $b^{(\dagger)}_{\bfq j}$ are the annihilation (creation) operators for the electrons and phonons respectively; $\epsilon(\bfk)$ and $\omega_j(\bfq)$ are the dispersion relations of the electrons and the phonons in branch $j$, respectively.  On a square lattice, the bare electronic dispersion is given by
\begin{align}
    \epsilon({\bf k})=-2t( \cos{k_x} + \cos{k_y} ) - 4t'\cos{k_x}\cos{k_y}\,. \label{eq:free_dispersion}
\end{align}
It describes electrons hopping on a lattice with nearest-neighbor and next-nearest-neighbor hopping
amplitude $t$ and $t'$.
The electrons interact locally and instantaneously via a repulsive  interaction~$U$, while the phonons in branch $j$ couple to an electron-hole bilinear with the coupling constant $g^j_{\mathbf{k}\bfq}$. 
In the following, we consider the paradigmatic case of a constant electron-phonon coupling $g$ and a 
single dispersionless phonon of frequency $\omega_0$, which is often referred to as the Hubbard-Holstein model~\cite{HOLSTEIN1959325,Hubbard1964}, see Fig.~\ref{fig:hubbard-holstein_graphic} for an illustration. 
The case of an acoustic phonon that couples to the charge density ($g^j_{\bfq \bfk} = g_{\bfq}$) is discussed in Appendix~\ref{app:acph}.

In a field theoretic treatment, the phonons appearing quadratically in the Hamiltonian~\eqref{eq:the_hamiltonian} can be integrated out.
This yields a frequency-dependent effective electron-electron interaction,
\begin{align}
\label{eq:bareint}
V_0(i\Omega) := U + V_H \frac{\omega_0^2}{(i\Omega)^2 - \omega_0^2}= U - V_H \frac{\omega_0^2}{\Omega^2 + \omega_0^2}\,, 
\end{align}
with bosonic Matsubara frequency $\Omega= 2\pi m/\beta$, $m \in \mathbb{Z}$, and a phonon-mediated electron-electron interaction strength given by the ratio
\begin{align}\label{eq:vh}
V_{H} := 2\frac{|g|^2}{\omega_0}\,;
\end{align}
see Appendix~\ref{app:hh} for details.
Physically, both the phonon frequency and the electron-phonon coupling depend on the ion mass $M_\mathrm{ion}$ as
\begin{align}\label{eq:massdep}
\omega_0 \sim \frac{1}{\sqrt{M_\mathrm{ion}}}\quad \text{and}\quad g^2 \sim \frac{1}{\sqrt{M_\mathrm{ion}}}\,.
\end{align}
Therefore, $V_H = 2|g|^2/\omega_0$ 
is independent of $M_\mathrm{{ion}}$.
As a consequence, a variation of $M_\mathrm{ion}$ affects only $\omega_0$ in the interaction~\eqref{eq:bareint}. 
Hence, at fixed $V_H$, two important 
limits for the influence of the phonons can be distinguished:
\begin{enumerate}
    \item The \textbf{anti adiabatic limit} for a vanishing mass $M_\mathrm{{ion}} \rightarrow 0$ with a diverging phonon frequency $\omega_0 \sim 1/\sqrt{M_\mathrm{ion}} \rightarrow \infty$. In this limit, Eq. \eqref{eq:bareint} yields an effective Hubbard interaction with
    \begin{align}
        V_0(i\Omega) \rightarrow U_{\textrm{eff}}:=U - V_H \,.
        \label{eq:aalim}
    \end{align}
\item The \textbf{adiabatic limit} for $M_\mathrm{ion} \rightarrow \infty$. Here, one has to consider the $i\Omega = 0$ contribution separately, for which the effect of the phonons does not vanish 
\begin{align}
     V_0(i\Omega) \rightarrow \begin{cases}
         U_\mathrm{eff} & \text{for }\; i\Omega = 0\\ 
         U & \text{otherwise}.
     \end{cases}
    \label{eq:alim}
\end{align}
\end{enumerate}
We see that besides two situations in which the phonons can be neglected in the adiabatic limit, i) at low temperatures $T \ll \omega_0$ and ii) for $V_H \ll U$, the lattice is soft to thermal fluctuations and phonons will proliferate even in presence of a large $M_\textrm{ion}$.

In this study, the adiabatic and anti adiabatic limits will be considered by very small and very large values of $\omega_0$ respectively, while keeping $V_H$ fixed. 
Note that $V_H=U$ represents a special situation, corresponding to $U_\mathrm{eff}=0$. 
We will thus distinguish between repulsive $U_\mathrm{eff}>0$ and attractive effective interactions $U_\mathrm{eff}<0$.

%---------
\section{Functional renomalization}
\label{sec:method}
%---------

Let $\psi$ and $\bar{\psi}$ denote the Grassmann fields corresponding to the fermion creation and annihilation operators $c$ and $c^\dagger$. 
For brevity, we use the combined frequency-momentum index $k = (i\nu, \bfk)$ and we denote the spin index by $\sigma$.
We denote the integration over this index $\int_{k} \equiv \int_{\bfk}\sum_{i\nu}$.  Here, the normalization by the Brillouin-zone volume is absorbed into the integral sign over momentum and the normalization by the factor of $\beta$ is absorbed in the sign representing the sum over the Matsubara frequencies. Integration over the spin indices just means a sum over the two spin orientations $\sigma_{l} \in \{\uparrow, \downarrow\}$.
The starting point is the Euclidean microscopic action
\begin{align}\label{eq:micac}
S[\bar{\psi}, \psi] = &-\int\limits_{\overset{k_1k_2}{\sigma_1\sigma_2}} \bar{\psi}_{\sigma_1k_1} G^{-1}_{0,\sigma_1\sigma_2}(k_1, k_2)\psi_{\sigma_2k_2} \nonumber \\ +\frac{1}{4}&\int\limits_{\overset{k_1k_2k_3k_4}{\sigma_1\sigma_2\sigma_3\sigma_4}}\hspace{-4mm} V_{0}\,\psi_{\sigma_1k_1}\bar{\psi}_{\sigma_2k_2}\psi_{\sigma_3k_3}\bar{\psi}_{\sigma_4k_4}\,,
\end{align}
where $G_0$ denotes the free fermionic propagator and $V_0=V_{0,\sigma_1\sigma_2\sigma_3\sigma_4}(k_1,k_2,k_3,k_4)$ represents the bare two-particle interaction.

In the Hubbard-Holstein model [cf. Eq.~\eqref{eq:the_hamiltonian}], by virtue of its translation invariance and spin-SU(2) symmetry, the free fermionic propagator is given by
\begin{align}
G_{0,\sigma_1\sigma_2}(k_1, k_2) &= G_0(i\nu_2 - i\nu_1, \bfk_2 - \bfk_1)\delta_{\sigma_1,\sigma_2}
\end{align}
with
\begin{align}
G_0(i\nu, \bfk) =& \frac{1}{i\nu -\epsilon({\bf k}) + \mu_0}\,.
\end{align}
Here, $\epsilon(\bfk)$ is the free electron dispersion [cf. Eq.~\eqref{eq:free_dispersion}], $\nu =(2m+1)\pi/\beta$,  $\ m \in \mathbb{Z}$ the fermionic Matsubara frequency, and $\beta=1/T$ the inverse temperature. 
The bare interaction is given by
\begin{align}
V_{0,\sigma_1\sigma_2\sigma_3\sigma_4}(k_1, &k_2, k_3, k_4) = \delta(k_1 - k_2 + k_3 - k_4)\nonumber \\ 
&\times \left[V_0(k_1, k_2, k_3, k_4) \delta_{\sigma_1\sigma_2} \delta_{\sigma_3\sigma_4}\right. \nonumber \\ 
&\left.\quad- V_0(k_1, k_4, k_3, k_2)  \delta_{\sigma_1\sigma_4} \delta_{\sigma_2\sigma_3}  \right]\,, 
\label{eq:spinconv} 
\end{align}
with $V_0(k_1, k_2, k_3, k_4) = V_0(i\nu_2 - i\nu_1)$ defined by Eq.~\eqref{eq:bareint}.

The quantum effective action includes all quantum and statistical fluctuations of the system defined in Eq.~\eqref{eq:micac} and can be written as
\begin{align}
\Gamma[\bar{\psi}, \psi] =& -\int\limits_{\overset{k_1k_2}{\sigma_1\sigma_2}} \bar{\psi}_{\sigma_1k_1} G^{-1}_{\sigma_1\sigma_2}(k_1, k_2) \psi_{\sigma_2k_2} \nonumber \\ 
&+ \frac{1}{4}\int\limits_{\overset{k_1k_2k_3k_4}{\sigma_1\sigma_2\sigma_3\sigma_4}}\hspace{-4mm} V\,\psi_{\sigma_1k_1}\bar{\psi}_{\sigma_2k_2}\psi_{\sigma_3k_3}\bar{\psi}_{\sigma_4k_4}+\ldots\,,
\end{align}
with $V=V_{\sigma_1\sigma_2\sigma_3\sigma_4}(k_1,k_2,k_3,k_4)$ and further higher-order vertices represented by the dots.
The full fermionic Green's function for the electron reads
\begin{align}
&G_{\sigma_1\sigma_2}(k_1, k_2) = G(i\nu_2 - i\nu_1, \bfk_2 - \bfk_1)\delta_{\sigma_1,\sigma_2}\,,\\
&\text{with}\ G(i\nu, \bfk) := \Big( i\nu -\epsilon({\bfk})  +\mu - \Sigma(i\nu, \bfk )\Big)^{-1}\,.
\end{align}
Here, we introduced the electron self-energy~$\Sigma(i\nu, \bfk )$. Further, we have redefined the chemical potential $\mu$ such that it includes the Hartree contribution of the electronic self-energy and that the self-energy decays to zero at large frequencies $\Sigma(i\nu \rightarrow \pm \infty ) = 0$. 
Thus $\mu = 0$ corresponds to half filling $n = 1/2$ for $t' = 0$ and the noninteracting Van Hove filling to $\mu = 4t'$. 

The fRG method  introduces a scale dependence, inducing a flow that interpolates between the microscopic action and the quantum effective action. 
To introduce the scale dependence we deform the bare Green's function and introduce a scale parameter~$\Lambda$ into the free propagator, i.e., $G_0 \rightarrow G_0^\Lambda$, such that the scale-dependent quantum effective action $\Gamma^\Lambda$ fulfills $\Gamma^{\Lambda_\textrm{init}} = S$ and $\Gamma^{\Lambda_\textrm{final}} = \Gamma$. 
Variation of $\Gamma^\Lambda$ with respect to $\Lambda$ leads to an exact functional flow equation with one-loop structure, describing the evolution from the microscopic initial action~$S$ to the final effective action $\Gamma$ as a function of $\Lambda$~\cite{Wetterich:1992yh,Metzner2012,Dupuis2021}.

In this work, we use a smooth frequency cutoff~\cite{PhysRevB.79.195125} supplemented by a scale-dependent shift in the chemical potential $\delta \mu^\Lambda$, which will be discussed below.
Then, the modified fermionic propagator is given by
\begin{align}
    G^{0, \Lambda}(\mathbf{k}, i\nu) := \frac{\Theta^\Lambda(i\nu)}{G_{0}^{-1}(\mathbf{k}, i\nu) + \delta \mu^\Lambda}\,,
\end{align}
with
\begin{align}
\Theta^\Lambda(i\nu) = \frac{\nu^2}{
\nu^2 + \Lambda^2}\,.
\end{align}
being the cutoff function.

In the context of strongly correlated fermion systems, an  expansion of $\Gamma^\Lambda$ in the Grassmann fields %is often invoked and 
yields an infinite hierarchy of ordinary differential equations for the one-particle irreducible vertex functions~\cite{Metzner2012}. 
In general, this cannot be solved exactly and one has to resort to truncation schemes. 
In the so-called level-2 truncation, only the flows of the self-energy $\Sigma^\Lambda(k)$ contributing to the scale-dependent propagator $G^\Lambda(k)$ and the spin-SU(2) resolved two-particle interaction vertex $V^\Lambda(k_1,k_2,k_3,k_4)$, cf.  Eq.~\eqref{eq:spinconv}, are considered and will be discussed below.
The ones of the three- and higher-order particle interaction vertices are neglected.
This approximation includes all corrections up to second order in the bare two-particle interaction~$\mathcal{O}(V_0^2)$, and partially further corrections up to infinite loop orders generated during the flow.

%------------
\subsection{Electronic self-energy flow and frequency dependence}
%------------

With SU$(2)$ spin symmetry, the flow equation for the self-energy is given by
\begin{align}
\hspace{-0.2cm}\mathrm{d}_{\Lambda}\Sigma^\Lambda(k) =\!\int_{p}\! S^\Lambda(p)\! \left( V^{\Lambda}(p, k, k, p)\!-\!2 V^{\Lambda}(k, k, p, p)\!\right), \label{eq:selfenergy_flow_eqn}
\end{align}
 where $S^\Lambda(p) = \mathrm{d}_{\Lambda}{G}^\Lambda(p)|_{\Sigma = \text{const.}}$ defines the  single-scale propagator
\begin{align}
\label{eq:ssp}
    S^\Lambda = G^\Lambda \left[  \dot{\Theta}^\Lambda(G_0^{-1} + \delta\mu^\Lambda) - \Theta ^\Lambda\mathrm{d}_\Lambda(\delta{\mu})^\Lambda \right]G^\Lambda\,,
\end{align}
(we have suppressed momentum indices for convenience of notation).
The inclusion of the electronic self-energy flow results in a Fermi-surface deformation along the flow, which leads to a flowing of the filling. 
The chemical potential shift is introduced to compensate for this, following a procedure given in Refs.~\cite{Vilardi2017,aleryani2024screeningeffectiverpalikecharge}. Specifically, the shift in the chemical potential is determined implicitly by inverting 
\begin{align}
  \!\!  n(\delta\mu^\Lambda)\! =\! \sum_{\nu \bfk}  \frac{e^{i\nu 0^+}}{G_0^{-1}(\mathbf{k}, i\nu)\!+\! \delta\mu^\Lambda \!-\! \Theta^\Lambda(i\nu) \Sigma^\Lambda(\mathbf{k},i\nu) }\,, \label{eq:n_of_mu}
\end{align}
where the filling $n$ is fixed.
Numerically, this is done by a root finding algorithm.
The $\Lambda$-dependent change of the chemical potential imposed by this condition has to be taken into account in the single-scale propagator \eqref{eq:ssp}. 
For this, the derivative in the $j$-th integration step $\Lambda_{j}$ of the flow is computed by the finite difference
\begin{align}
    \mathrm{d}_\Lambda (\delta \mu)^{\Lambda_{j}} = \frac{\delta \mu^{\Lambda_{j}} - \delta \mu^{\Lambda_{j-1}} }{\Lambda_{j} - \Lambda_{j-1}}\,.
\end{align}
This ensures that our results will have a consistent filling. 
We note that interactions generally  shift the Van Hove singularity, thereby altering the filling at which it occurs. 
In this study when we say we are at the Van Hove filling, we refer to the noninteracting Van Hove filling.

The self-energy is a relevant term in the RG flow, even at weak coupling, and we therefore include it in our approach.
Moreover, while power counting arguments suggest the frequency dependence to be irrelevant at weak coupling, it substantially affects the results on a quantitative level~\cite{Husemann2009,Vilardi2017}.
In the one-particle irreducible scheme, no flow in the imaginary part of the self-energy is generated when the frequency dependence of the vertex is neglected~\cite{Metzner2012}. 
Efforts have been made toward a full frequency treatment \cite{Vilardi2017} which includes the high-frequency asymptotics \cite{Rohringer2012,Wentzell2016}.
Its combination with the so-called "truncated-unity" (TU) fRG \cite{Husemann2009,Wang2012,Lichtenstein2017} 
using the channel decomposition in conjunction with a form-factor expansion for the fermionic momentum dependence brought the fRG for interacting fermions on 2D lattices to a quantitatively reliable level in the weak to moderate coupling regime~\cite{Hille2020}. Here, we employ its SBE formulation \cite{Bonetti_2022} which we will discuss 
in more detail below. 
We note that in the literature the flow of the self-energy and the frequency dependence are often neglected, i.e., only the flow of a static vertex is considered, and our present work therefore substantially expands upon that approximation.
%

%----------
\subsection{Flow of the phonon self-energy}
%----------

In the electron-phonon system, a bare bosonic Green's function can be ascribed to the phonon
\begin{align}
D_0(q) = \frac{2\omega_0}{(i\Omega)^2 - \omega_0^2}\,,
\label{eq:bosprog}
\end{align}
with $q = (i\Omega, \bfq)$ a compound bosonic index, where $i\Omega$ is a bosonic Matsubara frequency and $\omega_0$ the phonon frequency appearing in the Hamiltonian~\eqref{eq:the_hamiltonian}. 
Likewise, we define a full phonon Green's function
\begin{align}
D(q) = \frac{1}{D_0^{-1}(q) - \Xi(q)}\,,
\end{align}
with a phonon self-energy $\Xi$. Having integrated out the phonons in the problem, we expect the effects of the phonon renormalization to be implicitly present in the electron problem, as long as the frequency-dependent part of the vertex is allowed to flow.
Since we include the full frequency dependence of the vertex, we therefore implicitly include also the flow of the phonon self-energy. In fact, since the phonons in the Hubbard-Holstein model couple to the density operators, it is possible to extract the explicit phonon self-energy $\Xi$ from the charge density response function of the system $\chi^\D$ and thereby calculate also the renormalized phonon dispersion, see Sec.~\ref{sec:phonon_softening} for more details. 
We note that as a post-processing computation, this applies also for other approaches than the fRG.

%--------------
\subsection{Flow of the vertex}
\label{sec:flow_of_vertex}
%--------------

For the flow of the vertex function $V^\Lambda(k_1, k_2, k_3, k_4)$, it is convenient to perform a decomposition of the vertex and calculate the flow equations of the constituents.
In this work, we utilize the recently introduced SBE decomposition~\cite{Krien_SBE_original} in which the diagrams are classified in terms of their reducibility with respect to the bare interaction. 
Reducible diagrams can be interpreted as representing the exchange of a single boson corresponding to a collective fermionic excitation.
Diagrams that are not reducible, instead, can be further classified in terms of two-particle reducibility or reducibility with respect to the removal of a pair of Green-function lines ($GG$-reducibility).

Applying the SBE decomposition to systems with extended interactions requires special care~\cite{aleryani2024screeningeffectiverpalikecharge, Sarahthesis}.
In particular, we split the bare vertex into a bosonic part $\mathcal{B}$ and a fermionic part $\mathcal{F}$ as will be discussed now.

%--------------
\subsubsection{$\mathcal{B}+\mathcal{F}$ splitting}
%--------------

The generalization of the SBE to extended interactions requires to split the bare interaction $V_0$ into a bosonic and a fermionic part as
\begin{align}
    V^\X_0(q, k, k') = \mathcal{B}^\X(q) + \mathcal{F}^\X(q, k, k')\,, \label{eq:BF_splitting} 
\end{align}
in the spin-diagonalized basis, for the momentum-frequency conventions we refer to Appendix~\ref{app:cc}.
The index $\X\in \{\M, \D, \SC\}$ stands for the physical channel, i.e., the magnetic/spin channel (M), the density/charge channel (D), and the superconducting/pairing channel (SC).
The above splitting is determined uniquely by
\begin{align}
\lim_{i\nu,i\nu' \rightarrow 0}\int_{\bfk\bfk'}V^\X_0(q, k, k') = \mathcal{B}^\X(q).
\end{align} 
In particular, this ensures $\mathcal{B}$
(i)~depends only on a single bosonic transfer momentum-frequency $q$ and (ii)~contains the local and instantaneous part of the bare interaction $V_0$.
The bare interaction of our problem~\eqref{eq:bareint} is local in space, but retarded in time. 
In order to identify the instantaneous part, we Fourier transform and evaluate at $\tau = 0$,
\begin{align}
V_0(\tau\! =\!0) = \sum_{i\Omega} V_0(i\Omega) = U\! -\!V_H = U_\textrm{eff}\,, \label{eq:local_interaction}
\end{align}
where we have identified it with the effective interaction in the discussion of Section \ref{sec:model}. 
Isolating this local part, the bare interaction~\eqref{eq:bareint} can be written as 
\begin{align}
    V_0(i\Omega) = U_{\textrm{eff}}  + V_H\frac{\Omega^2}{\Omega^2 + \omega_0^2}\,,    \label{eq:nice_bare_interaction}
\end{align}
with $V_H$ and $U_{\textrm{eff}}$ as in Eqs.~\eqref{eq:vh} and~\eqref{eq:aalim}.
Translating this interaction into physical channels, we obtain the following natural splittings
\begin{subequations}
\begin{align}
    &\mathcal{B}_{}^\SC(q) =  U_\textrm{eff} \qq{} \\
    &\mathcal{B}_{}^\D(q) =  U_\textrm{eff} + 2V_{H} \frac{\Omega^2}{\Omega^2 + \omega^2_0} \qq{}\\
    &\mathcal{B}_{}^\M(q) =  -U_\textrm{eff}\label{eq:anti-adiabatic_splitting0}
\end{align}
\end{subequations}
and
\begin{subequations}
\begin{align}
    &\mathcal{F}_{}^\SC(k, k') = V_{H} \frac{(\nu - \nu' )^2}{(\nu - \nu' )^2 + \omega^2_0}  \\
    &\mathcal{F}_{}^\D(k, k') = -V_{H} \frac{(\nu-\nu')^2}{(\nu-\nu')^2 + \omega^2_0}  \\
    &\mathcal{F}_{}^\M(k, k') = -V_{H} \frac{(\nu - \nu')^2}{(\nu - \nu')^2 + \omega^2_0}\,.\label{eq:anti-adiabatic_splitting}
\end{align}
\end{subequations}

%------------
\subsubsection{Single-boson exchange formulation}
%------------

In the extension of the SBE to nonlocal interactions, the notion of $V_0$ reducibility is replaced by $\mathcal{B}$ reducibility.
The two-particle irreducible part is given by the bare interaction $V_{2\textrm{PI}} \approx V_0$, since the level-2 truncation does not generate two-particle irreducible diagrams.
The SBE decomposition of the vertex $V^\Lambda=V^\Lambda(k_1, k_2, k_3, k_4)$ then reads
\begin{align}
    V^\Lambda &=\nabla^\SC_{k_1 k_2}(k_1+k_3) - \nabla^\M_{k_1 k_2}(k_3-k_2) \nonumber \\
    &+\frac{1}{2}\left(\nabla^\D_{k_1 k_4}(k_2-k_1)  -
     \nabla^\M_{k_1 k_4}(k_2-k_1) \right) \nonumber \\
    & +M^\SC_{k_1 k_2}(k_1+k_3) - M^\M_{k_1 k_2}(k_3-k_2)  \nonumber \\
    &+\!\frac{1}{2}\!\left(M^\D_{k_1 k_4}(k_2\!-\!k_1)\!-\!M^\M_{k_1 k_4}(k_2\!-\!k_1)\right)\!-\!2U_\textrm{eff}\,, \label{eq:parquet_inSBE_physical}
\end{align}
where $\nabla^\X$ and $M^\X$ are scale-dependent quantities in the context of the renormalization group approach (we suppressed the $\Lambda$ index for convenience). 
More precisely, we defined the following quantities:
    $\nabla^\X_{kk'}(q)$ is the $\mathcal{B}^\X$-reducible vertex, representing the exchange of a single boson, and
    $M^\X_{kk'}(q)$ the $\mathcal{B}^\X$-irreducible vertex but $GG$-reducible in the channel $\X$. Diagrammatically, it represents multiboson exchanges.
    The term $-2U_\textrm{eff}$ results from summing the completely $GG$-irreducible part of the vertex (here taken to be $V_0$), the residual $\mathcal{F}$ diagrams that do not fall into either $\nabla$ nor $M$, and a contribution to account for double counting $\mathcal{B}$ diagrams in the bare part of $\nabla^\X$. Since the bare interaction $V_0$ of the problem is a pure density-density interaction, it was shown in Ref.~\cite{aleryani2024screeningeffectiverpalikecharge} that the result is simply $-2U_\textrm{eff}$ in analogy to the original SBE decomposition with $U_\textrm{eff}$ the \emph{local} part of the interaction~\eqref{eq:local_interaction}.
The important aspect of this decomposition is that the SBE parts can be written as 
\begin{align}
\nabla^\X_{kk'}(q) = \lambda_k^\X(q) w^\X(q) \lambda_{k'}^\X(q)\,,
\end{align}
where $\lambda_k^\X(q)$ and $w^\X(q)$ are scale dependent.
We stress that even though the bare interaction $V_0$ is nonlocal, $w^\X(q)$ depends only on one bosonic momentum-frequency and $\lambda^\X_k(q)$ only on one bosonic and one fermionic momentum-frequency~\footnote{If we had  considered reducibility with respect to $V_0$ like in the original SBE formulation instead of $\mathcal{B}^\X$, the object $w^\X$ would additionally depend on fermionic arguments $k$ and $k'$ which increases the numerical complexity and challenges its interpretation as a bosonic propagator.}.
This is because the decomposition is performed with respect to a purely bosonic object $\mathcal{B}^\X$.
In Hedin's formulation, $w^\X(q)$ and $\lambda_k^\X(q)$ correspond to the screened interaction and Hedin's fermion-boson vertex~\cite{PhysRev.139.A796}, respectively. 
Bosonic propagators are directly related to the $s$-wave susceptibilities
\begin{align}
w^\X(q) = \mathcal{B}^\X(q) - \left(\mathcal{B}^\X(q)\right)^2 \chi^\X(q)\,, \label{eq:swave_susceptibility_from_w}
\end{align}
which allows for a natural physical interpretation.

We note that the SBE decomposition provides a valuable tool in the quantum field-theoretic treatment of quantum many-body systems~\cite{Krien2020,Denz2020,Krien2020a,Krien2021,Krien2021b,Krien2022,Gievers_2022,Bonetti_2022,Adler2024,Kiese2024}.
It features a physically intuitive and also computationally efficient description of the relevant fluctuations in terms of processes involving the exchange of a single boson that describe a collective excitation and a residual part containing the 
multiboson processes.
At weak coupling, the effective bosonic interaction 
yields quantitatively accurate results, while the multiboson contributions are irrelevant and can be neglected~\cite{fraboulet2023singlebosonexchangefunctionalrenormalization}. 
This allows for a substantial reduction of the computational complexity of the vertex function: 
Since the multiboson processes are the only ones to depend on three independent momentum and frequency variables, neglecting them drastically reduces the computational complexity.
In contrast, the bosonic propagators and fermion-boson couplings depend on one and two independent arguments, respectively, and therefore their numerical treatment including full momentum and frequency dependence is much less demanding.

%-------------
\subsubsection{High-frequency asymptotics} 
%-------------

An important aspect in the numerical treatment and storage of the objects $\lambda^\X$ and $M^\X$ is their behavior at high frequencies. 
In general, in the $\mathcal{B}+\mathcal{F}$ splitted SBE it can be nontrivial, i.e., $\lambda^\X$ and $M^\X$  fail to decay to zero for large~$i\nu$ or~$i\nu'$ \cite{aleryani2024screeningeffectiverpalikecharge}. 
For $k^{(\prime)} = (i\nu^{(\prime)}, \bfk^{(\prime)})$, we have
\begin{subequations}
\label{eq:vertex_nontrivial_asymptotics}
    \begin{align}
\lambda^{\X,\text{asympt}}_{\bfk}(q) &= \lim_{i\nu \rightarrow \pm \infty} \lambda^\X_k(q)\\
M^{\X,\text{asympt}1}_{\bfk k'}(q) &= \lim_{i\nu \rightarrow \pm \infty} M^\X_{kk'}(q)\\
M^{\X,\text{asympt}1'}_{k \bfk'}(q) &= \lim_{i\nu' \rightarrow \pm \infty} M^\X_{kk'}(q)\\
M^{\X,\text{asympt}2}_{\bfk \bfk'}(q) &= \lim_{i\nu',i\nu \rightarrow \pm \infty} M^\X_{kk'}(q)\,,
\end{align}
\end{subequations}
where the last limit corresponds to taking the two limits  simultaneously. 
For the specific case of the Hubbard-Holstein interaction, a further simplification occurs: The asymptotics have only $s$-wave contributions; i.e., they are constant in $\bfk$ and $\bfk'$, because $\mathcal{F}^\X$ is. For the bosonic frequencies, the asymptotics are simply those of their bare values
\begin{subequations}
\begin{align}
&\lim_{i\Omega \rightarrow \pm \infty}w^\X(i\Omega) = \lim_{i\Omega \rightarrow \pm \infty} \mathcal{B}^\X(q) = U_\textrm{eff}^X\\
&\lim_{i\Omega\rightarrow \pm \infty}\lambda_k^\X(q) = 1\\
&\lim_{i\Omega\rightarrow \pm \infty}\hspace{-2mm}M^\X_{kk'}(q) = 0\,.
\end{align}
\end{subequations}
Knowing the asymptotic behavior of these objects is essential as it allows us to compute the quantities on finite frequency windows. 
Outside 
these windows, the objects are substituted by their asymptotes. 
The least complicated objects $w^\X$ are also the most important ones: They are directly related to the susceptibilities under investigation in this work.
They are also the most sizable contributions when $s$-wave fluctuations dominate. Therefore, the window for $w^\X$ is chosen to be the largest. 
The fermion-boson coupling $\lambda^\X$ is more complicated in comparison, depending on two frequencies, and therefore it comes second in the size of the window on which it is computed. 
Finally $M^\X$ is computed on the smallest frequency window. 
More details are presented in Appendix~\ref{app:technical_impelementation}.

%--------------
\subsubsection{Flow equations}
%--------------
\label{sec:flow_equations}
While the previous subsections apply in general and not only within the fRG, we now present the fRG specific flow equations.
The decomposition of the vertex into  channels has the advantage that the essential components 
depend most strongly on their native bosonic argument~\cite{Husemann2012}. 
In the TU approach~\cite{PhysRevB.101.155104,Lichtenstein2017}, the fermionic momenta are therefore expanded in terms of form-factors~$f_n(\bfk)$. 
In this study, we exclusively consider the $s$-wave and the $d$-wave form factors
\begin{subequations}
\label{eq:ff}
\begin{align}
f_{\swave}(\bfk) &= 1\,,\\ f_{\dwave}(\bfk) &= \cos(\bfk_x) - \cos(\bfk_y)\,,
\end{align}
\end{subequations}
and higher orders are neglected. 
The flow equations for $w^\X$, $\lambda^\X$, and $M^\X$ are given by
\begin{widetext}
\begin{subequations}
\label{eq:sbe_flow_equations}
    \begin{align}
        \mathrm{d}_\Lambda w^\mathrm{X}(q) &= -\mathrm{Sgn}\,\mathrm{\X}\left(w^\mathrm{X}(q)\right)^2\sum_{\mathclap{m,m',\nu}}\lambda^\mathrm{X}_m(q,i\nu)\mathrm{d}_\Lambda \Pi^\mathrm{X}_{m m'}(q,i\nu)\lambda^\mathrm{X}_{m'}(q,i\nu), \\
        \mathrm{d}_\Lambda \lambda^\mathrm{X}_n(q,i\nu) &= -\mathrm{Sgn}\,\mathrm{\X}\sum_{\mathclap{m,m',i\nu'}} \lambda^\mathrm{X}_m(q,i\nu')\mathrm{d}_\Lambda \Pi^\mathrm{X}_{m m'}(q,i\nu')\mathcal{I}^\mathrm{X}_{m'n}(q,i\nu',i\nu), \\
        \mathrm{d}_\Lambda M^\mathrm{X}_{nn'}(q,i\nu,i\nu')& = -\mathrm{Sgn}\,\mathrm{\X}\sum_{\mathclap{m,m',i\nu''}} \mathcal{I} ^\mathrm{X}_{nm}(q,i\nu,i\nu'')\mathrm{d}_\Lambda \Pi^\mathrm{X}_{mm'}(q,i\nu'')\mathcal{I} ^\mathrm{X}_{m' n'}(q,i\nu'',i\nu')\,,
        \end{align}     
        \end{subequations}
\end{widetext}
with $\mathrm{Sgn}\,\mathrm{SC}\! =\! -\mathrm{Sgn}\,\mathrm{D}\! =\! -\mathrm{Sgn}\,\mathrm{M}\! =\! +1$ and $\mathcal{I}^\X\! =\! V^\X - \nabla^\X$.
All quantities depend on the scale $\Lambda$.
The expression for the TU bubble reads
\begin{align}  
   \Pi^\X_{mm'}(q, i\nu) =& \mathrm{Sgn}\,\X  \int\limits_{k} G^\Lambda(q - (\mathrm{Sgn}\,\X) k) G^\Lambda(k)\nonumber\\
   &\qquad\qquad\times f_m(k) f^*_{m'}(k)\,.
 \label{eq:bubble_definition}
\end{align}
At the beginning of the flow, the initial conditions
for the vertex are given by its bare value $V_0$, i.e.
\begin{subequations}
\label{eq:flow_initial_values}
\begin{align}
    w^{\X,\Lambda = 0}(q) &= \mathcal{B}^\X(q),\\
    \lambda_n^{\X,\Lambda=0}(q, i\nu) &= \delta_{n,\swave},\\
    M^{\X,\Lambda = 0}_{nm}(q,i\nu, i\nu') &= 0\,.    \end{align}
\end{subequations}
The nontrivial asymptotic objects in Eq.~\eqref{eq:vertex_nontrivial_asymptotics} are calculated by including also a flow for them.
These flow equations are determined by the corresponding high-frequency limits of Eq.~\eqref{eq:sbe_flow_equations}, with 
the initial conditions obtained from the asymptotics of Eq.~\eqref{eq:flow_initial_values}.

In previous fRG applications in the SBE formulation, the flow of the rest functions $M^\X$ turned out to be negligible. 
The resulting SBE approximation provides an efficient computation scheme, allowing us to perform scans of large regions in parameter space.
In the pure Hubbard model the inclusion of the rest function leads only to small quantitative changes of the results away from the Fermi-surface instabilities, both at weak~\cite{fraboulet2023singlebosonexchangefunctionalrenormalization} and strong coupling~\cite{Bonetti_2022}. 
This finding can be directly transferred to the Hubbard-Holstein model only in the anti adiabatic limit.
For a systematic analysis of the relevance of the rest functions at a finite phonon frequency $\omega_0$, we refer to Appendix~\ref{app:technical_impelementation}.
We find that approaching the adiabatic limit, the impact of the rest functions becomes quantitatively more and more important.
However, qualitatively, e.g., on the type of the leading susceptibility, the results remain unaffected by their inclusion. 
This is in contrast to our findings for the extended Hubbard model~\cite{aleryani2024screeningeffectiverpalikecharge}, where the quality of the SBE approximation is comparable to the pure Hubbard model.

For this reason, we here take into account the rest function in almost all  calculations, the only exception is the large parameter scans presented in Figs.~\ref{fig:phase_diagram_halffilling} and~\ref{fig:phase_diagram_doped}.
For the latter, we expect small quantitative changes upon inclusion of the rest functions, but qualitative features will remain unaltered.

%-------------
\subsection{Contributions from the three-particle vertex and Migdal-Eliashberg theory}
%-------------

We now comment on the partial inclusion of three-particle vertex contributions in the so-called Katanin substitution~\cite{Katanin_substitution_2004}, which allows to recover the ME theory.
In the flow equation~\eqref{eq:sbe_flow_equations}, the full derivative of the bubble $\mathrm{d}_\Lambda \Pi$ involves the one of the propagator 
\begin{align}
    \mathrm{d}_\Lambda G^\Lambda(k) = S^\Lambda(k) + G^\Lambda(k)\mathrm{d}_\Lambda \Sigma^\Lambda(k)G^\Lambda(k)\,.
    \label{eq:katanin_substitution}
\end{align}
It is composed of the single-scale propagator and a term involving the self-energy derivative [cf. Eq.~\eqref{eq:selfenergy_flow_eqn}]. 
In the standard level-2 truncation, the derivative of the self-energy $\mathrm{d}_\Lambda \Sigma$ in the second term is neglected. 
Here, we consider the Katanin correction~\cite{Katanin_substitution_2004} including the self-energy insertions in Eq.~\eqref{eq:katanin_substitution}. 
It leads to an improved fulfillment of the Ward identities, since it partly takes into account the flow of the three-particle vertex $\mathrm{d}_\Lambda \Gamma^{(3)}$, otherwise considered only 
in the level-3 truncation of the flow equation hierarchy. 
The Katanin correction is particularly important in the fRG treatment of electron-phonon models in order to fully enclose the physics of ME theory~\cite{campbell_shankar_rg_me_theory,Honerkamp_2005} (see Appendix~\ref{app:ME}). 
In fact, the unbiased treatment of fluctuations from all channels on equal footing characteristic of the fRG will allow us
to go beyond ME theory.

%------------------
\section{Susceptibility analysis}
\label{sec:results}
%------------------

%------------------
\subsection{Definitions}
%------------------

%--------------
\begin{figure}[t]
    \centering
    \includegraphics[width=\linewidth]{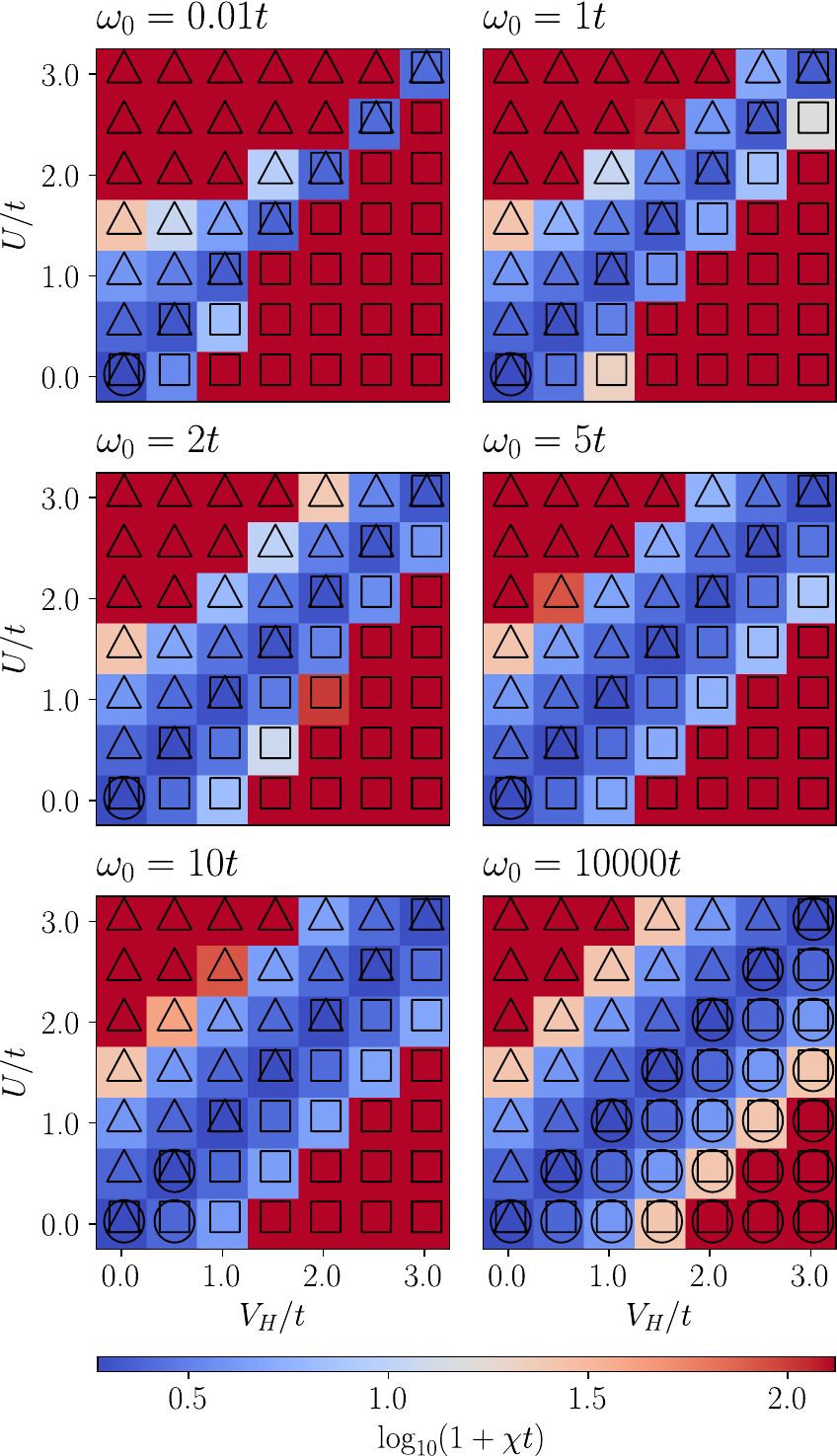}
    \caption{Leading $s$-wave susceptibilities at half filling for $t'=0$. The superconducting $s$-wave susceptibility
    $\chi^{\SC}$ is indicated by $\bigcirc$, the $s$-wave density susceptibility $\chi^\D$ by $\square$, and the $s$-wave magnetic susceptibility $\chi^{\M}$ by $\triangle$.
    Parameters are $\beta = 20/t$, $t' = 0$ and $\mu = 0$. Different phonon frequencies $\omega_0$ are indicated above the panels. Overlayed symbols 
    correspond to equal values.}\label{fig:phase_diagram_halffilling}
\end{figure}
%--------------

%--------------
\begin{figure}[t!]%[htbp]
    \centering
    \includegraphics[width=\linewidth]{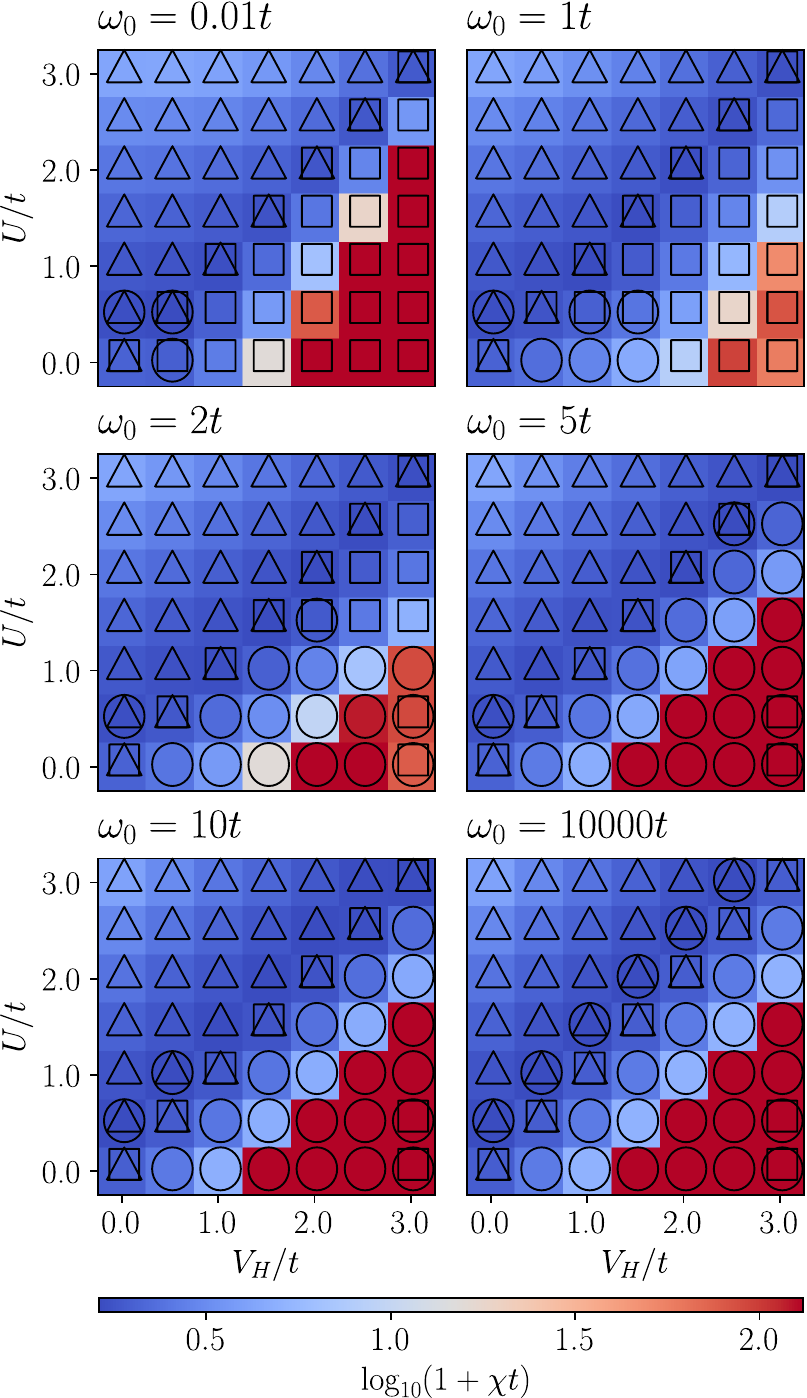}
    \caption{Leading $s$-wave susceptibilities at finite doping.
    We consider a filling $n=0.39$ with model parameters $t' = -0.25t$ and $\mu = 4t'$, corresponding to van-Hove filling of the noninteracting Fermi surface at $\beta = 20/t$.
    Symbols are chosen as in Fig.~\ref{fig:phase_diagram_halffilling}.}
    \label{fig:phase_diagram_doped}
\end{figure}
%--------------

In Matsubara-frequency space, the susceptibility, describing the linear response of a system to a small external perturbation, is defined via the Fourier transform with respect to imaginary time $\tau$ as 
\begin{align}
\chi^O(q) = &\int_0^\beta \dd{\tau } e^{i\Omega \tau}\left(\expval{T_{\tau} \hat{O}(\bfq, \tau) \hat{O}(\bfq, 0) } \right.\nonumber \\  &\qquad\qquad\qquad-\left. \expval{\hat{O}(\bfq, \tau)}\expval{\hat{O}(\bfq, 0)}\right)\,.
\end{align}
Here, the operator $\hat{O}$ is
the spin operator in the $z$~direction, the density operator, or the pairing operator, i.e.,
\begin{subequations}
\begin{align}
    \hat{s}_z(\bfq, \tau) =& \frac{1}{2}\left(\hat{n}_{\uparrow}(\bfq, \tau) - \hat{n}_{\downarrow}(\bfq, \tau)\right)\\
\hat{\rho}(\bfq, \tau) = &\frac{1}{2}\left(\hat{n}_{\uparrow}(\bfq, \tau) + \hat{n}_{\downarrow}(\bfq, \tau)\right)\\
\hat{\Delta}_{\swave/\dwave}(\bfq, \tau) =&\frac{1}{2} \int_\bfk f_{\swave/\dwave}(\bfk)(c^\dagger_{\uparrow}(\bfq - \bfk, \tau)c^{\dagger}_{\downarrow}(\bfk, \tau)\nonumber\\ 
& \qquad +c_{\uparrow}(\bfq - \bfk, \tau)c_{\downarrow}(\bfk, \tau))
\end{align}
\end{subequations}
for the magnetic (or spin) $\chi^\M$, density (or charge) $\chi^\D$, and the superconducting $\chi^\SC$ channel, 
respectively.

In form-factor notation, these susceptibilities are computed at the end of the flow~\footnote{The small quantitative differences with respect to the flowing results are due to the $1\ell$ 
truncation of the fRG~\cite{Hille2020}. This inconsistency at the two-particle level can be resolved by including multiloop corrections \cite{Kugler2018a,Kugler2018b,Tagliavini_2019}.} from
\begin{align}
\chi^\X_{nn'}(q) = \chi^{\X, \mathrm{bubble}}_{nn'}(q) + \chi_{nn'}^{\X, \mathrm{vertex}}(q) \,,\label{eq:susc_from_vertex}
\end{align}
determined by the bubble contribution
\begin{align}
\chi^{\X, \text{bubble}}_{n n'}(q) = \sum_{i\nu} \Pi^{\X}_{n n'}({q}, i\nu)\,,
\end{align}
cf. Eq.~\eqref{eq:bubble_definition} for~$\Pi^\X$, and the vertex correction
\begin{align}
 \chi^{\X, \mathrm{vertex}}_{n n'}(q) = -\sum_{\mathclap{m, m', i\nu, i\nu'}} \Pi^{\X}_{n m}({{q}}, i\nu) V^{\X}_{m m'}({{q}}, i\nu, i\nu') \Pi^{\X}_{m' n'}({q}, i\nu')\,.
 \label{eq:susc_vertex_contribution}
\end{align}
The channel-parametrized form of the vertex~$V^\X$ is explicitly shown in Appendix~\ref{app:cc}, see Eq.~\eqref{eq:physical_vertices}. 
Here, we focus on the $s$-wave susceptibilities $\chi^\X = \chi^\X_{\swave,\swave}$ for $\X = \M$, $\D$, and $\SC$ and the $d$-wave susceptibility $\chi^\dSC = \chi^\SC_{\dwave,\dwave}$.

%--------------
\subsection{Overview and evolution with $\omega_0$} 
%--------------

We first provide a descriptive overview of our numerical results as a function of the local Coulomb repulsion~$U$ and the phonon-mediated  electron-electron interactions~$V_H$, for different values of the phonon frequency~$\omega_0$.
In practice, a modification of the phonon frequency can be achieved by replacing certain atoms in the crystal by their isotopes due to the relation $\omega_0 \sim 1/ \sqrt{M_\textrm{ion}}$, cf. Eq.~\eqref{eq:massdep}.
The impact of such a change of ion mass on physical observables is known as the isotope effect.
Employing fRG methods, the isotope effect has been studied in Ref.~\cite{anomalous_isotope_effect} with a focus on its influence on the formation of order and the corresponding transition temperatures~$T_c$.
There, the frequency dependence of the vertex as well as the self-energy (both electron and phonon) were neglected.
Here, we include both.
Specifically, we performed calculations for a range of parameters $(U, V_H) \in [0, 3]\times[0, 3]$ at temperature $\beta = 20/t$ for two initial conditions of the Fermi surface: (1)~the half-filled case with perfect nesting, i.e, $t'=0$ and $\mu=0$ and (2)~the case of finite doping with $t'=-0.25$ and $\mu=4t'$, corresponding to a filling $n=0.39$.
In both cases, the noninteracting Fermi surfaces include van-Hove singularities.

In models where the interactions are uniform in space ($s$-wave in nature), e.g., the Hubbard-Holstein model, the $s$-wave fluctuations play a primary role in determining the physics of the system and in potentially driving higher harmonic fluctuations. 
Therefore, we will begin our analysis in Figs.~\ref{fig:phase_diagram_halffilling} and ~\ref{fig:phase_diagram_doped} focusing on the $s$-wave susceptibilities. 
The important $d$-wave superconducting susceptibility will be discussed later below.
In Fig.~\ref{fig:phase_diagram_halffilling}, for the case of half filling at perfect nesting, we show the leading $s$-wave susceptibilities, i.e., the maxima of the dominant susceptibilities [cf. Eq.~\eqref{eq:susc_from_vertex}]. 
%R
Similarly, in Fig.~\ref{fig:phase_diagram_doped}, we show the case of finite doping.
We note that, here, we have neglected the rest functions; see the discussion in Sec.~\ref{sec:flow_equations}.
In both figures, dark red areas represent regions where $\text{max}\, \chi^\X \gtrsim 10^2$ and hence signal an instability toward ordering in the channel $\X$. 
Blue areas, instead, correspond to smaller values where the different channels are typically of comparable size.
On the diagonal $U = V_H$, the effective local interaction vanishes $U_{\mathrm{eff}} = U - V_H = 0$ and thus one expects the physics of the region above with $U_{\mathrm{eff}} > 0$ to display a behavior characteristic of the repulsive Hubbard model where (antiferromagnetic) magnetic ordering prevails. 
Similarly, the region below with $U_{\mathrm{eff}} < 0$ will tentatively exhibit features  
of the attractive Hubbard model with dominant charge order and superconductivity. 
In the anti adiabatic limit, see Eq.~\eqref{eq:aalim}, this statement is exact as can be seen in the lower right panels corresponding to $\omega_0 = 10\,000t$. 
As a consequence, the only scale appears to be the distance to the diagonal.
At half filling, the degeneracy of $\chi^\SC(0,0)=\chi^\D(\pi,\pi)$  due to SU(2)$_P$ symmetry~\cite{general_shiba} is evident.
At finite doping, instead, the leading susceptibility is of superconducting type except for the largest values of $U_{\mathrm{eff}} = U - V_H$ at $U=3t$ and small $V_H$.

Decreasing $\omega_0$, the retardation induced by the phonons leads to the following effects: 
For $U_{\mathrm{eff}} > 0$, $\chi^\M$ is enhanced at half filling, where the red area in Fig. \ref{fig:phase_diagram_halffilling} becomes larger for smaller values of $\omega_0$ (the region spanned by the triangles is however not affected).
This is plausible by the overall larger effective interaction \eqref{eq:alim} with respect to the adiabatic limit, where $U_\textrm{eff}=U-V_H$. At finite doping, this effect is not visible in Fig. \ref{fig:phase_diagram_doped}.
Similarly, for $U_{\mathrm{eff}} < 0$ we observe  $\chi^\D$ to be enhanced upon reducing $\omega_0$. 
This is clearly visible at finite doping, while at half filling the region spanned by the squares is unaffected by $\omega_0$, but with a larger red area. 
More prominently, $\chi^\SC$ is suppressed. At half filling, the degeneracy between $\chi^\SC$ and $\chi^\D$ is lifted 
for smaller $\omega_0$ (since the phonons couple to the density fluctuations), leading to a reduced $\chi^\SC$ with respect to $\chi^\D$ except for $U=V_H=0$.
In contrast, at finite doping, the region spanned by the circles is gradually reduced until it vanishes in the adiabatic limit, except for the region around $0$.
A more detailed understanding of this behavior will be provided in Section \ref{sec:fluct}, where we carry out a detailed fluctuation diagnostics.

%-----------------
\begin{figure}[t!]
    \centering
    \includegraphics[width=\linewidth]{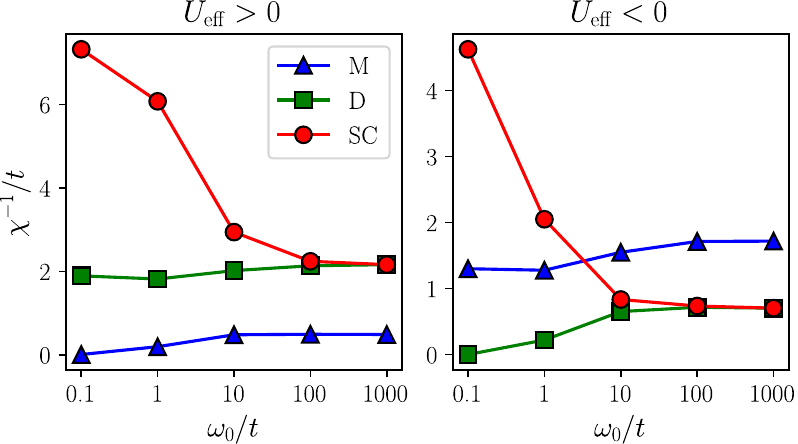} 
    \caption{Inverse ($s$-wave) susceptibilities, evaluated at the respective maxima, 
    as a function of $\omega_0$, at $\beta=20/t$ and half filling. Left panel: $U_\textrm{eff} > 0$ for $U=2.2t$ and $V_H = 1.4t$. Right panel: $U_\textrm{eff} < 0$ for $U=1.5t$ and $V_H = 2t$. 
    }
\label{fig:halffilling_inverse_susceptibility_omega0_influence}
\end{figure}
%-----------------

Next, in Fig.~\ref{fig:halffilling_inverse_susceptibility_omega0_influence} we examine the half-filling case more closely and focus on two representative values of the parameters for which $U_\textrm{eff}<0$ and $U_\textrm{eff}>0$ and trace the evolution of the different susceptibilities across a range of phonon frequencies from close to the adiabatic to near the anti adiabatic limit, i.e.,  $\omega_0 = 0.1t, 1t, 10t, 100t, 1000t$.
Here, we also take into account the rest function.
The inverse $s$-wave superconducting, density, and magnetic susceptibilities are evaluated at their respective maxima.
We observe that the density and magnetic susceptibilities exchange their roles in the two regimes of $U_\textrm{eff}>0$ and $U_\textrm{eff}<0$ shown in the left and right panels, respectively.
Moreover, the degeneracy of $\chi^\D$ and $\chi^\SC$ in the anti adiabatic limit is gradually lifted for decreasing $\omega_0$. 
Regardless of the sign of $U_{\mathrm{eff}}$, tuning down the value of $\omega_0$ toward the adiabatic regime, we find that the retardation with smaller $\omega_0$ leads to a tentative increase of the density and magnetic susceptibilities, i.e., a decrease of $(\chi^\D)^{-1}$ and $(\chi^\M)^{-1}$, in agreement with the overall increase of the effective interaction $U_\textrm{eff}$ from Eq.~\eqref{eq:aalim} in the anti adiabatic limit to Eq.~\eqref{eq:alim} in the adiabatic one. 
At the same time, $(\chi^\SC)^{-1}$ increases, as expected from BCS theory, where  
the critical temperature for the phonon-mediated superconductivity is proportional to~$\omega_0$.

%-----------------
\begin{figure}[t!]
    \centering
    \includegraphics[width=\linewidth]{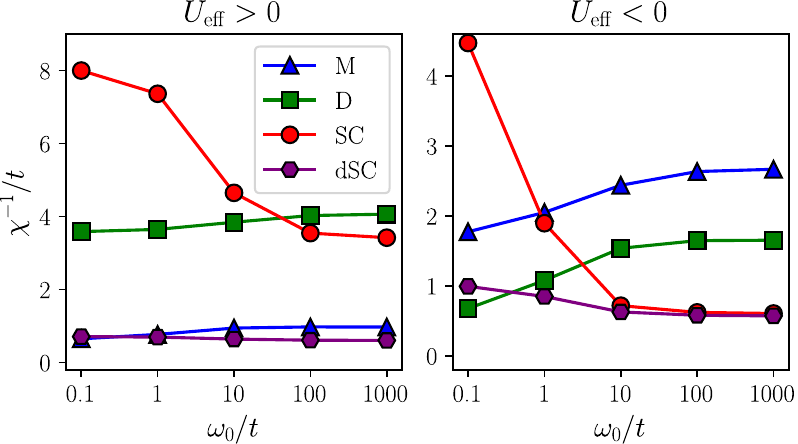} 
    \caption{Inverse susceptibilities as in Fig. \ref{fig:halffilling_inverse_susceptibility_omega0_influence}, but at finite doping with filling $n =0.39$ , for $t'=-0.25t$ and $\mu=4t'$. 
    Left panel: $U_\textrm{eff} > 0$ for $U=2.5t$ and $V_H = 1t$. Right panel: $U_\textrm{eff} < 0$ for $U=1.7t$ and $V_H = 2.2t$. 
    }
\label{fig:doped_inverse_susceptibility_omega0_influence}
\end{figure}
%-----------------

The corresponding results for finite doping are provided in Fig.~\ref{fig:doped_inverse_susceptibility_omega0_influence}. 
For this case, we also compute the $d$-wave superconducting susceptibility $\chi^\dSC$ (not reported in Fig.~\ref{fig:phase_diagram_doped}). 
While the general trends for the other susceptibilities are the same as at half filling except for the lifted degeneracy between $\chi^\D$ and  $\chi^\SC$ due to the broken SU$(2)_P$ symmetry, $\chi^\dSC$ appears to be the leading susceptibility
with $(\chi^\dSC)^{-1}$ being the smallest for $\omega_0\gtrsim 0.5$. 
Moreover, $(\chi^\dSC)^{-1}$ increases as $\omega_0$ is lowered, independently of whether $U_\textrm{eff}>0$ or $U_\textrm{eff}<0$. 
This result may seem counterintuitive considering that the magnetic fluctuations increase as $\omega_0$ is lowered and these are believed to drive $d$-wave pairing in the Hubbard model (note that for $\beta=20/t$ and $U=3t$, we are still far from a divergence of $\chi^\dSC$, i.e., a transition into a $d$-wave superconducting phase). 
Moreover, this finding is also in contrast with a recent fRG study on the isotope effect~\cite{anomalous_isotope_effect}, which reported an increase of $T_c$ as $\omega_0$ is decreased. 

%-------------
\begin{figure}[t]
    \centering
    \includegraphics[width=\linewidth]{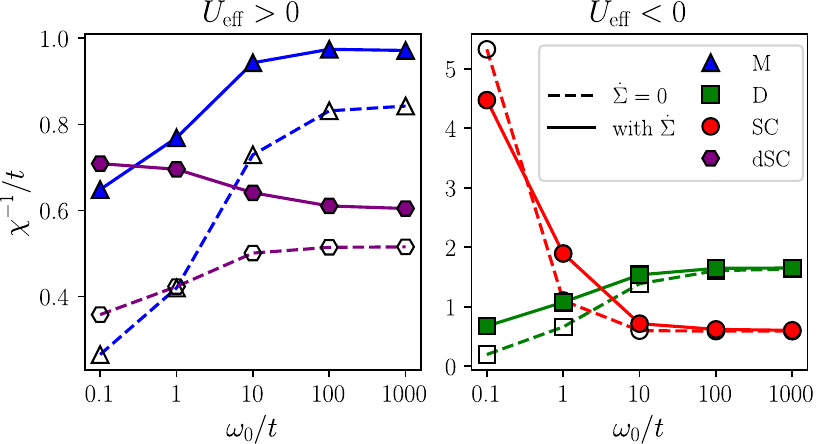}
    \caption{Close-up of inverse susceptibilities
    as in Fig.~\ref{fig:doped_inverse_susceptibility_omega0_influence}, with (solid lines) and without (dashed lines) self-energy flow. 
    Only the leading channels are displayed, i.e., $s$-wave and density superconducting for $U_\textrm{eff}<0$ (right panel) and $d$-wave superconducting and magnetic for $U_\textrm{eff}>0$ (left panel).
    Note the qualitative change of behavior in the $d$-wave superconducting susceptibility, leading to a sign change of the isotope effect.
    }    \label{fig:effect_of_selfenergy_on_isotope_effect}
\end{figure}
%-------------

This apparent contradiction is resolved by a careful analysis of the self-energy effects which were neglected in Ref.~\cite{anomalous_isotope_effect}:
For comparison, we show data for the susceptibilities as a function of $\omega_0$ with and without self-energy in Fig.~\ref{fig:effect_of_selfenergy_on_isotope_effect}.
We find that the growth of $(\chi^\dSC)^{-1}$ is induced by
the strong renormalization of the single-particle propagator; see also Ref.~\cite{heinzelmann2023entangledmagneticchargesuperconducting} where such an effect is observed for the magnetic susceptibility. 
In the absence of the self-energy corrections, $(\chi^\dSC)$ is enhanced in the adiabatic limit.
This qualitative change highlights the essential role played by the self-energy, which has to be taken into account for a correct description of the physical behavior, see also Ref.~\cite{macridin_suppression_2012}.
This is different for the $s$-wave superconducting and density susceptibilities, where the self-energy leads only to quantitative corrections: 
While in the anti adiabatic limit the self-energy feedback increases both $(\chi^\SC)^{-1}$ and $(\chi^\D)^{-1}$, we find opposite effects in proximity of the adiabatic limit where the self-energy increases $(\chi^\D)^{-1}$ but decreases $(\chi^\SC)^{-1}$. 
As we will argue later, this has a more profound reason due to charge density and $s$-wave superconducting fluctuations playing against one another in this limit, see Sec.~\ref{sec:fluct}.

%-------------
\subsection{Phonon-induced density fluctuations effects}
%-------------

We now explore the impact of the phonons on the electronic problem.
To that end, we denote the density susceptibility in absence of phonons ($V_H = 0$) as $\tilde{\chi}^\D=\tilde{\chi}^\D(q)$.
Then, introducing phonons induces a renormalization of the susceptibility, which in random phase approximation (RPA) is given by
\begin{align}
\chi^\D_\textrm{RPA} = \frac{\tilde{\chi}^\D}{1 + \abs{g}^2D_0 \tilde{\chi}^\D}\,,
\end{align}
with $D_0$ being the bare phonon propagator [cf. Eq.~\eqref{eq:bosprog}]. In the fRG, the RPA is recovered by neglecting the flow of the fermion-boson coupling, i.e., $\lambda^D\equiv 1$.
The inverse reads
\begin{align}
(\chi^\D_\textrm{RPA})^{-1} 
&= (\tilde{\chi}^\D)^{-1} + V_H \frac{\omega_0^2}{(i\Omega)^2 - \omega_0^2}\,.
\end{align}
For $i\Omega = 0$, the inverse density susceptibility hence has a linear dependence on the phonon-induced effective electron-electron coupling
\begin{align}
    (\chi^\D_\textrm{RPA})^{-1}= (\tilde{\chi}^\D)^{-1} - V_H \,.\label{eq:rpa_density_behaviour}
\end{align}
Deviations from the RPA can then be assigned to contributions from other channels, which feed back into the density channel through renormalization of~$\lambda^\D$. 
We discuss this interchannel feedback in the following. 

%----------------
\begin{figure}[t!]
    \centering
\includegraphics[width=\linewidth]{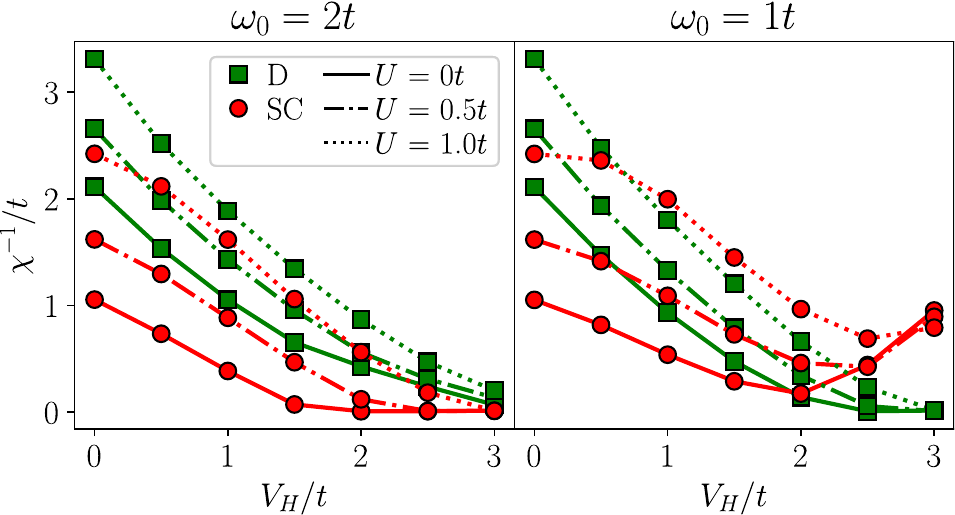}
    \caption{Evolution of the inverse density and $s$-wave superconducting susceptibility with $V_H$, for different values of $U$, corresponding to cuts in the phase diagram of Fig. \ref{fig:phase_diagram_doped} for finite doping. 
    The inverse density susceptibility departs from the linear behavior predicted by RPA for $V_H\gtrsim 1.5t$.
    Approaching the adiabatic regime at $\omega_0=1t$, the phonon-induced density fluctuations  
    suppress $s$-wave superconductivity, 
    i.e. lead to an increasing $(\chi^\SC)^{-1}$.
    }
    \label{fig:sc_d_suscs_vs_Vh_doped}
\end{figure}
%----------------

Figure~\ref{fig:sc_d_suscs_vs_Vh_doped} displays the inverse density and superconducting  
susceptibilities as a function of~$V_H$, for different values of~$U$. 
We consider $\omega_0 = 1t$ and $\omega_0 = 2t$ where the evolution with $V_H$ can be traced over the whole range of $V_H\in[0,3t]$ with no divergence occurring.
For all values of $U$, we find that for small $V_H\lesssim 1.5t$, the inverse of $\chi^\D$ exhibits the linear behavior expected from RPA~\eqref{eq:rpa_density_behaviour}. For larger values of $V_H$, we observe a clear deviation from the RPA prediction which is more pronounced for $\omega_0 = 1t$. 
We now turn to the $s$-wave susceptibility~$\chi^\SC$.
In the anti adiabatic regime ($\omega_0=2t$) where the system is effectively described by the attractive Hubbard model, doping the system destroys nesting and gives $\chi^\SC$ an edge over $\chi^\D$ (see also Figs.~\ref{fig:phase_diagram_doped} and~\ref{fig:doped_inverse_susceptibility_omega0_influence}).
Increasing $V_H$, i.e., increasing the effective attractive interaction $U_\textrm{eff} = U - V_H$,  
reduces both, the inverse of $\chi^{\SC}$ and $\chi^{\D}$.
A similar effect is observed upon lowering $U$.
Once $(\chi^{\D})^{-1}$ is sufficiently small, i.e., $\chi^{\D}$ is sufficiently close to the divergence, the strong density fluctuations appear to affect the behavior of $(\chi^{\SC})^{-1}$ by turning its decrease into an increase, i.e., by driving the observed suppression of $\chi^{\SC}$.
This sets in when approaching the adiabatic regime, where sufficiently small $\omega_0$ make $\chi^\D$ eventually prevail, and is clearly visible
for~$\omega_0=1t$.

%-----------------
\begin{figure}[t!]
    \centering
    \includegraphics[width=\linewidth]{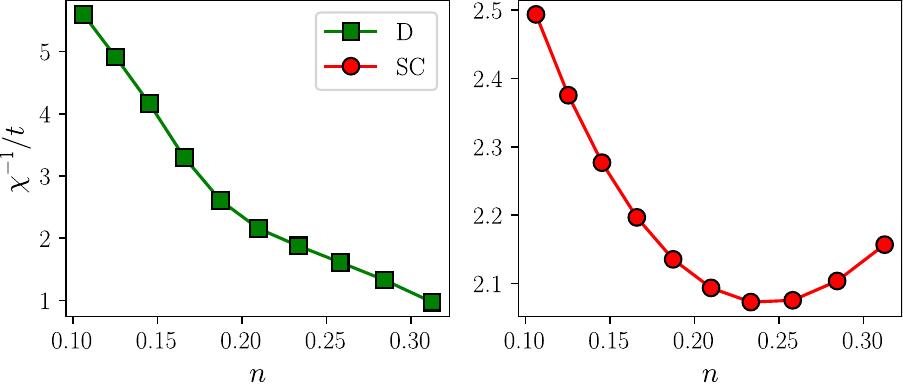}
    \caption{Evolution of the maxima of the superconducting and charge density susceptibilities as a function of filling, for $\beta = 5/t$, $V_H = 3t$, $U = 0$, $\omega_0 = 1.5t$ and $t' = 0$. As we approach half filling ($n = 0.5$), the charge density fluctuations grow uniformly, whereas the $s$-wave pairing exhibits 
    a maximum.}
    \label{fig:sc_susc_vs_filling}
\end{figure}
%-----------------

Interestingly, this effect has also been observed in quantum Monte Carlo simulations at $U = 0$ and for $t' = 0$, see Ref.~\cite{MARSIGLIOreview2020168102}. 
We here demonstrate this to be a general feature also at finite~$U$ and in presence of a nearest-neighbor hopping amplitude.
In the anti adiabatic regime, $(\chi^{\D})^{-1}$ and $(\chi^{\SC})^{-1}$ are of the same size and it is difficult to see the crossover from the RPA behavior as in Fig.~\ref{fig:sc_d_suscs_vs_Vh_doped} for $\omega_0=1t$.
A more detailed analysis through a fluctuation diagnostics will be provided in Sec.~\ref{sec:fluct}.

Another way to observe the phonon-induced suppression of $\chi^\SC$ is by tuning the doped system toward half filling~\cite{MARSIGLIOreview2020168102}.
For fRG results in a similar parameter regime see Fig.~\ref{fig:sc_susc_vs_filling}. 
While $\chi^\D$ increases as expected, $\chi^\SC$ exhibits a nonmonotonic behavior.
This is in contrast to the conventional 
ME theory, for which the $s$-wave superconducting critical temperature  prediction, $T_c \sim \sqrt{V_H}$~\cite{MARSIGLIOreview2020168102}, entails a monotonic increase of $\chi^\SC$ with~$V_H$.
This breakdown of ME theory~\cite{bounds_on_superconducting_Tc_2018,Sadovskii2025} observed in various QMC studies~\cite{flat_bands_phonon_non_monotonic_SC_Tc,MARSIGLIOreview2020168102} was shown to originate from the phonon renormalization~\cite{marsiglio2021phononselfenergyeffectsmigdaleliashberg}. 
In renormalized ME theory this is taken into account by an additional equation for the phonon self-energy which allows to recover the downturn of $\chi^\SC$, see Ref.~\cite{berger_two-dimensional_1995}. 
In the language where phonons have been integrated out, the phonon self-energy effects are encoded in the density fluctuations that renormalize the phonon-mediated bare interaction.

%------------
\subsection{Phonon softening and  
lattice instability}
\label{sec:phonon_softening}
%------------

Lattice vibrations affect electronic correlations, and vice versa.
Although, in our treatment, we have integrated out the phonons, we can still extract information on phonon renormalization from a solution of the purely electronic problem. 
A general effect induced by phonon renormalization is the phenomenon of softening, where the phonon dispersion $\omega(q)$ is renormalized to $\omega_\textrm{eff} (q)$.  
In the RPA, $\omega_\textrm{eff}(\bfq)$ can be estimated \cite{manybody_wonderful_book_flensberg,PhysRevResearch.5.013218} through the electronic corrections to the bare phonon propagator, cf. Eq.~\eqref{eq:bosprog}, as
\begin{align}
\label{eq:D_RPA}
D_\textrm{RPA}
&= \frac{D_0}{1 + \abs{g}^2 \tilde{\chi}^\D D_0 }= \frac{2\omega_0}{(i\Omega)^2 - \omega^2_\textrm{RPA,eff}}
\end{align}
with the effective dispersion
\begin{align}
\omega_\textrm{RPA,eff} := \omega_0 \sqrt{1 - 2\abs{g}^2 \tilde{\chi}^\D(q)/\omega_0}\,. \label{eq:RPA_dispersion}
\end{align}
The effective dispersion $\omega_{\textrm{RPA,eff}}(i\Omega=0, \bfq)$ vanishes if 
\begin{align}
    2\abs{g}^2 \tilde{\chi}^\D(i\Omega= 0, \bfq)/\omega_0 = 1
    \label{eq:lattinst}
\end{align}
or equivalently $V_H\tilde{\chi}^\D(i\Omega = 0, \bfq)= 1$. 
This is indeed possible as $V_H\tilde{\chi}^\D$ is unbounded, implying that for sufficiently large $g$ (or small $\omega_0$), $\omega_\textrm{RPA,eff}$ will become imaginary. 
Therefore, it has been suggested~\cite{PhysRevResearch.5.013218,PhysRevB.84.104506} that Eq.~\eqref{eq:lattinst} corresponds to a lattice instability even if $\tilde{\chi}^\D$ is finite. 
However, a more accurate treatment of the phonon renormalization within renormalized ME theory~\cite{berger_two-dimensional_1995} was shown to prevent its occurrence.
Here, we illustrate that this problem does not appear for the exact expression of the phonon self-energy for which we present a consistent derivation in the following.

%----------------
\begin{figure}[t!]
    \centering   \includegraphics[width=1.0\linewidth]{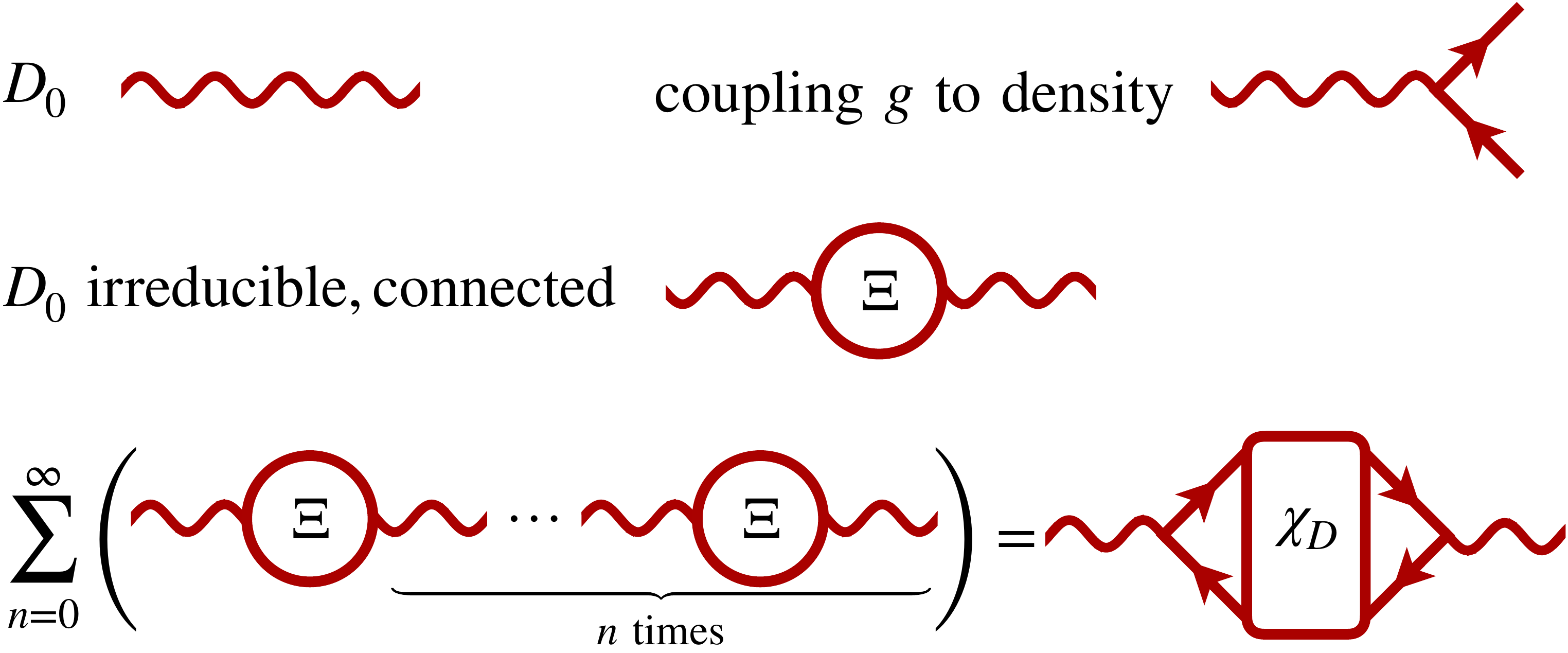}
    \caption{Density susceptibility determined diagrammatically from the phonon self-energy:
The phonon self-energy sums up all diagrams that are $D_0$ irreducible, i.e., the chain of $\Xi$ with $D_0$ to produce all $D$ amputated diagrams that are $D_0$ reducible [see Eq.~\eqref{eq:xirpa}].
The diagrams generated this way 
describe a process beginning and ending with a phonon. But since the diagrams are $D$-amputated, that can only be if the process begins and ends with bare electron-phonon vertex $g$. If we strip off the $g$ factors, we end up with the full set of diagrams that make up the density 
susceptibility, as the phonons couple to the density operator.}
 \label{fig:phonon_se_diagrammatic}
\end{figure}
%-------------

The central observation is that the density susceptibility~$\chi^D$ can be determined diagrammatically from the phonon self-energy~$\Xi$ as
\begin{align}
-\chi^\D &= \abs{g}^{-2} (\Xi + \Xi \circ D_0 \circ \Xi 
+ \cdots) = \frac{\abs{g}^{-2}\Xi}{1-D_0 \Xi}\,,
\label{eq:xirpa}
\end{align}
see Fig.~\ref{fig:phonon_se_diagrammatic} for a diagrammatic representation.
Inverting the above relation, we obtain an exact expression for the phonon self-energy from~$\chi^D$~\cite{berger_two-dimensional_1995, gunnarson_sum_rulesPhysRevB.75.035119}
\begin{align}
\Xi = \frac{-|g|^2 \chi^\D}{1-|g|^2 \chi^\D D_0}\,. \label{eq:phonon_self_energy_from_chi_D}
\end{align}
In the Dyson equation for the full phonon propagator $D^{-1}=D_0^{-1}-\Xi$, we can then identify the renormalized phonon dispersion by
\begin{align}
D= \frac{2\omega_0}{(i\Omega)^2 - \omega_\textrm{eff}^2}
\end{align}
with
\begin{align}
\omega_\textrm{eff}(q) := \omega_0\sqrt{1+2\Xi(q)/\omega_0}\,. \label{eq:eff_phonon}
\end{align}
Note that the phonon self-energy is a monotonic function of $\chi^\D(i\Omega=0, \bfq)>0$. 
In particular, we have $\Xi(\chi^\D \rightarrow 0) = 0$ and $\Xi(\chi^\D \rightarrow \infty) = -\omega_0/2$. 
Hence, the phonon self-energy is bounded in the normal phase
\begin{align}
  -\frac{\omega_0}{2} < \Xi(i\Omega = 0, \bfq) \leq 0\,.
\end{align}
For the renormalized phonon dispersion~\eqref{eq:eff_phonon}, this implies that (i)~$\omega_\textrm{eff}(i\Omega = 0, \bfq)\leq\omega_0$, a property termed as phonon softening, and (ii)~for finite $g$, it vanishes only in correspondence to $\chi^\D \rightarrow \infty$.
This means that in the Hubbard-Holstein model there is no dissociation between a lattice instability and an electronic density instability.

In contrast, for the RPA phonon self-energy which can be read off from Eq. \eqref{eq:D_RPA}, we obtain
\begin{align}
\Xi_\textrm{RPA} = -|g|^2 \tilde{\chi}^\D_\textrm{RPA}=\Xi+ \mathcal{O}(g^4)\,,
\end{align}
since $\chi^\D = \tilde{\chi}^D$ in lowest order in $g$.
As $\Xi_\textrm{RPA}$ is not bounded, the RPA renormalized phonon dispersion becomes imaginary for large electron-phonon couplings.
Evidently, this is an artifact of the RPA. 

%------------
\begin{figure}[t]
    \centering
    \includegraphics[width=\linewidth]{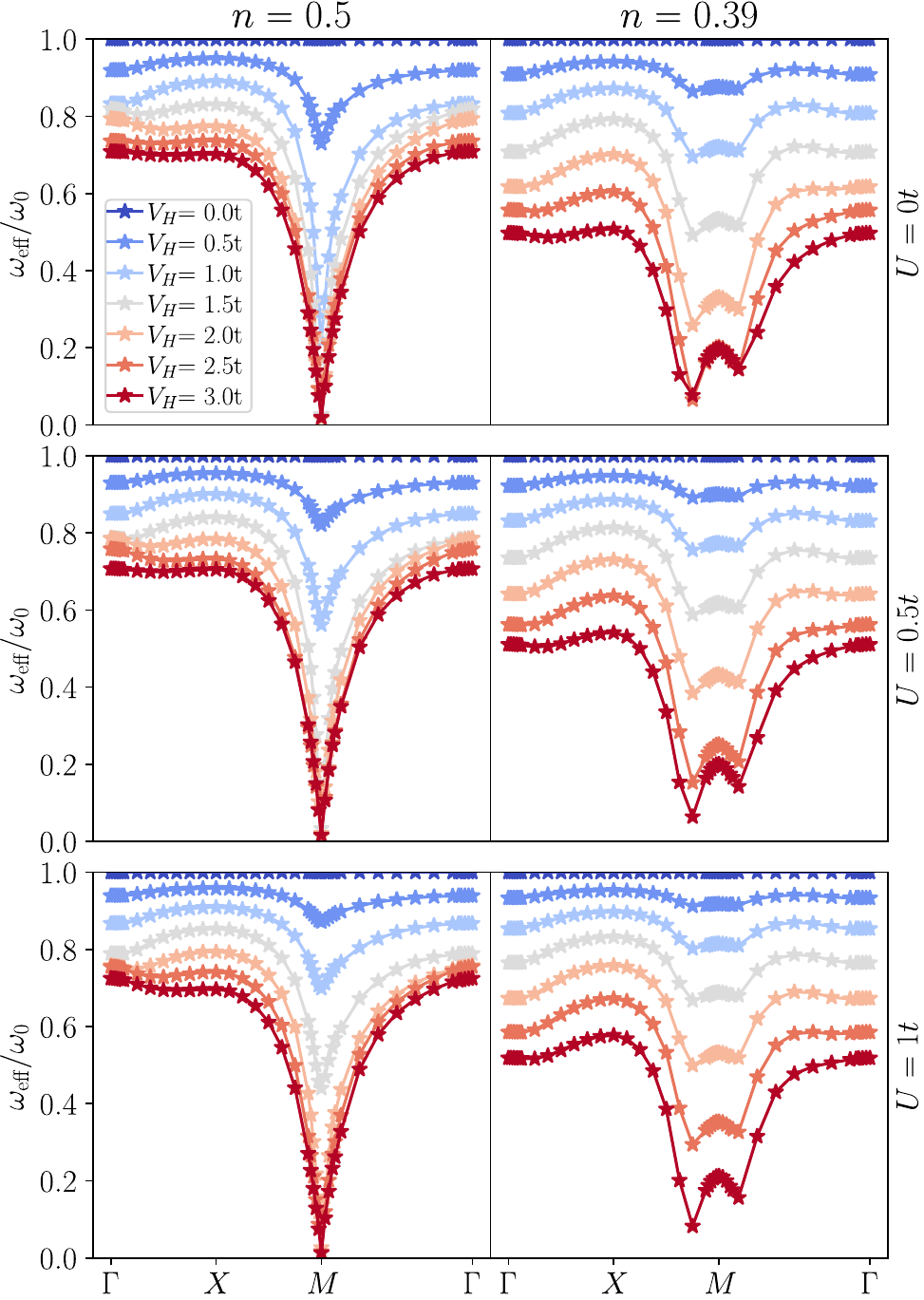} 
    \caption{Renormalization of the phonon dispersion $\omega_\textrm{eff}(\bfq)$ 
     along a path in the Brilouin zone, at half filling (left panels) and finite doping with filling $n=0.39$ (right panels), for $\omega_0 = 1t$, $\beta=20/t$, and different values of $U$ and $V_H$.}
    \label{fig:phonon_softening}
\end{figure}
%------------

Combining Eq.~\eqref{eq:phonon_self_energy_from_chi_D} with Eq.~\eqref{eq:eff_phonon} we can extract the phonon dispersion from the density susceptibility $\chi^\D$.
In Fig.~\ref{fig:phonon_softening}, we show fRG results for the phonon dispersion along a path on the Brillouin zone both at half filling and finite doping, for $\omega_0 = 1t$ and different values of $U$ and $V_H$. 
Approaching the charge density instability, at half filling $\omega_\textrm{eff}$ rapidly drops to zero in correspondence of the ordering vector $(\pi, \pi)$, while at finite doping we have two minima at incommensurate wave vectors.
Increasing $U$ reduced the phonon softening.

The proper inclusion of the phonon renormalization is relevant not only for a quantitatively accurate, but also for a correct qualitative description. 
On one hand, the phonon softening may affect the low energy description of a system. 
For example, it has been shown that a system which from purely electronic considerations can host a spin liquid, may not be able to do so in the presence of soft phonon modes~\cite{Seifert_2024}. On the other hand, it allows to capture the suppression of~$\chi^\SC$ by density fluctuations which is absent without phonon renormalization~\cite{marsiglio2021phononselfenergyeffectsmigdaleliashberg}.\bigskip

%-------------
\section{Intertwined fluctuations and isotope effects}
\label{sec:fluct}
%-------------

%---------------
\subsection{Fluctuation diagnostics}
\label{sec:fluctuation_diagnostics}
%---------------

The magnitude of an order-parameter propagator (minus its bare value) can be taken as a measure of the strength of fluctuations.
Hence, to systematically analyze the fluctuations of the model, we introduce magnetic, density, and superconducting order-parameter fields, i.e., $\vec{m}, \rho$, and $\Delta$, respectively.
To that end, one may split the SU(2)-resolved bare interaction as
\begin{align}
V_{0}(k_1, k_2&, k_3, k_4)   \nonumber =   V_{0,k_1k_2}^{\pp}(k_1+k_3)+V_{0,k_1k_4}^{\ph}(k_2-k_1)\\
&+V_{0,k_1k_2}^{\xph}(k_3-k_2)- 2 V_0(k_1, k_2, k_3, k_4)\,.
\end{align}
Writing the first three terms each in physical channels as in Eq.~\eqref{eq:physical_vertices}, splitting $V_{0}^\X = \mathcal{B}^\X + \mathcal{F}^\X$, and finally applying three Hubbard-Stratonovich transformations for each of the $\mathcal{B}^\X$ yields the following action~\cite{Denz2020}
\begin{widetext}
\begin{align}
S_{\text{bos}}\left[\psi, \bar{\psi}, \Delta, \Delta^*, \rho, \vec{m}\right] =& - \int_{k, \sigma} \bar{\psi}_\sigma(k) (i\nu + \mu - \epsilon_k) \psi_{\sigma}(k)- 2U_\textrm{eff} \int_{k, k', q} \bar{\psi}_{\uparrow}(k+q)\bar{\psi}_{\downarrow}(k'-q)\psi_{\downarrow}(k')\psi_{\uparrow}(k),\notag\\
& +\int_q \Delta^*(q) \frac{1}{\mathcal{B}^\SC(q)}\Delta(q) +\frac{1}{2} \int_q \rho(-q) \frac{1}{\mathcal{B}^\D(q)} \rho(q) - \frac{1}{2}\int_q  \vec{m}(-q) \cdot \frac{1}{\mathcal{B}^\M(q)}  \vec{m}(q) \notag\\
& + \int_{k, q} \Delta(q) \bar{\psi}_\uparrow\left(\frac{q}{2} + k\right)\bar{\psi}_{\downarrow}\left(\frac{q}{2} - k\right) + \int_{k, q} \Delta^*(q) \psi_\downarrow\left(\frac{q}{2} - k\right)\psi_{\uparrow}\left(\frac{q}{2} + k\right)\notag\\
& + \int_{k, q, \sigma} \rho(q) \bar{\psi}_\sigma\left(k+\frac{q}{2}\right)\psi_\sigma\left(k-\frac{q}{2}\right) + \int_{k, q, \sigma, \sigma'} \vec{m} \cdot \bar{\psi}_\sigma\left(k+\frac{q}{2}\right)\vec{\tau}_{\sigma, \sigma'} \psi_{\sigma'}\left(k-\frac{q}{2}\right) .\label{eq:bosonic_action}
\end{align}
\end{widetext}

Adding a regulator to the fermionic part of the theory and deriving the equations for the one-particle irreducible %(1PI) 
vertices while truncating the flow of vertices with $N_{f} + 2N_b > 4$, where $N_f$, $N_b$ is the number of fermionic,  bosonic legs, yields the SBE flow equations~\cite{Bonetti_2022,Denz2020} reported in Eq.~\eqref{eq:sbe_flow_equations}. 
These flow equations are equivalent to the fermionic flow equations in the level-1 truncation. 
Thus the flow equations in this bosonisation scheme are free of ambiguities in the decoupling and treat all channels on equal footing, see also Ref.~\cite{Denz2020}. 
Note that the bare values of the bosonic propagators $w_0^\X = \mathcal{B}^\X$ and fermion-boson couplings $\lambda^\X_0 = 1$ in Eq.~\eqref{eq:bosonic_action} coincide with the initial values of the SBE flow Eq.~\eqref{eq:flow_initial_values}.
Interactions between the different bosons are mediated by the fermions through the fermion-boson couplings in Eq.~\eqref{eq:bosonic_action} and thus the interchannel correlations between the different types of bosons are reflected in the renormalization of the fermion-boson couplings.
Indeed, freezing their flow to their initial value $\lambda^{\X, \Lambda=0} = 1$ yields single-channel ladder resummations for the bosonic propagators~$w^\X$. 

The importance of bosonic fluctuations is determined by the difference $w^\X - w^\X_0$ which is directly proportional to the $s$-wave susceptibility $\chi^\X$ by Eq.~\eqref{eq:swave_susceptibility_from_w}. 
The question we address in this section is the following: 
How do the different fluctuations, encoded in the susceptibilities, change with the phonon frequency $\omega_0 \sim 1/\sqrt{M_\textrm{ion}}$? 
The straightforward answer is provided already in Figs.~\ref{fig:halffilling_inverse_susceptibility_omega0_influence} and~\ref{fig:doped_inverse_susceptibility_omega0_influence}. 
However, in the bare action $\omega_0$ enters only the density bare bosonic propagator $w^\D_0 = \mathcal{B}^\D$, whereas $w^\SC_0 = \mathcal{B}^\SC$ and $w^\M_0 = \mathcal{B}^\M$ do not depend on it. 
Therefore, the influence of $\omega_0$ on the superconducting and magnetic susceptibilities can only  arise from the interplay between the channels, requiring a more nuanced answer to the above question.

%----------------
\begin{figure*}[t!]
    \centering    \includegraphics[width=0.99\linewidth]{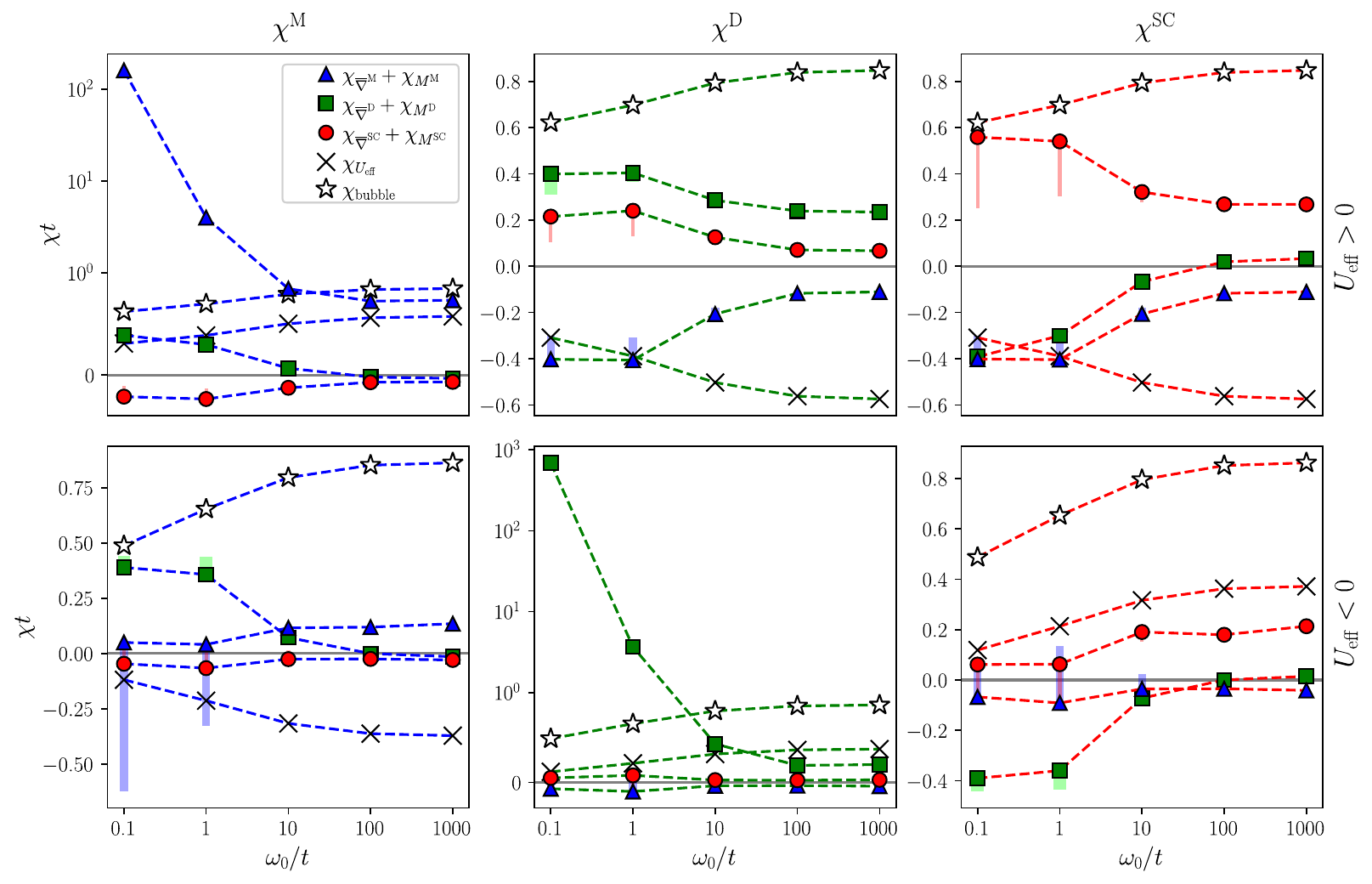}
    \caption{Fluctuation diagnostics of the ($s$-wave) superconducting, density, and magnetic susceptibilities for the same parameters as in Fig. \ref{fig:halffilling_inverse_susceptibility_omega0_influence}. The large values in the adiabatic limit indicate the onset of an antiferromagnetic or a charge-density wave instability; see also the logarithmic scale in the respective panels (note that for better visibility the scale in the region  $[-10^0, 10^0]$ is linear). 
    The blue, green, and red bars represent the contributions of the rest functions $\chi^{\X'}_{M^\X}$, with the end points indicating the result for  $\chi^{\X'}_{\overline{\nabla}^\X}$ only. }    \label{fig:halffilling_susceptibility_fluctdiag}
\end{figure*}
%----------------

%----------------
\begin{figure*}[t!]
    \centering
    \includegraphics[width=0.99\linewidth]{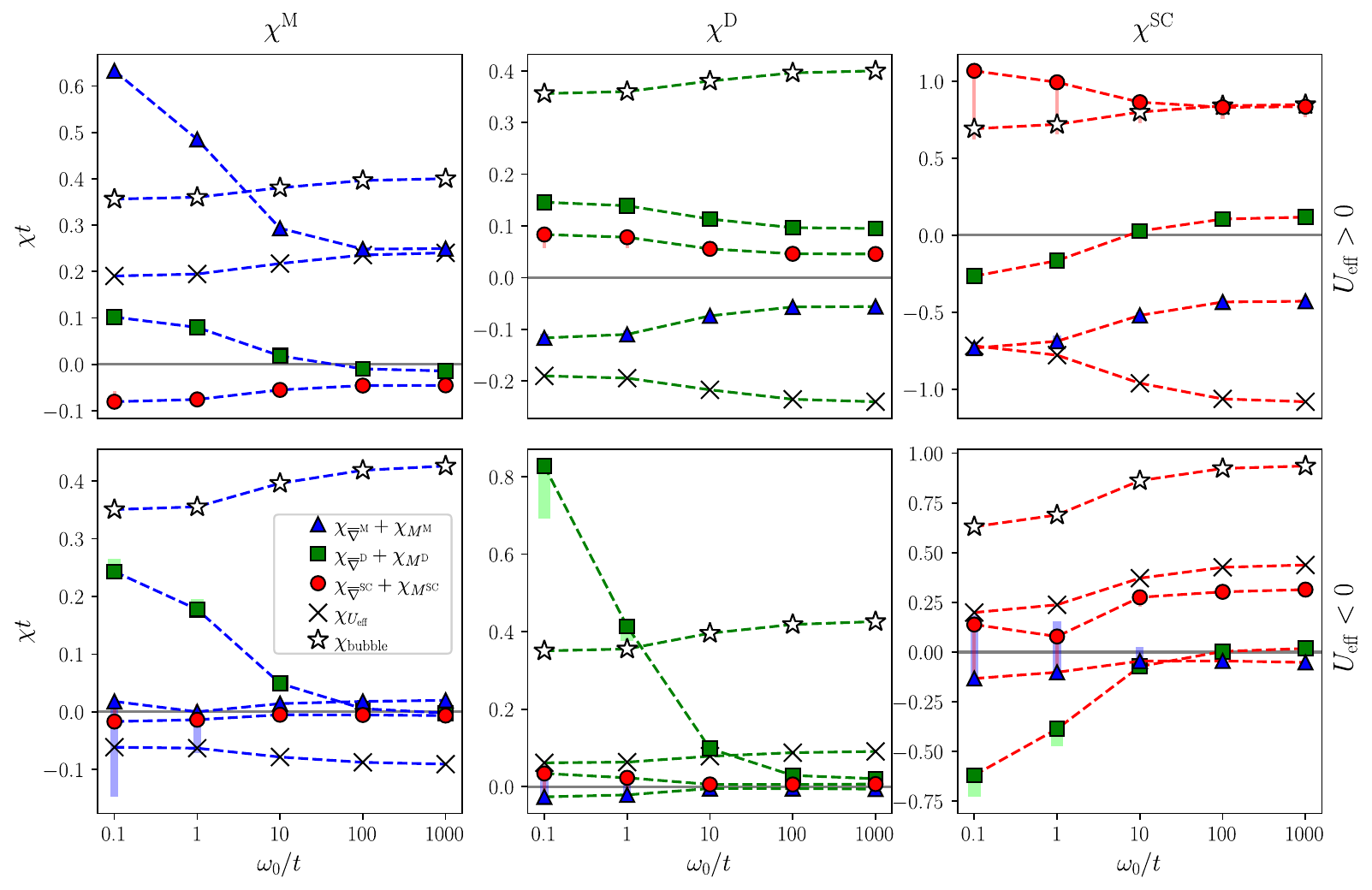} 
    
    \caption{Fluctuation diagnostics of the ($s$-wave) superconducting, density, and magnetic susceptibilities as in Fig. \ref{fig:halffilling_susceptibility_fluctdiag} but at finite doping, for the parameters as in Fig. \ref{fig:doped_inverse_susceptibility_omega0_influence}. 
    }    \label{fig:finitedoping_susceptibility_fluctdiag}
\end{figure*}
%----------------

\begin{figure}[t!]
    \centering    \includegraphics[width=\linewidth]{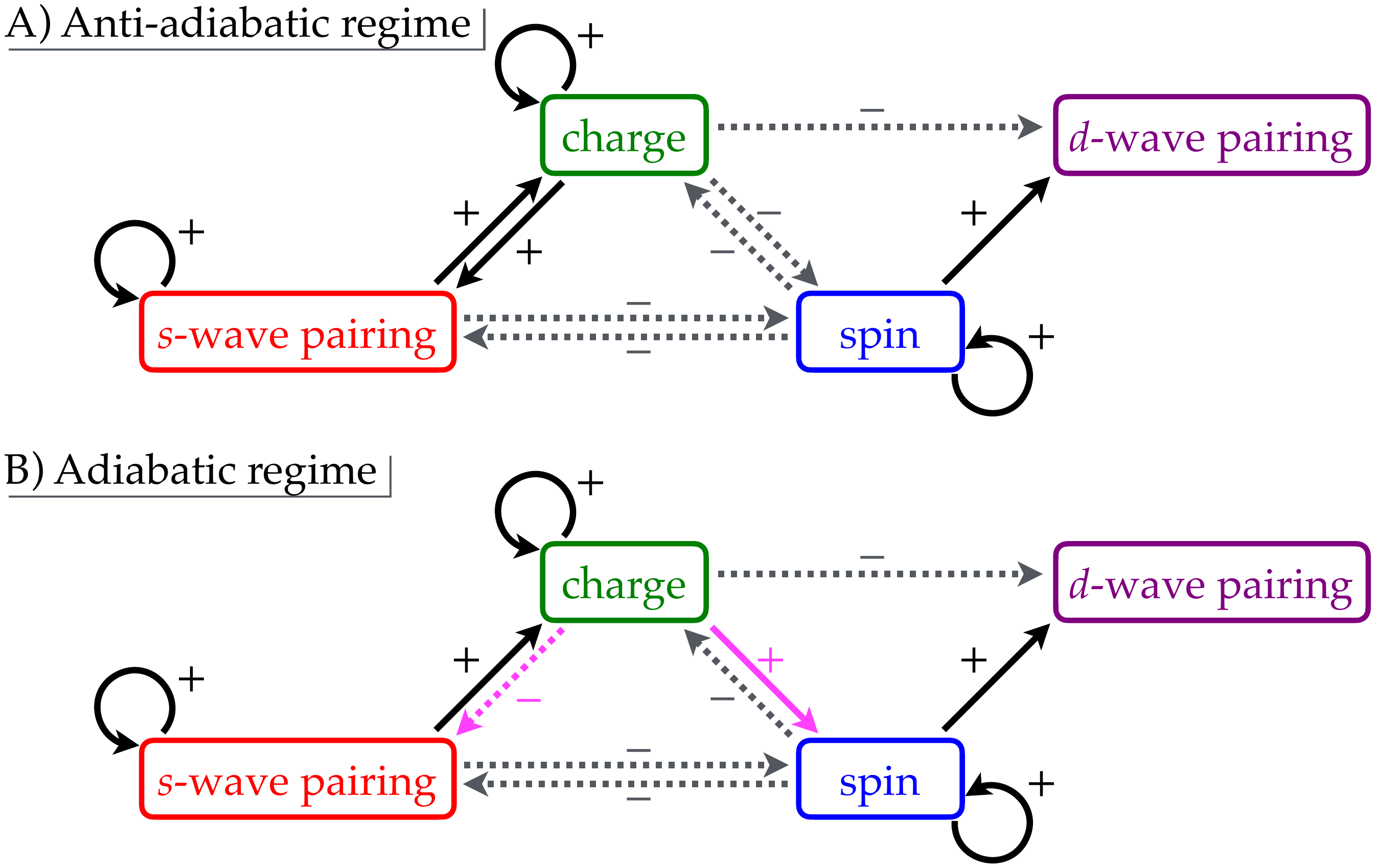}
    \caption{Panel~A) Interplay of fluctuations in the different contributions to the susceptibilities 
    of the effective Hubbard model in the anti adiabatic regime  
    as described by the diagnostic matrix in Eq.~\eqref{eq:diagnostic_matrix_chi}. 
    Solid arrows indicate an enhancement corresponding to a $+$, whereas dashed arrows indicate a suppression with a $-$ in $D_\chi$. 
    Panel~B) Interplay of fluctuations in the different contributions to the susceptibilities near the adiabatic regime. Note that in contrast to the effective Hubbard model in the anti adiabatic limit, the charge density fluctuations suppress $s$-wave superconducting pairing fluctuations while enhancing magnetism.
    }
    \label{fig:interplay_purehubbard}
\end{figure}

Combining exact many-body relations with the vertex decomposition, allows one to glean information about the intertwining of fluctuations in the  fluctuation diagnostics~\cite{Schaefer2021b,Gunnarsson2015}.
Specifically, inserting the SBE decomposition, Eq.~\eqref{eq:parquet_inSBE_physical}, into the expression for the vertex corrections, Eq.~\eqref{eq:susc_vertex_contribution}, we obtain a splitting of the susceptibility
\begin{align}
\chi^\X = \chi^{\X}_{\mathrm{bubble}} + \sum_{\X'} \left(\chi^\X_{\bar{\nabla}^{\X'}} + \chi^\X_{M^{\X'}} \right) + \chi^\X_{U_{\text{eff}}}\,.
\label{eq:fluctuation_diagnostics_of_chi}
\end{align}
Here, the double counting contribution of the local part of the interaction is absorbed in the definition~\cite{heinzelmann2023entangledmagneticchargesuperconducting,aleryani2024screeningeffectiverpalikecharge}
\begin{align}
    \bar{\nabla}^\X := \nabla^\X - U^\X_\textrm{eff}\,.
    \label{eq:defnablabar}
\end{align} 
The results for the $s$-wave susceptibilities are shown in Fig.~\ref{fig:halffilling_susceptibility_fluctdiag} for half filling and in Fig.~\ref{fig:finitedoping_susceptibility_fluctdiag} for finite doping, for the same parameters as in Figs.~\ref{fig:halffilling_inverse_susceptibility_omega0_influence} and \ref{fig:doped_inverse_susceptibility_omega0_influence}, respectively.
(In Fig.~\ref{fig:finitedoping_dwave_susceptibility_fluctdiag}, we also report the fluctuation diagnostics for the $d$-wave superconducting susceptibilities at finite doping which we discuss in the next section.) 
In these figures, the bars indicate the contribution from the rest function, with the end point at  
the result for $\chi_{\bar{\nabla}^{\X'}}^\X$.
While the rest function appears to have almost negligible effects for $\omega_0>1t$, substantial corrections affect the results in proximity of the adiabatic regime, see also Appendix~\ref{app:technical_impelementation} for a more detailed discussion. 
At half filling, as $\omega_0$ is decreased, we observe a large enhancement of the density contribution to itself for $U_\textrm{eff} < 0$ and of the magnetic one to itself for $U_\textrm{eff} > 0$. 
This trend survives also at finite doping, where both fluctuation channels are suppressed due to the absence of Fermi-surface nesting. 
In contrast, the contributions from the bubble decrease uniformly as $\omega_0$ is decreased.
Another important finding is the change of sign of the density contributions to both the superconducting and magnetic susceptibilities. 
A schematic representation of the effects different contributions have on the susceptibilities is shown in  Fig.~\ref{fig:interplay_purehubbard}. 
In particular, the $+/-$ signs indicate an enhancement or a suppression as determined from the numerical results.
In the following, we illustrate how the SBE decomposition allows to infer most of these signs providing an analytic understanding.

%-----------------
\subsection{Diagnostic matrices}
\label{sec:diagnostic_matrices}
%-----------------

%-----------------
%-----------------

The tendencies induced by the different channel contributions can be inferred using the method of diagnostic matrices introduced for the Hubbard model 
\cite{aleryani2024screeningeffectiverpalikecharge}. 
Specifically, the diagnostic matrix of a quantity 
$A$ is defined as the sign matrix $D_A$ with entries
\begin{align}
D_A^\X = \begin{cases}
    + & \text{if fluctuations in 
    $\X$ \emph{encourage} $A$}\\
    - & \text{if fluctuations in  
    $\X$ \emph{suppress} $A$\,.}
\end{cases}
\end{align}
%

%-----------------
\subsubsection{$s$-wave fluctuations}
%-----------------

In this section, we focus on the uniform ($s$-wave) quantities. 
Since we are merely interested in the signs, we can neglect all momentum and frequency dependencies. 
We will further neglect the self-energy, and revisit its effects later. 
Starting from the pure Hubbard model retrieved in the anti adiabatic limit, we will subsequently include the effects of a decreasing~$\omega_0$ to account for the  retarded interaction of the Hubbard-Holstein model.
Equations~\eqref{eq:physical_vertices} for the SBE decomposition~\eqref{eq:parquet_inSBE_physical} of the vertices then read to leading order
\begin{subequations}
\begin{align}
V^\SC \approx& \bar{\nabla}_{\pp}^{\SC} + \frac{1}{2}\bar{\nabla}_{\pp}^{\D} - \frac{3}{2}\bar{\nabla}_{\pp}^{\M} + U_\textrm{eff}\\
V^\D \approx& \left(2\bar{\nabla}_{\ph}^{\SC} - \bar{\nabla}_{\xph}^{\SC}\right) + \frac{1}{2}\left(2\bar{\nabla}_{\ph}^{\D} - \bar{\nabla}_{\xph}^{\D}\right) \nonumber\\
&- \frac{3}{2}\left(2\bar{\nabla}_{\ph}^{\M} - \bar{\nabla}_{\xph}^{\M}\right) + U_\textrm{eff} \label{eq:vertex_d_rough}\\
V^\M \approx& -\bar{\nabla}_{\xph}^{\SC} - \frac{1}{2}\bar{\nabla}_{\xph}^{\D} + \frac{3}{2}\bar{\nabla}_{\xph}^{\M} - U_\textrm{eff}\,,
\end{align}
\end{subequations}
where we focus on the SBE contributions.
In the absence of frequency and momentum dependencies, the diagrammatic parametrisations are trivial and we have
\begin{align}
V^\SC \sim  V^\D \sim -V^\M \sim \bar{\nabla}^{\SC} + \frac{1}{2}\bar{\nabla}^{\D} - \frac{3}{2}\bar{\nabla}^{\M} + U_{\textrm{eff}}\,. \label{eq:skeleton_of_physical_vertices}
\end{align}
Assuming $\abs{\lambda^\X} \lessapprox 1$ (see Ref.~\cite{fraboulet2023singlebosonexchangefunctionalrenormalization,Bonetti_2022,Krien_SBE_original} for the Hubbard model), we observe that at weak coupling 
\begin{align}
\bar{\nabla}^\X = \lambda^X w^\X \lambda^\X - U_\textrm{eff}^X \lessapprox 0\,, \label{eq:nabla_bar_is_negative}
\end{align}
obtained from Eq.~\eqref{eq:swave_susceptibility_from_w} that implies $0\leq\chi^\X \sim \mathcal{B}^\X - w^\X = U_\textrm{eff} - w^\X$ in the anti adiabatic limit.
With $\mathrm{Sgn}\,\bar{\nabla}^\X = -$, the diagnostic matrix $D_V$ can be read off from 
\eqref{eq:skeleton_of_physical_vertices}
\begin{align}
D_V = - \begin{pmatrix}
+  & + & - \\
+ & +  & - \\
-  & - & +  
\end{pmatrix}=-D_\chi = -\left(D^{\X'}_{\chi^\X}\right)
, \label{eq:diagnostic_matrix_chi}
\end{align} 
where here and in the following the matrix components with indices $\X,\X'$ appear in the order $\X,\X'\in\{\SC,\D,\M\}$.
The diagnostic matrix $D_\chi$ for the susceptibilities is inferred from the vertex corrections~\eqref{eq:susc_vertex_contribution} (we note that the contribution of the bare bubble is not considered).  
Figure \ref{fig:interplay_purehubbard} shows a graphic illustration.
We can also extract the influence of the bare interaction $U_{\mathrm{eff}}$
\begin{align}
D_{\chi
}^{U_{\mathrm{eff}}} &= \begin{pmatrix}
 -\mathrm{Sgn}\,U_\textrm{eff}\\
 -\mathrm{Sgn}\,U_\textrm{eff}\\
 \mathrm{Sgn}\,U_\textrm{eff} 
\end{pmatrix}. \label{eq:diagnostic_matrix_chi_U}
\end{align}
Similarly, we can derive the diagnostic matrix for the interaction irreducible vertex $\mathcal{I}^\X = V^\X - \nabla^\X$ and use it to determine the one for the fermion-boson coupling
\begin{align}
    D_\lambda &= - D_{\mathcal{I}} = 
    \begin{pmatrix}
0  & + & - \\
+ & -  & - \\
-  & - & +  
\end{pmatrix},
\end{align}
which will be useful later.

Coming back to the diagnostic matrix for the susceptibilities $D_{\chi}$ provided in Eq.~\eqref{eq:diagnostic_matrix_chi} for the anti adiabatic limit, we can extract the following trends  
induced by $s$-wave fluctuations: i)~superconducting and density fluctuations go hand in hand, while both play against $\M$, and vice versa, and ii)~$\M$ encourages itself. 
All fluctuations are seeded by the bare interaction $U_{\mathrm{eff}}$, see Eq.~\eqref{eq:diagnostic_matrix_chi_U} for $D_{\chi}^{U_{\mathrm{eff}}}$ which reflects the common wisdom that $s$-wave pairing and density fluctuations are enhanced only in presence of a negative bare local interaction, whereas magnetic fluctuations are suppressed in this case, and opposite for repulsive interactions~\cite{aleryani2024screeningeffectiverpalikecharge}. 
Comparing with the numerical results of the full calculation shown in Figs.~\ref{fig:halffilling_susceptibility_fluctdiag} and~\ref{fig:finitedoping_susceptibility_fluctdiag}, we find a perfect agreement with the predictions of $D_{\chi}$ and $D_{\chi}^{U_{\mathrm{eff}}}$ in the anti adiabatic limit, i.e., with the signs of the contributions at large~$\omega_0$ or equivalently small~$M_\textrm{ion}$. 

Reducing $\omega_0$ (or increasing $M_\textrm{ion}$), we expect deviations from the behavior of an effective Hubbard model with a purely local interaction. In particular, some contributions may change their signs in presence of a finite $\omega_0$ and a sizable $V_H$, for which the retarded part of the Hubbard-Holstein interaction can attain large values at finite frequencies and the frequency dependence must be taken into account. 
In order to make progress, we expand the susceptibility around $M_\textrm{ion} \gtrapprox 0$,
\begin{align}
\chi^\X &= \chi^\X(M_\textrm{ion} = 0) + M_\textrm{ion} \left.\frac{\partial \chi^\X}{\partial{M_\textrm{ion}}}\right|_{M_\textrm{ion} = 0}  +\mathcal{O}(M_\textrm{ion}^2) \label{eq:expansion_of_susc_in_M_ion}
\end{align}
and derive the diagnostic matrix for its derivative, see Appendix~\ref{app:diagm} for the details,
\begin{align}
    D_{\partial \chi} &= \left(D_{\partial \chi^\X}^{\X'}\right) = \begin{pmatrix}
+\text{Sgn}\,U_\textrm{eff}  & - & -\text{Sgn}\,U_\textrm{eff} \\
+\text{Sgn}\,U_\textrm{eff} & +  & -\text{Sgn}\,U_\textrm{eff} \\
-\text{Sgn}\,U_\textrm{eff}  & + & +\text{Sgn}\,U_\textrm{eff} 
\end{pmatrix}. \label{eq:diagma}
\end{align}
It tells us how --  starting from the anti adiabatic limit --  the fluctuations vary with $M_\textrm{ion}$;
i.e. if the matrix element is $+/-$, the contribution from channel $\X'$ to $\chi^\X$ increases/decreases when $M_\textrm{ion}$ is increased or equivalently when $\omega_0$ is decreased. We note that the effects of the fluctuations in one channel on another one do not have to follow the same trend for the reverse, as indicated by the asymmetry.

Let us investigate the resulting retardation effects,
both for $U < V_H$ ($U_\textrm{eff} < 0$) and $U > V_H$ ($U_\textrm{eff} > 0$). 
For $U_\textrm{eff} < 0$, 
we have seen that the region dominated by charge fluctuations increases. 
This can be understood by looking at the diagnostic matrix that provides the sign of the different contributions to the variation of the susceptibilities near the anti adiabatic limit
\begin{align}
D_{\partial\chi}(\mathrm{Sgn}\,U_\textrm{eff} = -) = \begin{pmatrix}
 \boxed{\begin{matrix}- & -\\ - & + \end{matrix}} & \begin{matrix} + \\ + \end{matrix} \\ \begin{matrix} + & + \end{matrix} & -
\end{pmatrix}.
\end{align}
Since the magnetic fluctuations are suppressed due to $D_{\chi^\M}^\SC = D_{\chi^\M}^\D = D_{\chi^\M}^{U_\textrm{eff}} = -$, we thus focus  on the $2\times2$ block involving $\SC$ and $\D$. 
We see that decreasing $\omega_0$ suppresses the $s$-wave superconducting susceptibility by suppressing both the contributions to itself as well as the contributions of the charge fluctuations to it. 
The situation for the charge  susceptibility differs in that the contributions to itself increase, with the net effect to tip the scales toward leading charge fluctuations.
For $U_\textrm{eff} > 0$, instead, both $s$-wave superconducting and charge fluctuations are small in the anti adiabatic limit, since the bare interaction does not seed them $D_{\chi^\SC}^{U_\textrm{eff}} = D_{\chi^\D}^{U_\textrm{eff}} = -$. 
Moreover, magnetism will also play against them 
$D_{\chi^\SC}^\M = D_{\chi^\D}^\M = -$. Thus we here focus  
on the 
entry for 
the magnetic susceptibility 
and its feedback on itself
\begin{align}
D_{\partial\chi}(\mathrm{Sgn}\,U_\textrm{eff} = +) = \begin{pmatrix}
+ & - & - \\ + & + & - \\ - & + & {\boxed{+}}
\end{pmatrix}.
\end{align}
From the diagnostic matrix emerges that decreasing $\omega_0$ enhances the positive feedback of the magnetic fluctuations on itself. 
This explains the observed results in Figs.~\ref{fig:phase_diagram_halffilling} and~\ref{fig:phase_diagram_doped}. 

Finally, we comment on the 
adiabatic limit $\omega_0 \to 0$ (or equivalently $M_\textrm{ion} \to\infty$).
A na\"ive application of Eq.~\eqref{eq:diagma} fails to correctly describe the results in Figs.~\ref{fig:halffilling_susceptibility_fluctdiag} and \ref{fig:finitedoping_susceptibility_fluctdiag}  for $\omega_0 \lesssim 1t$, where a nonmonotonic behavior appears in some contributions. This is also the regime in which we observe important contributions from the rest function, neglected in the derivation of~$D_{\partial\chi}$. 
A thorough analysis reveals that  $M^\X$ contributions become large due to ladders in the fermionic part of the interaction $\mathcal{F}^\X$ contained in $M^\X$ which become local in the adiabatic limit, see Appendix~\ref{app:technical_impelementation} for further details. 

%---------------
\subsubsection{$d$-wave fluctuations}
\label{sec:dwave_fluctuation_diagnostics}
%---------------

We now show how the diagnostic matrices can be applied also to $d$-wave fluctuations, where we focus on the $d$-wave superconducting susceptibility $\chi^\dSC(0)$ evaluated at $q = 0$. 
In analogy to the $s$-wave  analysis, we identify the bubble contribution and the vertex corrections
\begin{align}
\chi^\dSC(q) = \chi^{\dSC}_{\mathrm{bubble}}(q) + \chi^\dSC_{\mathrm{vertex}}(q)\,,
\end{align}
with 
\begin{subequations}
\begin{align}
 \chi^{\dSC}_{\mathrm{bubble}}(q) =& \sum_{i\nu} \Pi^\dSC(q, i\nu)= 
 \sum_{i\nu} \Pi^\SC_{\dwave \dwave}(q, i\nu)\\
  \chi^{\mathrm{dSC}}_{\mathrm{vertex}}(q) =& -\sum_{\mathclap{m, m', i\nu, i\nu'}} \Pi^{\SC}_{\dwave m}(q, i\nu) 
 V^{\SC}_{m m'}(q, i\nu, i\nu') \nonumber \\ &\qquad \times \Pi^{\SC}_{m' \dwave}(q, i\nu')\,. \label{eq:vertd}
\end{align}
\end{subequations}
In the static limit at $i\Omega = 0$, the mixed bubbles involving an $s$- and a $d$-wave form factor vanish at $\bfq = 0$ \cite{heinzelmann2023entangledmagneticchargesuperconducting,fraboulet2023singlebosonexchangefunctionalrenormalization}. Therefore, the sum over the form factors in Eq. \eqref{eq:vertd} restricts to $m = m' = \dwave$. 
Furthermore, we note that the $d$-wave component of $\nabla^{\SC}$ vanishes in its native channel at $\bfq = 0$: Projecting the $\bfk$ and $\bfk'$ dependence onto the $d$-wave form factor
\begin{align}
\nabla&_{\dwave\dwave}^\SC(q) = \sum_{\bfk, \bfk'} f^*_{\dwave}(\bfk) f_{\dwave}(\bfk') \nabla^\SC_{k, k'}(q)  \nonumber\\&= \left(\sum_{\bfk} f^*_{\dwave}(\bfk)\lambda^\SC_k(q) \right) w^\SC(q) \left(\sum_{\bfk'} f_{\dwave}(\bfk')\lambda^\SC_{k'}(q) \right),
\end{align}
the two integrals vanish at $\bfq = 0$ since $\lambda^\SC$ is invariant under all point group rotations whereas $f_{\dwave}$ changes sign under a $\pi/2$ rotation. 
Inserting the SBE decomposition, we hence obtain
\begin{align}
\chi^{\dSC}_{\mathrm{vertex}} = \sum_{\X \neq \SC} \chi^{\mathrm{dSC}}_{\bar{\nabla}^{\X}}  + \sum_{\X} \chi^{\mathrm{dSC}}_{M^{\X}}\,.
\end{align}
There is no direct contribution from $U_\textrm{eff}$ since $U_\textrm{eff}$ is constant in space and thus has no $d$-wave component. 
Note that the superconducting channel enters only the rest function and not the effective interaction. 
This implies that the SBE approximation which neglects the rest functions cannot capture an instability driven by the divergence of $M^\SC$. 
For a proper description of $d$-wave superconductivity, the multiboson contribution is essential and has to be included~\cite{Bonetti_2022}.
At the same time, the $M^\SC$ contribution is seeded by the effective interactions (like the $s$-wave orders by the bare interaction), see Appendix~\ref{app:multibosoncont}. 
For the diagnostic matrix analysis, we thus focus on the SBE contributions and write
\begin{align}
\chi^{\mathrm{dSC}} \simeq \chi^{\mathrm{dSC}}_{\bar{\nabla}^{\dD}} + \chi^{\mathrm{dSC}}_{\bar{\nabla}^{\dM}}\,,
\end{align}
where $\dM$ and $\dD$ refer the $d$-wave projections of the fermionic momentum $\bfk$, $\bfk'$ dependencies.
As these now enter through $f_{\dwave}$, the channel projections in momentum are no longer trivial and can not be ignored.
For the derivation of the diagnostic matrix analysis where we consider only the signs, we focus on the dominant momentum contributions $\nabla^\X_{k, k'}(q) \sim \text{Sgn}\,\nabla^\X \,\delta(\bfq - \bfq^* )$, with $\bfq^*\approx (\pi,\pi)$ for the $s$-wave density and magnetic channels.
Projecting the magnetic and density effective interactions onto the $\pp$ channel, we obtain (see Eqs.~\eqref{eq:q_xph_to_pp} and ~\eqref{eq:q_ph_to_pp})
\begin{widetext}
\begin{subequations}
\begin{align}
P^{\rightarrow \pp}[\nabla^M]_{\dwave, \dwave}(q_{\pp} = 0) &= \sum_{\bfk_\pp, \bfk'_\pp} f_{\dwave}(\bfk_{\pp})f_{\dwave}(\bfk_{\pp}') \nabla^\M_{\bfk_\pp, \bfk'_\pp}(\bfq_\pp - \bfk_\pp - \bfk'_\pp)\nonumber\\ 
&\sim \sum_{\bfk_\pp, \bfk'_\pp} f_{\dwave}(\bfk_{\pp})f_{\dwave}(\bfk_{\pp}') \delta(\bfq_\pp - \bfk_\pp - \bfk'_\pp - \bfq^* )  \,\mathrm{Sgn}\,\nabla^\M \nonumber\\
&= \mathrm{Sgn}\,\nabla^\M \sum_{\bfk_\pp} f_{\dwave}(\bfk_\pp) f_{\dwave}(-\bfk_{\pp} - \bfq^*)= - \mathrm{Sgn}\,\nabla^\M \sum_{\bfk_\pp} (f_\dwave(\bfk_\pp))^2 \sim - \mathrm{Sgn}\,\nabla^\M\\
P^{\rightarrow \pp}[\nabla^D]_{\dwave, \dwave}(q_{\pp} = 0) &= \sum_{\bfk_\pp, \bfk'_\pp} f_{\dwave}(\bfk_{\pp})f_{\dwave}(\bfk_{\pp}') \nabla^\D_{\bfk_\pp, \bfq_\pp -\bfk'_\pp}(\bfk'_\pp - \bfk_\pp )\nonumber\\
&\sim \sum_{\bfk_\pp, \bfk'_\pp} f_{\dwave}(\bfk_{\pp})f_{\dwave}(\bfq_\pp - \bfk_{\pp}') \delta(\bfq_\pp - \bfk_\pp - \bfk'_\pp - \bfq^* )  \,\mathrm{Sgn}\,\nabla^\D ~\sim - \mathrm{Sgn}\,\nabla^\D\,,
\end{align}
\end{subequations}
\end{widetext}
where we used that the $d$-wave form factor changes sign $f_{\dwave}(\bfq \pm \bfq^*) \approx -f_{\dwave}(\bfq)$ under a shift of $\bfq^* \approx (\pi, \pi)$.
The reversed sign with respect to the $s$-wave superconducting susceptibility
implies the opposite trend that magnetic fluctuations encourage $d$-wave superconductivity whereas density fluctuations suppress it \footnote{Note: this reversal would not occur if the ordering vector was $\bfq^* = (0, 0)$ implying that ferromagnetic fluctuations suppress $d$-wave superconductivity and uniform charge fluctuations encourage it.}, see also Fig.~\ref{fig:interplay_purehubbard}.

%---------------
\begin{figure}[b]
    \centering
    \includegraphics[width=1.\linewidth]{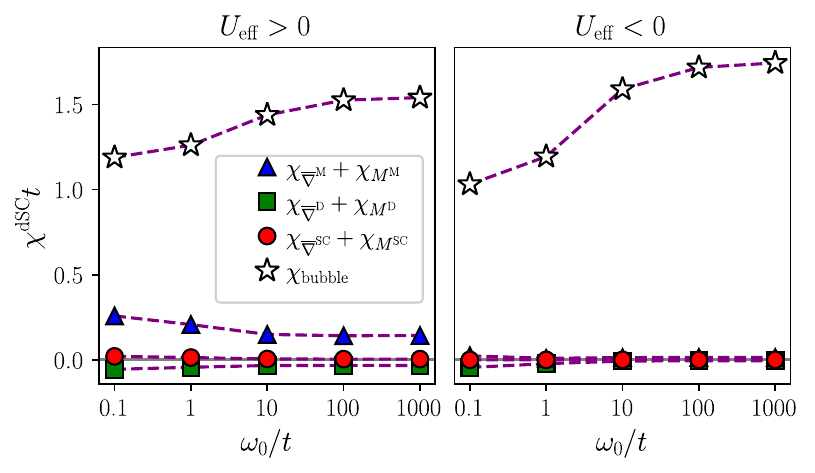}
    \caption{Fluctuation diagnostics of the $d$-wave superconducting susceptibility as in Fig. \ref{fig:finitedoping_susceptibility_fluctdiag}. 
     }   \label{fig:finitedoping_dwave_susceptibility_fluctdiag}
\end{figure}
%---------------

\begin{figure}
    \centering
    \includegraphics[width=\linewidth]{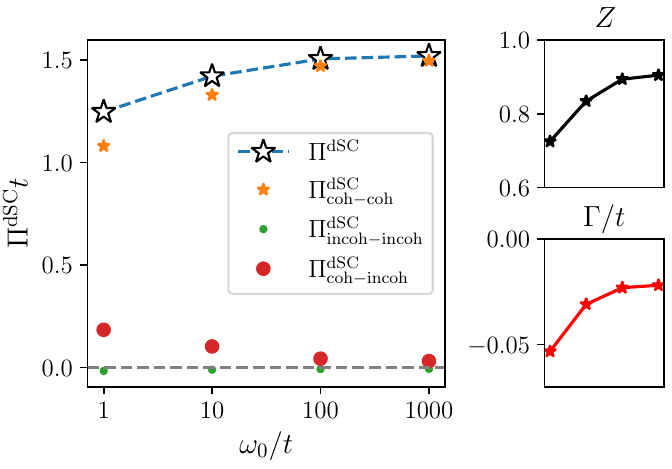}
    \caption{Different contributions to the $d$-wave bubble according to the decomposition~\eqref{eq:G_coherent_plus_incoherent}, for the same parameters as in Fig.~\ref{fig:finitedoping_dwave_susceptibility_fluctdiag} ($U_\textrm{eff}>0$).
    The corresponding values for the quasiparticle residue $Z$ and damping $\Gamma$ are shown on the right. 
    }
    \label{fig:bubble_fluct_diag}
\end{figure}

Concerning the influence of $\omega_0$ on the $d$-wave fluctuations, there are two competing effects: 
as seen in Sec.~\ref{sec:results}, the antiferromagnetic fluctuations are enhanced as $\omega_0$ is decreased, which encourages $d$-wave superconductivity. 
On the other hand, also charge-density wave fluctuations are enhanced which would lead to a suppression of $d$-wave superconductivity.
For $U_\textrm{eff}>0$, the density fluctuations are suppressed and we expect an overall enhancement of $\chi^\dSC$ as the bare interaction becomes more retarded. 
This is indeed found in a previous fRG study~\cite{anomalous_isotope_effect}, in which the flow of the self-energy is neglected. 
However, this is in disagreement with our results of Fig.~\ref{fig:doped_inverse_susceptibility_omega0_influence}, where $\chi^\dSC$ decreases as $\omega_0$ is decreased. 
The numerical fluctuation diagnostics results, displayed in Fig.~\ref{fig:finitedoping_dwave_susceptibility_fluctdiag}, reveal that $\chi^\dSC$ is essentially determined by the $d$-wave bubble which decreases when approaching the adiabatic limit. 
Therefore, we conclude this to be a self-energy effect, as we have seen already in Fig.~\ref{fig:effect_of_selfenergy_on_isotope_effect}. 
This corroborates findings obtained by dynamical cluster Monte Carlo calculations~\cite{macridin_suppression_2012} that the self-energy competes with magnetic fluctuations against driving $d$-wave superconductivity. 
As $\omega_0$ is decreased, $U_{\text{eff}} > 0$ is increased to $U > U_\textrm{eff}$ except at zero frequency. 
This increase of the overall interaction induces a larger self-energy which regularizes the Green's function and thus reduces the size of the bubble. 
To diagnose this further, we consider 
\begin{align}
G(i\nu, \bfk) = G_{\text{coh}}(i\nu, \bfk) + G_{\text{incoh}}(i\nu, \bfk)\,, \label{eq:G_coherent_plus_incoherent}
\end{align}
where $G_{\text{coh}}$ represents the part of the Green's function encoding the contribution due to coherent Fermi-liquid-like quasiparticles
\begin{align}
G_{\text{coh}}(i\nu, \bfk) := \frac{Z(\bfk)}{i\nu - \epsilon^\textrm{coh}_\bfk + iZ(\bfk)\Gamma(\bfk)}\,,
\end{align}
with the quasiparticle residue defined by 
$Z(\bfk) = ( 1 - \partial_{(i\nu)} \Sigma(i\nu, \bfk)|_{i\nu \rightarrow 0})^{-1}$,
the renormalized dispersion 
$\epsilon^{\text{coh}}(\bfk) =  Z(\bfk)(\epsilon - \mu + \delta \mu + \Re \Sigma(i\nu \rightarrow 0, \bfk) )$, 
and the quasiparticle damping 
$\Gamma(\bfk) = \Im \Sigma(i\nu\rightarrow0, \bfk)$. 
The incoherent part $G_{\text{incoh}}$ represents the other excitations
\begin{align}
G_{\text{incoh}}:= G - G_{\text{coh}}\,.
\end{align} 
Inserting this splitting of the Green's function~\eqref{eq:G_coherent_plus_incoherent} into the definition of the bubbles determines three terms 
\begin{align}
\Pi^\dSC = \Pi^\dSC_{\text{coh-coh}} + \Pi^\dSC_{\text{incoh-incoh}} + \Pi^\dSC_{\text{coh-incoh}}\,,
\end{align}
where $\Pi^\dSC_{\text{coh-coh}}$ represents the contribution from coherent quasiparticles, $\Pi^\dSC_{\text{incoh-incoh}}$ the one from other excitations, and $\Pi^\dSC_{\text{coh-incoh}}$ contains mixed contributions. 
Approximating the limit $i\nu \rightarrow 0$ by %for the zeroth frequency 
the first Matsubara frequency~\footnote{This approximation applies as long as the analytically continued $\Sigma(z = i\nu)$ does not present any features below the first Matsubara frequency.
Otherwise, $G_{\text{coh}}$ has to be determined by an analytic continuation.},
we find that the self-energy exhibits a minimum at $i\nu \sim \omega_0$. Therefore, we restrict this analysis to $\omega_0 > 2\pi/\beta$. 
The results for $\Pi^\dSC$ are shown in Fig.~\ref{fig:bubble_fluct_diag}, from which emerges that the observed decrease for low $\omega_0$ is due to the reduction of the coherent part, reflected also in the 
quasiparticle residue.
We attribute this behavior to the increased incoherent scattering with the phonons as they become softer and softer. 
We conclude that the quasiparticle damping due to phonons drives the suppression of $d$-wave pairing in the Hubbard-Holstein model. 

%-------------
%-------------

%-------------
\section{Conclusions and outlook}
\label{sec:concl}
%-------------

Using the SBE formulation of the fRG, we have presented a detailed analysis of the Hubbard-Holstein model on the square lattice, both at half filling and finite doping.
In contrast to previous studies, the full frequency dependence of the vertex as well as the electron and phonon self-energies are accounted for. 
We computed the ($s$- and $d$-wave) superconducting $\chi^\SC$, density $\chi^\M$, and magnetic $\chi^\M$ susceptibilities in the $(U,V_H)$ plane (in the SBE approximation) and investigated their evolution with $\omega_0$ from the adiabatic to the anti adiabatic limit.
We found that the screening induced by the self-energy is responsible for changing the sign of the isotope effect in the $d$-wave superconducting susceptibility $\chi^\dSC$, see Fig.~\ref{fig:effect_of_selfenergy_on_isotope_effect}. 
The role of the phonon-self energy, implicitly included through the renormalization of the density susceptibility, has been shown to be responsible for the nonmonotonicity of the $s$-wave $\SC$ susceptibility as a function of the electron-phonon coupling, indicating a regime where ME theory is not valid, see Figs.~\ref{fig:sc_d_suscs_vs_Vh_doped} and \ref{fig:sc_susc_vs_filling}. 
We also showed how using the exact expression of the phonon self-energy prevents the occurrence of an unphysical lattice instability without a charge-density wave.
Performing a fluctuation diagnostics, we furthermore analyzed the influence of phonons on electronic fluctuations and their interplay. 
We have shown how the retardation of the phonon-mediated interaction reverses the role of the density fluctuations from being supportive to being destructive to $s$-wave superconductivity. 
In addition, we supplemented the numerical results by an analytical derivation of the diagnostic matrices which allowed us to identify the key diagrams driving the influence of phonons in the different fluctuation channels.
One important conclusion that we can draw from these fluctuation-diagnostics results is 
%In relation to the experimental "bounds on superconducting $T_c$" for phonon-based  superconductors, we have clearly shown how 
that phonon-induced charge fluctuations can suppress superconductivity even for increasing electron-phonon coupling.

On the methodological side, we illustrated that the SBE formulation of the fRG provides a suitable tool to study the complex interplay of the different fluctuation channels in the various regimes of the Hubbard-Holstein model.
In particular, the fRG description goes beyond ME theory. 
Despite the substantial reduction of the computational effort provided by the SBE approximation, the application to multi-orbital systems remains challenging within a full frequency-dependent treatment of both, the vertex as well as the self-energy. 
Here, promising candidates for a more efficient frequency treatment are  the IR~\cite{shinaoka2017compressing,chikano2018performance} or DLR~\cite{kaye2022discrete,kaye2022libdlr} bases, in addition to the recently introduced quantic tensor trains ~\cite{Shinaoka2023} and tensor cross interpolation ~\cite{Rohshap2024}. 

Future directions of this work include the extension to the strong-coupling regime by exploiting the dynamical mean-field theory ~\cite{Metzner1989,Georges1996}  solution as a starting point for the fRG flow~\cite{Taranto2014,Vilardi2019,Bonetti_2022}, the investigation of Peierls bond phonons \cite{PhysRevLett.42.1698,RevModPhys.60.781} which couple to the hopping instead of the charge density, and the exploration of hexagonal-lattice models as relevant to many correlated moir\'e materials. 

%-----------
\section*{Acknowledgements}
%-----------

The authors acknowledge A.~Baum, D.~Campbell, L.~Classen,
F.~Domizio, M.~Capone, I.~Eremin, K.~Fraboulet, S.~Giuli, P.~Kopietz, F.~Krien, E.~Moghadas, M.~Patricolo, L.~Philoxene, J.~Profe, and  A.~Toschi for fruitful 
discussions. 
We acknowledge financial support from the Deutsche Forschungsgemeinschaft~(DFG) within the research unit
FOR 5413/1 (Grant No. 465199066).
This research was also funded by the Austrian Science Fund (FWF) 10.55776/I6946.
A.A.-E. and M.M.S acknowledge funding from the Deutsche Forschungsgemeinschaft (DFG, German Research Foundation) under Project No.~277146847 (SFB 1238, project C02). MMS also acknowledges funding from DFG Project No.~452976698 (Heisenberg program).
Computational resources provided by the Paderborn Center for Parallel Computing (PC2), specifically the Noctua 2 high-performance computing system, have been used for conducting this research.
In addition, we also used the Austrian Scientific Computing (ASC) infrastructure. 
For open access purposes, the author has applied a CC BY public copyright license to any author accepted manuscript version arising from this submission.

\appendix

%------------
\section{Acoustic phonons}
\label{app:acph}
%------------

%------------
In the seminal work \cite{morel_anderson1962}, it was argued within the framework of (Unrenormalized) Migdal-Eliashberg theory that it is the short-wavelength phonons that primarily mediate the electron-electron attraction relevant for $s$-wave superconductivity. Consequently, a model of optical/Einstein phonons was argued to be a better approximation for superconductivity in metals. As we have seen in the main text, the role of inter-channel feedback, particularly the influence of density fluctuations, plays a crucial role. A question that remains then is, to what extent would considering an acoustic/Debye phonon instead, that is gapless at $\bfq = 0$, influence our conclusions?

To that end, we consider the additional case of an acoustic phonon and briefly comment on it.
We will show that the results are qualitatively the same as the ones discussed in the main text.

We consider an acoustic phonon with dispersion
\begin{align}
\omega(\bfq) =  \frac{\omega_0}{2}\left(\abs{\sin(\bfq_x)} + \abs{\sin(\bfq_ y)}\right).
\label{eq:acoustic_phonon_dispersion}
\end{align}
From Eq.~\eqref{eq:electron_phonon_coupling}, it follows that the electron phonon coupling $g(\bfq)$ takes the form
\begin{align}
g(\bfq) \sim \frac{(\bfq_x + \bfq_y) V_\textrm{el-ion}(\bfq)}{\sqrt{\omega(\bfq)}}\,.
\end{align}
Integrating out this phonon as in appendix~\ref{app:hh} yields an effective electron-electron interaction with a coupling strength  $V_H(\bfq) = 2g^2(\bfq)/\omega(\bfq)$. Assuming that the electron-ion interaction is strongly screened such that $V_\textrm{el-ion}$ can be approximated by a constant, we have $g(\bfq) \sim \sqrt{|\bfq|}$. In the proximity of $\bfq = 0$, we have $\omega(\bfq) \sim \abs{\bfq}$, and thus we obtain $V_H(\bfq) \sim \text{const.}$ Therefore, the gaplessness at $\bfq = 0$ will not show up as a feature for $V_H$ but only for $\omega(\bfq)$. Thus, also for the treatment of acoustic phonons, we will consider $V_H$ to be a (constant) model parameter.

We performed calculations at half filling, for the same parameters as in the main text, see  Figs.~\ref{fig:halffilling_susceptibility_fluctdiag_acoustic} and  \ref{fig:halffilling_inverse_susceptibility_omega0_influence_acoustic} for the results. We find a qualitative agreement with the ones of Figs.~\ref{fig:halffilling_inverse_susceptibility_omega0_influence} and \ref{fig:halffilling_susceptibility_fluctdiag} for the optical phonons. In particular, the picture provided in Fig.~\ref{fig:interplay_purehubbard} applies as well for acoustic phonons.

\begin{widetext}\ 
\begin{figure*}[h!]
    \centering
    \includegraphics[width=0.99\linewidth]{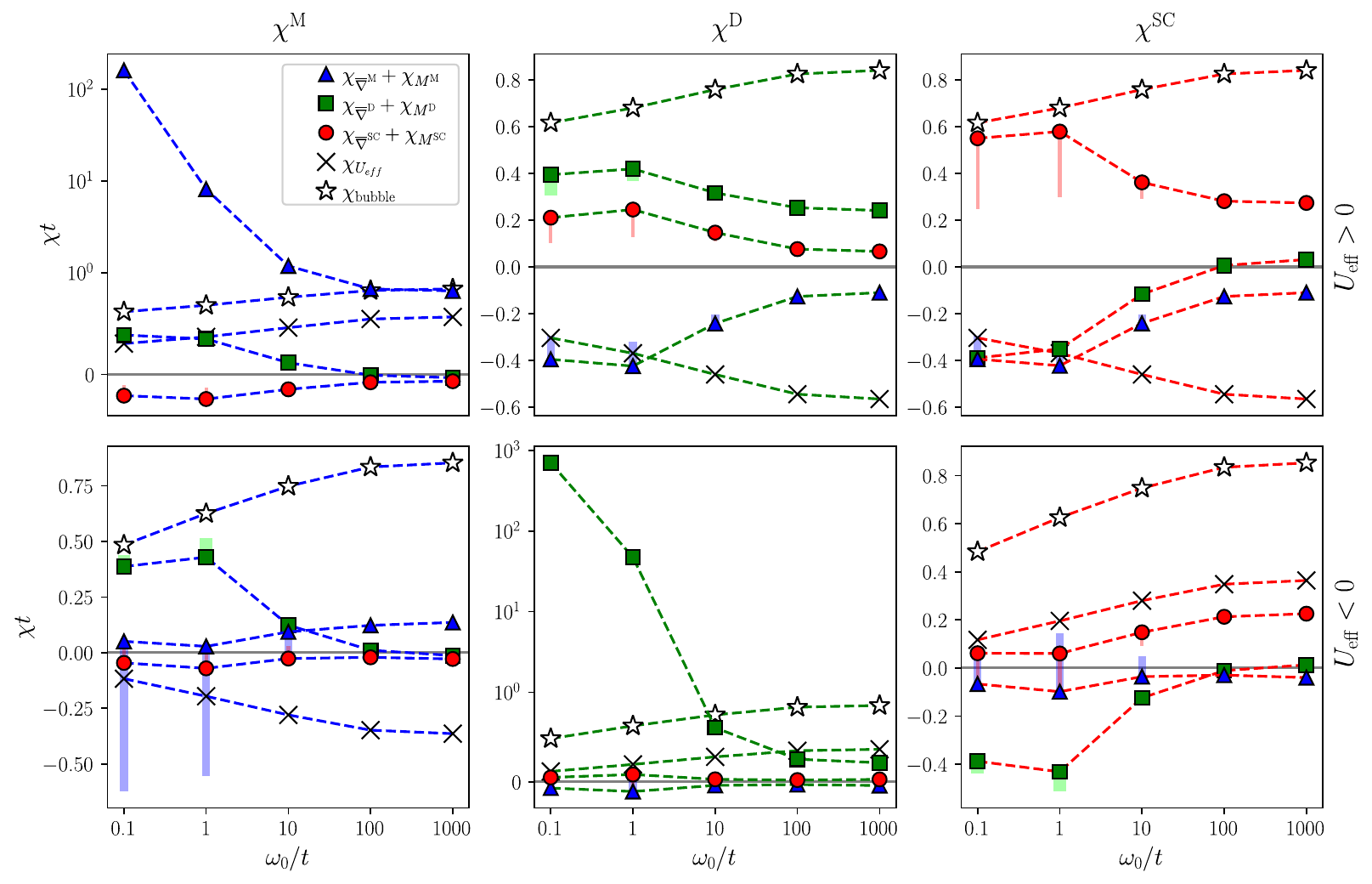}
    \caption{Results for fluctuation diagnostics 
    as in Fig.~\ref{fig:halffilling_susceptibility_fluctdiag}, for the same parameters but for the acoustic phonon dispersion \eqref{eq:acoustic_phonon_dispersion}.}    \label{fig:halffilling_susceptibility_fluctdiag_acoustic}
\end{figure*}
\end{widetext}

\begin{figure}
    \centering
    \includegraphics[width=1.0\linewidth]{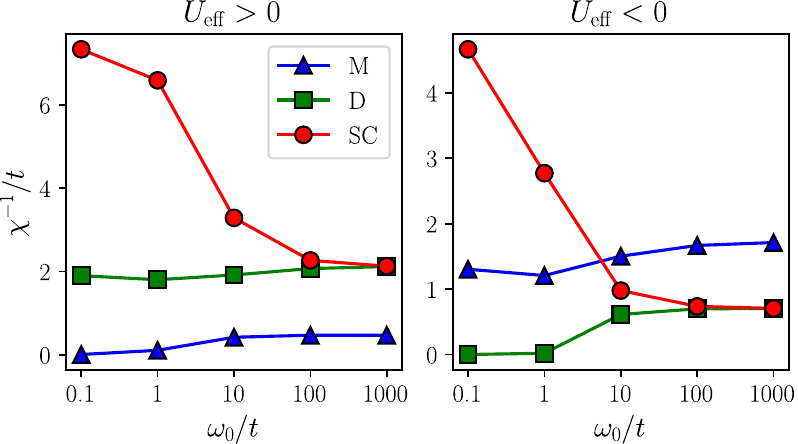} 
    \caption{Inverse susceptibilities as in Fig.~\ref{fig:halffilling_inverse_susceptibility_omega0_influence} but with the dispersionless phonon replaced by the acoustic phonon dispersion of Eq. \eqref{eq:acoustic_phonon_dispersion}.}    \label{fig:halffilling_inverse_susceptibility_omega0_influence_acoustic}
\end{figure}

\section{Effective interaction of the Hubbard-Holstein model}
\label{app:hh}
In this section, we briefly recall the derivation of the phonon-mediated effective electron-electron interaction in the Hubbard-Holstein model.
As a starting point one considers the ions at positions $\bfR$ oscillating around their equilibrium positions $\bfR^*$ with displacement $\bfu$ in analogy to a system of coupled harmonic oscillators. The linear mode analysis of the equations of motion in momentum space yields the eigenvalue problem
\begin{align}
D(\bfq) \bfe_j(\bfq) = \omega_j^2 \bfe_j(\bfq)\,, 
\end{align}
where the eigenvectors $\bfe_j$ describe the normal modes or the polarization of the ion oscillations identified by the phonon type or branch and~$D$
is the so-called dynamical matrix. 
Its components are given by
\begin{align}
 D_{ab}(\bfq) := \sum_{{\bfR} \in \Lambda} \frac{1}{M_\textrm{ion}} \Phi_{ab}({\bfR}) \exp(i\bfq\cdot {\bfR})\,, \label{eq:dynamical_matrix}
\end{align}
with $\Phi_{ab}$ the components of the Hessian of the ion-ion potential $V_{\textrm{ion-ion}}$. 
Note that Eq.~\eqref{eq:dynamical_matrix} implies that $\omega_j \propto 1/\sqrt{M_\textrm{ion}}$.

The components of the ion displacement from the equilibrium position are related to the phonon operators by 
\begin{align}
\bfu = \sum_{\bfq j}\frac{1}{\sqrt{2M_\textrm{ion}\omega_j(\bfq)}} \bfe_j(\bfq) e^{i\bfq \cdot \bfR^*} \left( b_{\bfq j} + b^\dagger_{-\bfq j} \right).
\end{align}
The Holstein electron-phonon interaction is then determined by the linear expansion of the electron-ion interaction [neglecting terms of $\mathcal{O}(\bfu^2)$ beyond the the phonon dispersion]
\begin{align}
\!\!V_\textrm{el-ion}({\bf{r}} \!-\! \bfR) =& V_\textrm{el-ion}({\bf{r}}\! -\! \bfR^*)\!-\! \bfu \cdot \nabla 
V_\textrm{el-ion}({\bf{r}} \!-\! \bfR)\,, 
\end{align}
where $\bf{r}$ indicates the position of the electron.
In second quantization, the Hamiltonian reads
\begin{align}
H_\textrm{el-ph} =& \sum_{\mathbf{k}\sigma\mathbf{q}j} \frac{\bfe_j(\bfq)}{\sqrt{2M_\textrm{ion} \omega_{j}(\bfq) }}\cdot\bra{\bfk + \bfq, \sigma}\nabla 
V_\textrm{el-ion}\ket{\bfk, \sigma}\nonumber\\& 
\qquad\times
c^\dagger_{\bfk + \bfq \sigma} c_{\bfk \sigma} \left(b_{\bfq j} + b^\dagger_{-\bfq j}\right).
\end{align}
The final expression is obtained by identifying
\begin{align}
g^j_{\mathbf{k}\bfq}  :=  \frac{ \bfe_j(\bfq)}{\sqrt{2M_\textrm{ion} \omega_{j}(\bfq)}} \cdot \bra{\bfk + \bfq, \sigma}\nabla 
V_\textrm{el-ion}\ket{\bfk, \sigma}\,, \label{eq:electron_phonon_coupling}
\end{align}
with $g^j_{\mathbf{k}\bfq} \propto 1/\sqrt{M_\textrm{ion} \omega_{j}(\bfq)} \propto 1/\sqrt[4]{M_\textrm{ion}}$.

Integrating out the phonons, that couple to the electrons through the (modified) 
displacement operators
 $   A_{\bfq j} = b_{\bfq j} + b^\dagger_{-\bfq j}$
and $A^\dagger_{\bfq j} = A_{-\bfq j}$, one obtains a purely electronic Hamiltonian.
In the path integral formulation of the coherent-state fields for $c^\dagger_{\bfk \sigma}, c_{\bfk \sigma}, A^\dagger_{\bfq j}, A_{\bfq j}$, the action reads
\begin{align}
    S[\bar{\psi}, \psi, \phi^*, \phi] =& -\int_{k\sigma} \bar{\psi}_{k\sigma} G_{0}^{-1}(k\sigma)\psi_{k\sigma}\nonumber\\ &+ \int_{qj} \phi^*_{qj} D_0^{-1}(qj) \phi_{qj}\nonumber\\ &+ \int_{k\sigma qj} g^j_{\mathbf{k}\bfq} \bar{\psi}_{k + q \sigma} \psi_{k \sigma} \phi_{qj} + \textrm{c.c.},  
\end{align}
where we introduced $q = (\bfq, i\Omega)$ and $k = (\bfk, i\nu)$. For simplicity, we restrict ourselves to a single phonon mode and drop the branch index in the following. The electronic bare propagator is determined by  
\begin{align}
\label{eq:action}
G_{0}(k) = \int_\tau e^{i\nu \tau} \expval{\mathrm{T} c_k(\tau) c^\dagger_k}_0 = \frac{1}{i \nu - e(\bfk)},
\end{align}
while the phononic propagator is given by
\begin{align}
D_{0}(q) =& -\int_\tau e^{i\Omega \tau} \expval{\mathrm{T} A_{q}(\tau) A_{-q}(0)}_0 \nonumber\\=& -\int_\tau e^{i\Omega \tau} \expval{\mathrm{T} (b_{\bfq j}(\tau) + b^\dagger_{-\bfq j}(\tau))(b_{-\bfq j} + b^\dagger_{\bfq j}) } \nonumber\\
=& -\int_\tau e^{i\Omega \tau}\left( \expval{\mathrm{T} b_{\bfq j}(\tau)b^\dagger_{\bfq j}}_0 - \expval{\mathrm{T} b^\dagger_{-\bfq j}(\tau)b_{-\bfq j}}_0\right) \nonumber\\
=& \frac{1}{i\Omega - \omega(\bfq)} + \frac{1}{-i\Omega - \omega(\bfq)} = \frac{2\omega(\bfq)}{(i\Omega)^2 - \omega^2(\bfq)}, \label{eq:phonon_propagator}
\end{align}
where the terms violating the particle number conservation vanish. 
We can hence integrate out the $\phi$ fields by completing the square in the action.
Holstein phonons arising from lattice distortions do not depend on $k$ in general. 
Neglecting the $k$-dependence, the phonon-mediated electron-electron interaction reads 
\begin{align}
V_{H}(q) = \frac{2 \abs{g_{\bfq}}^2 \omega(\bfq)}{(i\Omega)^2 - \omega^2(\bfq)}\,. 
\end{align}
For Einstein or optical phonons, a dispersionless phonon mode $\omega(\bfq) = \omega_0$ and constant electron-phonon coupling $g_\bfq = g$ is used. Then, the above expression simplifies to 
\begin{align}
V_H(i\Omega) = - V_{H} \frac{\omega^2_0}{\Omega^2 + \omega^2_0}\,,
\label{eq:bare_interaction2}
\end{align}
with  $V_H = 2\abs{g}^2/\omega_0$ controlling the coupling strength. The phonon frequency $\omega_0 > 0$ sets the relevant scale of the phonon dynamics with respect to the electron one. 
 
We finally note that different conventions for the definition of the electron-phonon coupling $g$ exist in the literature. In our definition we have considered $g$ to be the coupling constant of the electrons to the modified displacement operators $A$ which are directly related to the phonon operators $b^{(\dagger)}$. In some works, $\tilde{g}$ was taken to be the coupling constant of the electrons to the positions $\bfX := \bfu/\sqrt{M_\textrm{ion}}$. The two conventions are related simply by $\tilde{g} = g \sqrt{2\omega_0}$, with $V_H = \tilde{g}^2/\omega_0^2$ in that case. 

%-------------
\section{Channel conventions}
\label{app:cc}
%-------------

The two-particle vertex operator of a U$(1)$ invariant system in the quantum effective action 
reads
\begin{align}
\frac{1}{4}\sum V_{\sigma_1\sigma_2\sigma_3\sigma_4}(k_1, k_2, k_3, k_4) c_{\sigma_1}(k_1) c^\dagger_{\sigma_2}(k_2) c_{\sigma_3}(k_3) c^\dagger_{\sigma_4}(k_4)\,,
\end{align}
where $k_1$, $k_3$ represent the ingoing and $k_2$, $k_4$ the outgoing legs. If the system has also SU$(2)$ and lattice translation symmetry, the  
full vertex can also be written in the form of Eq.~\eqref{eq:spinconv}. 
Therefore, there is only a single independent coupling function 
from which the full vertex can be reconstructed: 
\begin{align}
V_{\uparrow\uparrow\downarrow\downarrow}(k_1, k_2, k_3, k_4) =: V(k_1, k_2, k_3, k_4)\,, \label{eq:full_vertex_function_definition}
\end{align}
with the only nonvanishing terms corresponding to $k_4 = k_1 - k_2 + k_3$ for a system with translation invariance. 

In the following, we consider the spin structure. 
Focusing on charge sectors, we distinguish two ways to view the vertex as a two body operator 
\begin{subequations}
\begin{align}
V^{\mathrm{PH}}_{\sigma_\textrm{in}\sigma_\textrm{in}' | \sigma_\textrm{out} \sigma_\textrm{out}'} &:= V_{\sigma_\textrm{in}, \sigma_\textrm{in}', \sigma_\textrm{out}, \sigma_\textrm{out}'}  \qq{\ (\text{charge = 0})}\\V^{\mathrm{PP}}_{\sigma_\textrm{in}\sigma_\textrm{in}' | \sigma_\textrm{out} \sigma_\textrm{out}'} &:= V_{\sigma_\textrm{in}, \sigma_\textrm{out}, \sigma_\textrm{in}', \sigma_\textrm{out}'} \;\qq{(\text{charge = 1})}
\end{align}
\end{subequations}
where the omitted frequency-momentum arguments follow the spin arguments. 
In the spin-diagonal basis
\begin{subequations}
\begin{align}
\ket{\D} &= \frac{1}{\sqrt{2}}\sum_{\sigma, \sigma'}\delta_{\sigma\sigma'} c^\dagger_{\sigma} c_{\sigma'} \ket{} = \frac{1}{\sqrt{2}}(\ket{\uparrow\uparrow} + \ket{\downarrow\downarrow})\\
\ket{\M} &= \sum_{\sigma\sigma'} \mathbf{\sigma^z}_{\sigma\sigma'}c^\dagger_{\sigma} c_{\sigma'}\ket{} = \frac{1}{\sqrt{2}}(\ket{\uparrow\uparrow} - \ket{\downarrow\downarrow})\,,
\end{align}
\end{subequations}
Expressing $V^{\mathrm{PH}}$ in this basis diagonalises it and yields
\begin{subequations}
\begin{align}
V^{\D}(k_1, k_2, k_3, k_4) &:= \left[V^{\mathrm{PH}}_{\uparrow\uparrow|\uparrow\uparrow} +V^{\mathrm{PH}}_{\uparrow\uparrow|\downarrow\downarrow}\right](k_1, k_2 | k_3, k_4)\nonumber\\
&= 2V(k_1, k_2, k_3, k_4) - V(k_1, k_4, k_3, k_2)   \\
V^{\M}(k_1, k_2, k_3, k_4) &:= \left[V^{\mathrm{PH}}_{\uparrow\uparrow|\uparrow\uparrow} - V^{\mathrm{PH}}_{\uparrow\uparrow|\downarrow\downarrow}\right](k_1, k_2|k_3, k_4)\nonumber\\
&=  - V(k_1, k_4, k_3, k_2)\,.
\end{align}
\end{subequations}
$V^\D$ is the density or charge component of the vertex that acts on the $S = 0$ sector and $V^\M$  the magnetic or spin component 
in the $z$-direction acting on the $S = 1$, $S_z = 0$ sector. The $S_z = \pm 1$ triplet components represented by the $x$ and $y$ magnetic vertices are degenerate with $V^\M$ by SU(2) symmetry.
Similarly, the singlet and triplet $S_z = 0$ particle-particle states \begin{subequations}
\begin{align}
\ket{s\SC} &= \frac{1}{\sqrt{2}}(c^\dagger_{\uparrow} c^\dagger_{\downarrow} - c^\dagger_{\downarrow} c^\dagger_{\uparrow}) \ket{} = \frac{1}{\sqrt{2}}(\ket{\uparrow\downarrow} - \ket{\downarrow\uparrow})\\ \ket{t\SC} &= \frac{1}{\sqrt{2}}(c^\dagger_{\uparrow} c^\dagger_{\downarrow} +c^\dagger_{\downarrow} c^\dagger_{\uparrow})\ket{} = \frac{1}{\sqrt{2}}(\ket{\uparrow\downarrow} + \ket{\downarrow\uparrow})\,.
\end{align}
\end{subequations}
can be used to diagonalize $V^{\mathrm{PP}}$ to yield 
\begin{subequations}
\begin{align}
V^{s\SC}(k_1, k_2, k_3, k_4) &= [V^{\mathrm{PP}}_{\uparrow\downarrow|\uparrow\downarrow} - V^{\mathrm{PP}}_{\uparrow\downarrow|\downarrow\uparrow}](k_1, k_3| k_2, k_4)\nonumber\\
&= V(k_1, k_2, k_3, k_4) + V(k_1, k_4, k_3, k_2)   \\
V^{t\SC}(k_1, k_2, k_3, k_4) &= [V^{\mathrm{PP}}_{\uparrow\downarrow|\uparrow\downarrow} + V^{\mathrm{PP}}_{\uparrow\downarrow|\downarrow\uparrow}](k_1, k_3 | k_2, k_4)\nonumber\\
&=  V(k_1, k_2, k_3, k_4) - V(k_1, k_4, k_3, k_2)\,.
\end{align}
\end{subequations}
For a purely local interaction in a one-band model, the triplet component vanishes. However, this is not the case for an extended or retarded interaction.

\begin{figure}[b]
\centering
    \includegraphics[width=0.7\linewidth]{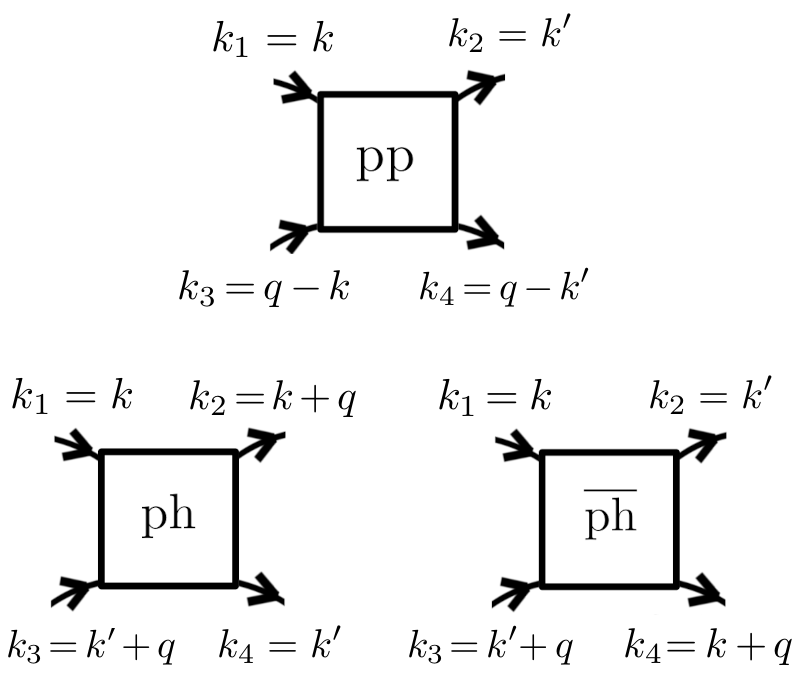}
    \caption{The three diagrammatic channel parametrizations of the two-particle vertex  
    $V(k_1, k_2, k_3, k_4)$.}
    \label{fig:diagrammatic_channel_conventions}
\end{figure}

At this stage, a parametrization based on a bosonic transfer momentum-frequency $q$ turns out to be useful. The diagrammatic channel parametrization is shown in Fig.~\ref{fig:diagrammatic_channel_conventions}. In terms of this parametrization, the spin diagonalized components of the vertex are given by
\begin{subequations}
\label{eq:physical_vertices}
\begin{align}
 V^\M_{k, k'}(q) &= -V^\xph_{k, k'}(q) = - V(k, k', k'+q, k + q)\\ 
    V^\D_{k, k'}(q) &=  2V^\ph_{k, k'}(q)- V^\xph_{k, k'}(q)\nonumber\\ &= 2V(k, k+q, k'+q, k') - V(k, k', k'+q, k+q) \\
        V^\SC_{k, k'}(q) &=  V^\pp_{k, k'}(q) = V(k, k', q - k, q - k')\,.
\end{align}
\end{subequations}
Note that we combined the singlet and triplet superconducting components to a single channel $\SC$ from which the former two can be extracted via 
\begin{subequations}
\begin{align}
V^{\mathrm{sSC}}_{k, k'}(q) &= V^{\SC}_{k, k'}(q) + V^{\SC}_{k, q-k'}(q)\\
V^{\mathrm{tSC}}_{k, k'}(q) &= V^{\SC}_{k, k'}(q) - V^{\SC}_{k, q-k'}(q)\,.
\end{align}
\end{subequations}
In the main text, we use the conversion from the $\ph$ and $\xph$ to the $\pp$ parameterization
\begin{align}
q_{\ph} = k'_{\pp} - k_{\pp}\,, \  k_{\ph} = k_{\pp}\,, \  k'_{\ph} = q_{\pp} - k'_{\pp}\,,\label{eq:q_ph_to_pp}\\
q_{\xph} = q_{\pp} - k_{\pp} - k'_{\pp}\,, \  k_{\xph} = k_{\pp}\,, \  k'_{\xph} = k'_{\pp}\,.\label{eq:q_xph_to_pp}
\end{align}

\section{Technical details on the numerical implementation}
\label{app:technical_impelementation}

We here provide the technical parameters used for  obtaining the presented numerical results. We also assess the quality of the SBE approximation neglecting the flow of the rest function and discuss the importance 
of the self-energy flow. The flow equations were integrated numerically using the Runge-Kutta Cash-Karp 5(4) method with adaptive step size as implemented in the Boost C++ library \texttt{odeint} (\texttt{runge\_kutta\_kashkarp\_54}).

\subsection{Momentum and frequency dependencies}

\begin{figure}
    \centering
    \includegraphics[width=0.6\linewidth]{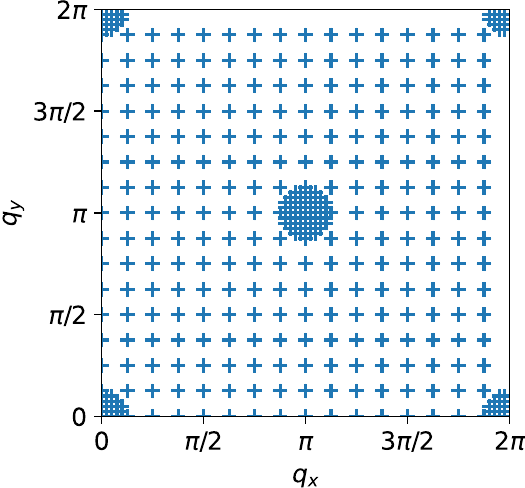}
    \caption{The momentum grid used for  
    the transfer momentum dependence of $w^\X$, $\lambda^\X$, and $M^\X$. We insert a refinement of $N_{k, \delta}$ additional points around the $\Gamma$ and $M$ points. These were chosen by including 
    $2\%$ of the number of shells
    around the centers of the refinement on a finer grid with $(N_kN_{\delta})^2$ (with $N_k = 16$ and $N_{\delta}=5$) points.}
    \label{fig:enter-label}
\end{figure}

\begin{table}
    \centering
    \begin{tabular}{c|cccc}
 & \parbox{60pt}{Bosonic \\frequencies}& \parbox{60pt}{Fermionic frequencies}\ &\parbox{50pt}{Bosonic momenta}\ & \parbox{50pt}{Fermionic momenta}\\
 \\
 \hline
 \\
         $w$&  $128 N_w + 1$&  $\varnothing$& $N_k^2+N_{k,\delta}$ & $\varnothing$\\
         $\lambda$&  $4N_w + 1$&  $4N_w$& $N_k^2+N_{k,\delta}$ & $\text{Form factors}$\\
         $M$&  $4N_w + 1$&  $2N_w$& $N_k^2+N_{k,\delta}$ & Form factors \\
         $\Sigma$&  $\varnothing$&  $20N_w$& $\varnothing$ & $N_k^2$\\
         $\Pi$&  $128N_w+1$&   $128N_w$& $N_k^2N_\delta^2$ & Form factors\\
    \end{tabular}
    \caption{Number of frequency and momentum components used for the parametrization, where $N_\delta = 5$ for the bubble.
    For the treatment of the momenta within the TU-fRG, the bosonic momenta of the vertices live on a discretized Brillouin zone, whereas the fermionic momentum dependence is restricted to (a few) form factors. 
    Specifically, we use only an $s$-wave form factor at half filling, and include an additional $d$-wave one at finite doping. The fermionic momenta of the self-energy live on a discretized Brillouin zone. 
     }
    \label{tab:technical_parameters}
\end{table}

We use several Brillouin zone grid discretizations for different objects. The simplest is a uniform discretization into $N_k\times N_k$ points, where we have taken $N_k = 16$. This discretization was used for the momentum dependence of the self-energy $\Sigma$. A more refined momentum discretization was used for the bosonic momentum dependencies of $w^\X$, $\lambda^\X$, and $M^\X$, where the vicinity of the high symmetry points $\Gamma = (0, 0)$ and $M = (\pi, \pi)$ were refined as shown in Fig.~\ref{fig:enter-label}. A denser momentum grid of size $N_kN_{\delta}$ was used for the bosonic momentum dependence of the TU bubbles~\eqref{eq:bubble_definition}. 
The momentum integration in the bubble 
has been performed via fast Fourier transforms as specified in Ref.~\cite{Hille2020a}. As introduced in the main text, the fermionic momentum dependencies of $w^\X$, $\lambda^\X$, and $M^\X$ were expanded in form factors and then truncated to just the $s$-wave form factor at half filling and an additional $d$-wave form factor at finite doping. In the latter case, a further approximation was used, namely that the mixed bubbles vanish $\Pi^\X_{nm}(q) = 0 \qq{for} n\neq m$. This is in general true for $\X = \SC$ and $\bfq = 0$, but amounts to an approximation otherwise. For the discussed regime where only $\dSC$ at zero momentum is relevant, this is justified. 

Also the number of frequencies varies for the different objects, see Table~\ref{tab:technical_parameters}. Specifically, we used $N_w = 8$. For large fermionic frequencies outside the frequency windows, we have further approximated the nontrivial fermionic frequency asymptotics~\eqref{eq:vertex_nontrivial_asymptotics} by $\lambda^{\X,\text{asympt}} = 1$ and $M^{\X,\text{asympt1}} = M^{\X,\text{asympt1}'} = M^{\X,\text{asympt2}} = 0$, as from the definition of static susceptibilities~\eqref{eq:susc_from_vertex}. At $i\Omega = 0$, the large fermionic frequency contributions in the integrand will be suppressed by the decay of one of the two bubbles.
In the calculation of all the objects, lattice symmetries as well as algebraic symmetries were used where possible, see Ref.~\cite{Sarahthesis}.

%----------------
\subsection{SBE approximation}
\label{app:sbe_approximation}
%----------------

\begin{figure*}[t!]
    \centering
\includegraphics[width=\linewidth]{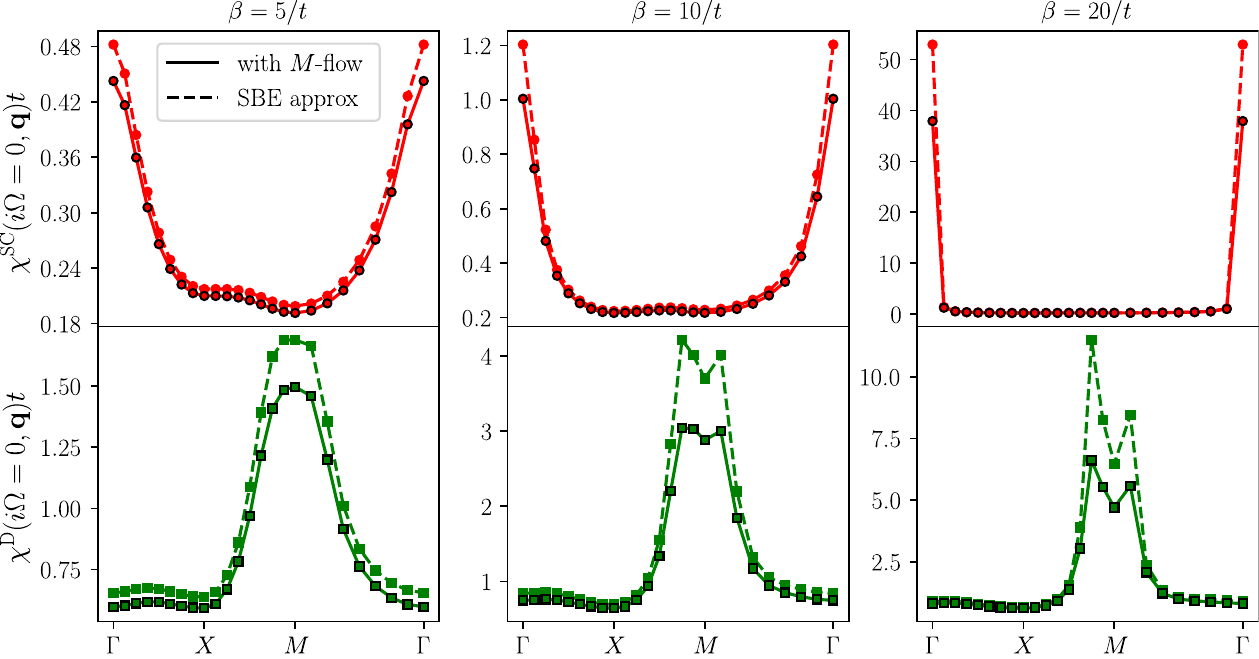}
    \caption{Effect of the SBE approximation on the density and superconducting susceptibilities at finite doping, 
    for  $V_H = 2.5t$, $U = 0$, $\omega_0 = 1.5t$, $t'=-0.25t$, and $\mu=-1t$, and various temperatures around the boundary of the reduced Brillouin zone.
    }
    \label{fig:sbeapp}
\end{figure*}

We here investigate the validity of the SBE approximation, i.e. of neglecting the flow of $M^\X$ in Eq.~\eqref{eq:sbe_flow_equations}. We stress here that this approximation does not correspond to neglecting all multiboson diagrams in the vertex $V$ \footnote{In contrast to the SBEb approximation which neglects all multiboson diagrams, the SBEa neglects only the flow of $M^\X$ and is less severe; multiboson insertions into the single-boson part of the vertex can still occur \cite{fraboulet2023singlebosonexchangefunctionalrenormalization}.}, but only the flow of $M^\X$. The effect of this approximation is shown in Fig.~\ref{fig:sbeapp} for various temperatures. 
%
%---------------
%---------------
%
The calculations were performed for a representative value of $\omega_0 = 1.5t$ far away from the anti-adiabatic limit, but not deep in the adiabatic limit. In order to focus on the phonon part of the problem, we set $U = 0$ and $V_H = 2.5t$. Furthermore, we consider finite doping with $n = 0.39$ and $t' = -0.25$. We observe that the SBE approximation is qualitatively successful, although quantitatively it suffers, in particular in correspondence of the ordering vectors. Thus, to lighten the numerical effort, we have opted to use the SBE approximation only for the calculation of the phase diagrams in Figs.~\ref{fig:phase_diagram_halffilling} and \ref{fig:phase_diagram_doped} and the two figures that analyzed cuts of its data (Figs.~\ref{fig:sc_d_suscs_vs_Vh_doped} and \ref{fig:phonon_softening}). In all the other calculations, we have included the flow of $M^\X$. For the purposes of the fluctuation diagnostics, the contributions of $M^\X$ were found to be important, as we saw that its inclusion can break the monotonicity of some of the different contributions as a function of $\omega_0$ as we approach the adiabatic limit. This can be seen by looking at the faint error bars in Figs.~\ref{fig:halffilling_susceptibility_fluctdiag} and \ref{fig:finitedoping_susceptibility_fluctdiag}, which show where the data points would be placed if only the contributions from $\bar{\nabla}^\X$ were considered. 

In the following, we provide an explanation for the quantitative failure of the SBE approximation near the adiabatic limit. The systematic increase of the importance of the multiboson contributions in that regime can be understood from the fact that the bosonic parts of the bare interaction essentially vanish 
\begin{align}
\mathcal{B}^\X \sim V_H - V_H \frac{\Omega^2}{\Omega^2 + \omega_0^2} \rightarrow -\delta_{\Omega, 0}\,,
\end{align}
whereas the fermionic parts become essentially constant
\begin{align}
\mathcal{F}^\X \sim V_H \frac{(\nu - \nu')^2}{(\nu - \nu')^2 + \omega^2_0} \rightarrow V_H (1-\delta_{\nu, \nu'})\,.
\end{align}
In particular, for finite frequencies $\mathcal{F}^\SC = -\mathcal{F}^\D = -\mathcal{F}^\M \sim V_H$. Thus, the importance of ladder diagrams of $\mathcal{F}$ will exceed the importance of ladders of $\mathcal{B}$. Since these ladders are contained in $M^\X$,  it becomes important in proximity of the adiabatic limit. A possible way out without including the flow of the full rest functions is to use simplified flow equations for $M^\X$ that generate only the $\mathcal{F}$ ladders 
\begin{align}
\mathrm{d}_\Lambda{M}^{\X,{\mathcal{F}\text{-lad}}}_{nn'}(q, i\nu, i\nu') = &-\text{Sgn}\X\sum_{\mathclap{i\nu'',m,m'}}\mathcal{F}_{nm}^\X(i\nu - 
i\nu'') \nonumber \\
&\!\!\!\!\!\times\mathrm{d}_\Lambda{\Pi}^\X_{mm'}(q, i\nu'')\mathcal{F}_{m'n'}^\X(i\nu'' - i\nu')\,.
\end{align}
If we approximate $\mathcal{F}^\SC(i\nu - i\nu') =- \mathcal{F}^\D(i\nu - i\nu')\sim - \mathcal{F}^\M(i\nu - i\nu') \sim V_H \Theta_{\omega_0}(|i\nu - i\nu'|)$, with $\Theta_{\omega_0}$ a step function at $\omega_0$, the above equation simplifies near the adiabatic limit
\begin{align}
\mathrm{d}_\Lambda{M}^{\X,{\mathcal{F}\text{-lad}}}_{\mathrm{sw},\mathrm{sw}}(q, i\nu, i\nu') \approx -\text{Sgn}\X (V_H)^2\sum_{\mathclap{i\nu'' \in \mathcal{R}^{\omega_0}_{\nu,\nu'}}}\mathrm{d}_\Lambda{\Pi}^\X_{\mathrm{sw},\mathrm{sw}}(q, i\nu'')\,,
\end{align}
where $\mathcal{R}^{\omega_0}_{\nu,\nu'} = \{i\nu'' \in [-i\nu_{\text{max}}, i\nu_{\text{max}}]\  |\ \ \omega_0< \abs{i\nu''-i\nu}  \ \text{and}\ \omega_0 < \abs{i\nu''-i\nu'} \}$ with $[-\nu_{\text{max}}, \nu_{\text{max}}]$ the maximum frequency window. Note that $\mathcal{R}^{\omega_0}_{\nu, \nu'}$ is the set of all frequencies within this window except for a few with a distance of $\omega_0$ from $\nu$ and $\nu'$. Thus, sufficiently close to the adiabatic limit $\omega_0 < \pi T$, the sums for different $\nu$ and $\nu'$ can be calculated cumulatively
\begin{align}
&\ \ \ \sum_{\mathclap{i\nu''\in \mathcal{R}^{\omega_0}_{\nu,\nu'}}}\mathrm{d}_\Lambda{\Pi}^\X_{\mathrm{sw},\mathrm{sw}}(q, i\nu'') = \left(\sum_{\mathclap{i\nu''}}\mathrm{d}_\Lambda{\Pi}^\X_{\mathrm{sw},\mathrm{sw}}(q, i\nu'') \right)- \nonumber \\ &\mathrm{d}_\Lambda{\Pi}^\X_{\mathrm{sw},\mathrm{sw}}(q, i\nu)   -\mathrm{d}_\Lambda{\Pi}^\X_{\mathrm{sw},\mathrm{sw}}(q, i\nu')  +\mathrm{d}_\Lambda{\Pi}^\X_{\mathrm{sw},\mathrm{sw}}(q, i\nu)\delta_{\nu, \nu'}\,,
\end{align}
where the first term containing the sum is independent of $\nu$ and $\nu'$. 
With an efficient calculation of the fermionic $\mathcal{F}^\X$-ladders contributions we expect that the quality of the SBE approximation can be recovered also in presence of soft phonons.

\subsection{Electronic self-energy flow}

We finally illustrate the importance of including the self-energy flow and of fixing the filling. Its inclusion can affect the results not only quantitatively, but also qualitatively. In particular, the comparison between the results obtained with and without the flow of the self-energy shows that the leading instability may be different. In Fig.~\ref{fig:self_energy_and_fixed_filling_influence}, we show the inverse of the maxima of the bosonic propagators $w^\D$ and $w^\SC$ along the RG flow, similarly to Fig. 4 of Ref.~\cite{Reckling_2018}. We chose the same parameters, with $V_H = 2.5t$, $\omega_0 = 1.5t$, $t'=-0.25t$, $\mu = -0.95t$, and $\beta = 20/t$. We show results for three cases: i) without self-energy flow, ii) with self-energy flow but without the flowing counter term of the chemical potential, and iii) the full calculation including both the self-energy and the chemical potential counter term flow to keep the filling fixed. As the scale is lowered, $w^\D$ begins to lead. In the absence of the self-energy flow, this continues and eventually emerges as an incommensurate charge-density wave instability at a finite scale. In the presence of a flowing self-energy, the growth of $w^\D$ is hampered at a later scale and finally overtaken bt $w^\SC$ which eventually diverges instead. In the insets, we show the effect of fixing the filling. In particular, in the absence of a chemical potential counter term, the filling is seen to change by more than $50\%$.

\begin{widetext}\ 
\begin{figure}[h!]
    \centering
    \includegraphics[width=\linewidth]{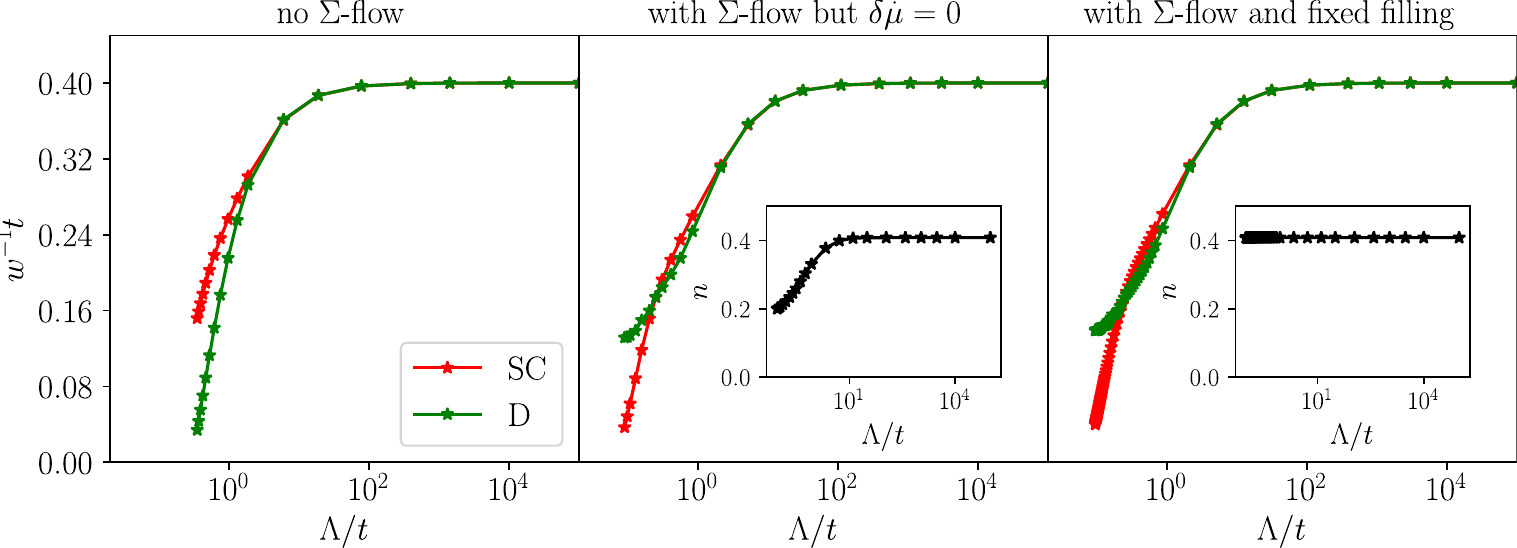} 
    \caption{ Evolution of the effective interaction with and without 
    self-energy flow, for  $V_H = 2.5t$, $U=0$, $\omega_0 = 1.5t$, $t'=-0.25t$, $\mu = -0.95t$, and $\beta = 20/t$.
    Note how the leading instability towards the end of the flow can change if the flow of the self-energy is taken into account.}
    \label{fig:self_energy_and_fixed_filling_influence}
\end{figure}
\end{widetext}

%---------------
\section{Migdal-Eliashberg theory} 
\label{app:ME}
%---------------

Here, we review the derivation of the 
ME~equation from the fRG flow equations, see Refs.~\cite{campbell_shankar_rg_me_theory} for the symmetric and Ref~\cite{Honerkamp_2005} for the symmetry-broken state.
In the symmetric phase, the unrenormalized ME~equation for the self-energy reads
\begin{align}
\Sigma_{\text{ME}}(k) = \int_{k'} G(k') V_0(k' - k) \label{eqn:Eliashberg_equation}
\end{align}
and is solved self-consistently for $V_0$ defined in Eq.~\eqref{eq:bareint}.
In the fRG, the unrenormalised ME is obtained by approximating the vertex to contain only forward scattering contributions in the $\xph$~channel
\begin{align}
V(k_1, k_2, k_3, k_4) \approx  &\delta(k_2 - k_3)\delta(k_1 - k_4)\times\notag\\
&V^\xph_{k_1, k_2}(q = k_3 - k_2) \,.
\end{align}
Plugging this ansatz for the vertex into the self-energy flow equation, Eq.~\eqref{eq:selfenergy_flow_eqn}, yields
\begin{align}
\mathrm{d}_{\Lambda}\Sigma(k) &=  \int_{k'} S(k') V(k', k, k, k') \,, \label{eq:selfenergy_forward_scattering} 
\end{align}
where the other term in the flow equation
vanishes 
(the nonvanishing elements have zero measure). 

The flow equation for the vertex reads
\begin{align}
\mathrm{d}_\Lambda V^{\xph}_{k_1, k_2}(q_\xph = 0) =&\ \mathrm{d}_\Lambda V(k_1, k_2, k_2, k_1) \nonumber \\ 
= \int_p V(k_1, p, p, k_1) &(\mathrm{d}_\Lambda{G})(p)G(p) V(p, k_2, k_2, p)\,. \label{eq:simple_xph_fs_flow_equation}
\end{align} 
In terms of the SBE decomposition and the flow Eqs.~\eqref{eq:sbe_flow_equations}, this equation is equivalent to
\begin{widetext}
\begin{subequations}
    \begin{align}       \mathrm{d}_{\Lambda}w^{\xph,\textrm{FS}} &=\left(w^{\xph,\textrm{FS}}\right)^2\sum_{m,m',\nu}\lambda_m^{\xph,\textrm{FS}}(i\nu)\mathrm{d}_\Lambda{\Pi}^\ph_{mm'}(q = 0,\nu)\lambda^{\xph,\textrm{FS}}_{m'}(i\nu)
                \\
        \mathrm{d}_\Lambda \lambda^{\xph,\textrm{FS}}_n(i\nu) &= \sum_{m,m',\nu'} \lambda^{\xph,\textrm{FS}}_m(\nu')\mathrm{d}_{\Lambda}{\Pi}^\ph_{mm'}(q = 0,i\nu')\mathcal{I}^{\xph,\textrm{FS}}_{m'n}(i\nu',i\nu) \\
        \mathrm{d}_{\Lambda}{M}^{\xph,\textrm{FS}}_{nn'}(i\nu,i\nu') &= \sum_{m,m',\nu''} \mathcal{I}^{\xph,\textrm{FS}}_{nm}(i\nu,i\nu'')\mathrm{d}_{\Lambda}{\Pi}^{\xph,\mathrm{FS}}_{mm'}(i\nu'')\mathcal{I}^{\xph,\textrm{FS}}_{m'n'}(i\nu'',i\nu')\,,
    \end{align}
\end{subequations}
\end{widetext}
where we use form-factor notation and ``$\xph$,FS'' stands for ``forward scattering in the $\xph$ channel''; i.e. (suppressing fermionic arguments) $w^{\xph,\textrm{FS}} = -w^{\M}(q=0)$, $\lambda^{\xph,\textrm{FS}} = -\lambda^\M(q = 0)$, $M^{\xph,\textrm{FS}} = -M^{\M}(q = 0)$, $\Pi^{\xph,\textrm{FS}} = -\Pi^\M(q = 0)$ and the irreducible vertex 
$\mathcal{I}^{\xph,\textrm{FS}} :=  V^\xph(q = 0) - \nabla^{\xph,\textrm{FS}} = M^{\xph,\textrm{FS}} - \mathcal{F}^\M$.
Besides these, the flow of the other components and other $q$~values is neglected.

The derivative appearing in the right hand side of Eq.~\eqref{eq:simple_xph_fs_flow_equation} is a total derivative (due to the Katanin-substitution). 
Therefore it admits a ladder solution
\begin{align}
V&(k_1, k_2, k_2, k_1) = V_0(k_1, k_2, k_2, k_1)\nonumber\\ &+ \int_p V(k_1, p, p, k_1) G(p)G(p) V_0(p, k_2, k_2, p), \label{eq:forward_scattering_vertex}
\end{align}
which we may insert into Eq.~\eqref{eq:selfenergy_forward_scattering} to find
\begin{widetext}
\begin{align}
\mathrm{d}_{\Lambda}{\Sigma}(k) &= \int_{k'} S(k')V_0(k', k, k, k') + \int_{k', p} S(k') V(k', p, p, k') G(p)G(p) V_0(p, k, k, p) \nonumber\\
&= \int_{k'} S(k') V_0(k', k, k, k')  + \int_{p}  \mathrm{d}_{\Lambda}{\Sigma}(p) G(p)G(p) V_0(p, k, k, p) = \int_{k'} \left(S(k') +  G(k')\mathrm{d}_{\Lambda}{\Sigma}(k')G(k') \right) V_0(k', k, k, k')\nonumber \\
&= \int_{k'} \mathrm{d}_{\Lambda}{G}(k') V_0(k', k, k, k')  \,.\label{eqn:Eliashberg_equation_derivative}
\end{align}
\end{widetext}
Integrating this equation from the initial scale $\Lambda_\textrm{init}$ to the final scale $\Lambda_\textrm{final}$, we recover Eq.~\eqref{eqn:Eliashberg_equation}.

%-------------------
\section{Derivation of the diagnostic matrix $D_{\partial \chi}$}
\label{app:diagm}
%-------------------

The key to determine the signs in Eq. \eqref{eq:diagma} is to note that within the SBE formulation, the $\omega_0$-dependence enters only through the density bosonic bare interaction in a diagrammatic
expansion of the susceptibilities
\begin{align}
\mathcal{B}^\D &= U_\textrm{eff} + U_{\omega_0}(\Omega)\, ,
\end{align}
with
\begin{align}
&U_{\omega_0}(\Omega) := 2V_H\frac{\Omega^2}{\Omega^2 + \omega_0^2}\,,
\end{align}
where $U_{\omega_0}$ contains the frequency-dependent retarded 
part of the interaction. 
In the weak coupling regime, both $V_H$ and $U_\textrm{eff}$ are small compared to 
the bandwidth allowing for a perturbative treatment. 
In the following, we determine the diagrammatic contributions to first order in $U_{\omega_0}$ and lowest order in $U_\textrm{eff}$.

\begin{figure}[h!]
    \centering
    \includegraphics[width=0.6\linewidth]{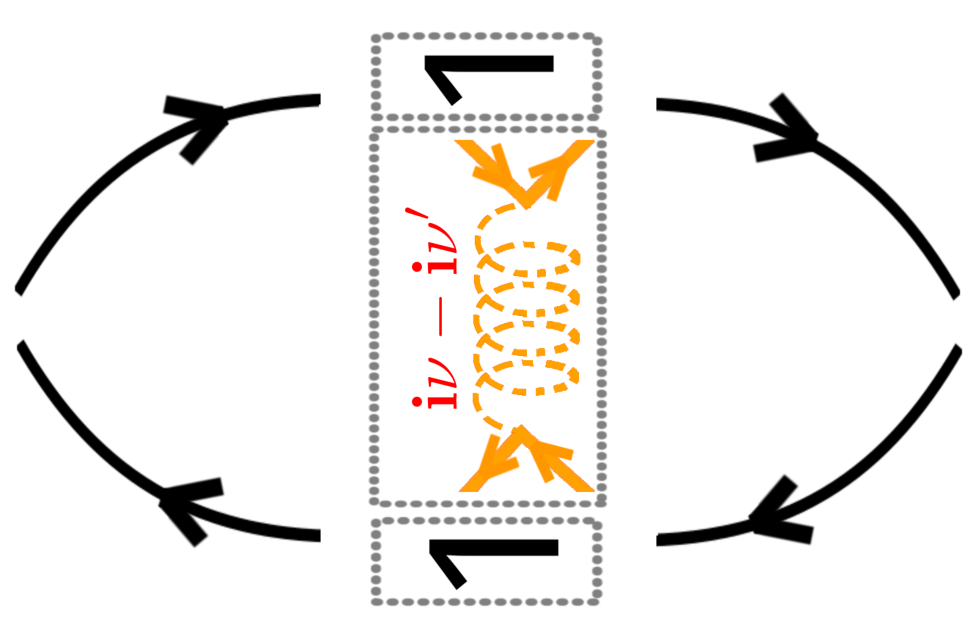}
    \caption{The leading-order diagram in $\chi^\M_{\nabla^\D}$. The yellow line represents the retarded part of the interaction $U_{\omega_0}$ which enters vertically as $\nabla^\D = \lambda^\D w^\D \lambda^\D$ is expressed in the $\xph$ channel. The three gray boxes emphasize the origin of the contributions: $1 \in \lambda^\D$ and $U_{\omega_0} \in \mathcal{B}^\D \in w^\D$. A similar diagram is obtained for $\chi^\SC_{\nabla^\D}$, where  $\nabla^\D$ is expressed in the $\pp$ channel instead
    and the left and right $\ph$ bubbles are replaced by $\pp$ bubbles.}
    \label{fig:leading_density_contributions_m_sc_omega_dependent}
\end{figure}

\begin{figure}[h!]
    \centering 
    \includegraphics[width=1.0\linewidth]{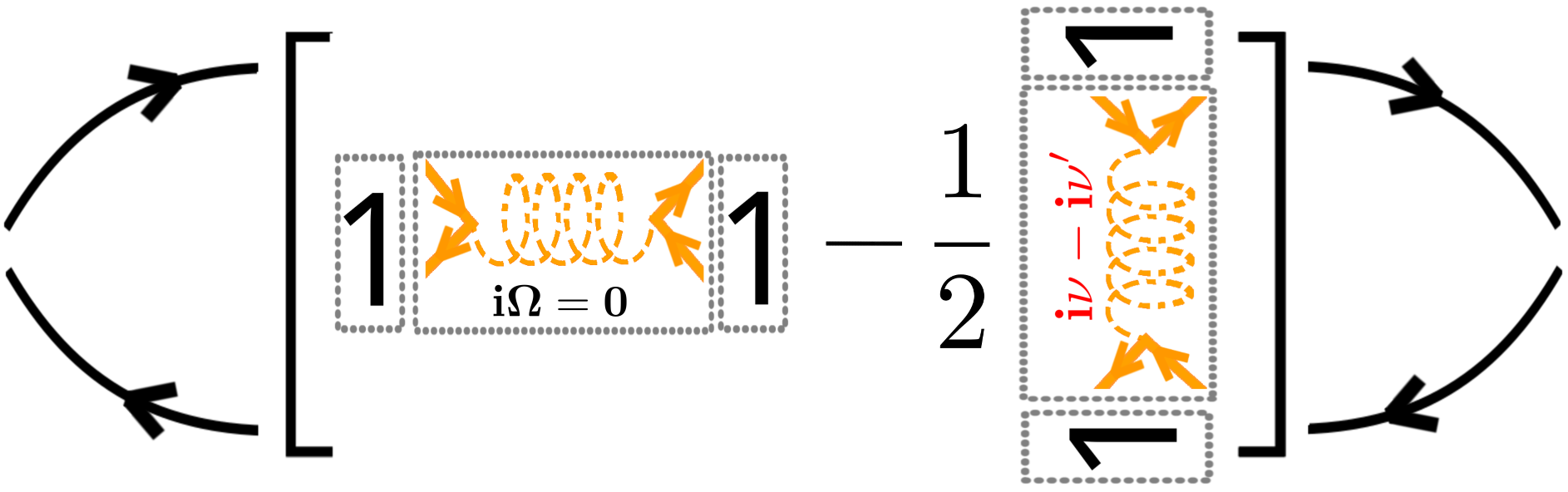}
    \caption{The leading-order diagram in $\chi^\D_{\nabla^\D}$, with contributions from the $\ph$ (horizontal) and $\xph$ (vertical) channels. Note that the bubble frequency varies only in the vertical insertion.}
    \label{fig:leading_density_contributions_d_omega_dependent}
\end{figure}

We first consider the density contributions to $\chi^\X$. 
The lowest order contributions of $\bar{\nabla}^\D$ are of first order in $U_{\omega_0}$ and do not contain $U_\textrm{eff}$. 
For $\chi^{\SC/\M}$, the $U_{\omega_0}$ interaction line is antiparallel. Therefore, the integration over the frequencies in the loops samples all values of $U_{\omega_0}$  
leading to an increase with increasing $M_\textrm{ion}$, see Fig.~\ref{fig:leading_density_contributions_m_sc_omega_dependent}. 
The diagnostic matrix elements are thus given by
\begin{align}
D_{\partial \chi^{\SC/\M}}^{\D} = &- \text{Sgn}\,\Pi^{\SC/\M} \cdot (-D_{V^\D}^{\SC/\M}) \cdot (+) \cdot \text{Sgn}\,\Pi^{\SC/\M}\nonumber\\
=&D_{V^\D}^{\SC/\M}\,.
\end{align}
We therefore obtain
\begin{align}
D_{\partial \chi^\X}^{\D} = \mqty(-\\ +  \\ +).
\end{align}
The situation for the density contribution to $\chi^\D$ is a little different. Here,  
in writing the zeroth order term we have to include the $\xph$ contribution from the bare contribution to $\nabla^\D$ given by $~(\mathcal{B}^\D)_{\ph} - (\frac{1}{2}\mathcal{B}^\D)_{\xph}$ [cf. Eq. \eqref{eq:vertex_d_rough}]. The first term is parallel, while the second is antiparallel. Since we calculate the static susceptibility at $i\Omega = 0$, the parallel part is also be evaluated at $i\Omega = 0$, which vanishes: $(\mathcal{B}^\D)_{\ph}(i\Omega = 0, \nu, \nu') \sim U_{\omega_0}(i\Omega = 0) = 0$. Thus, only the second term contributes $-\frac{1}{2}(\mathcal{B}^\D)_{\xph}(i\Omega = 0, \nu, \nu') \sim -\frac{1}{2}U_{\omega_0}(\nu - \nu')$, with integrated frequencies $\nu$ and $\nu'$. This additional minus sign implies that the zeroth order term will decrease as $\M_\textrm{ion}$ increases (see Fig.~\ref{fig:leading_density_contributions_d_omega_dependent}). The final diagnostic matrix element then reads
\begin{align}
D_{\partial \chi^{\D}}^{\D} &= - \text{Sgn}(\Pi^{\D} )\cdot \left(-D_{V^\D}^{\D}\right) \cdot (-) \cdot \text{Sgn}(\Pi^{\D})=-D_{V^\D}^{\D} \,.
\end{align}

\begin{figure}[h!]
    \centering
    \includegraphics[width=0.6\linewidth]{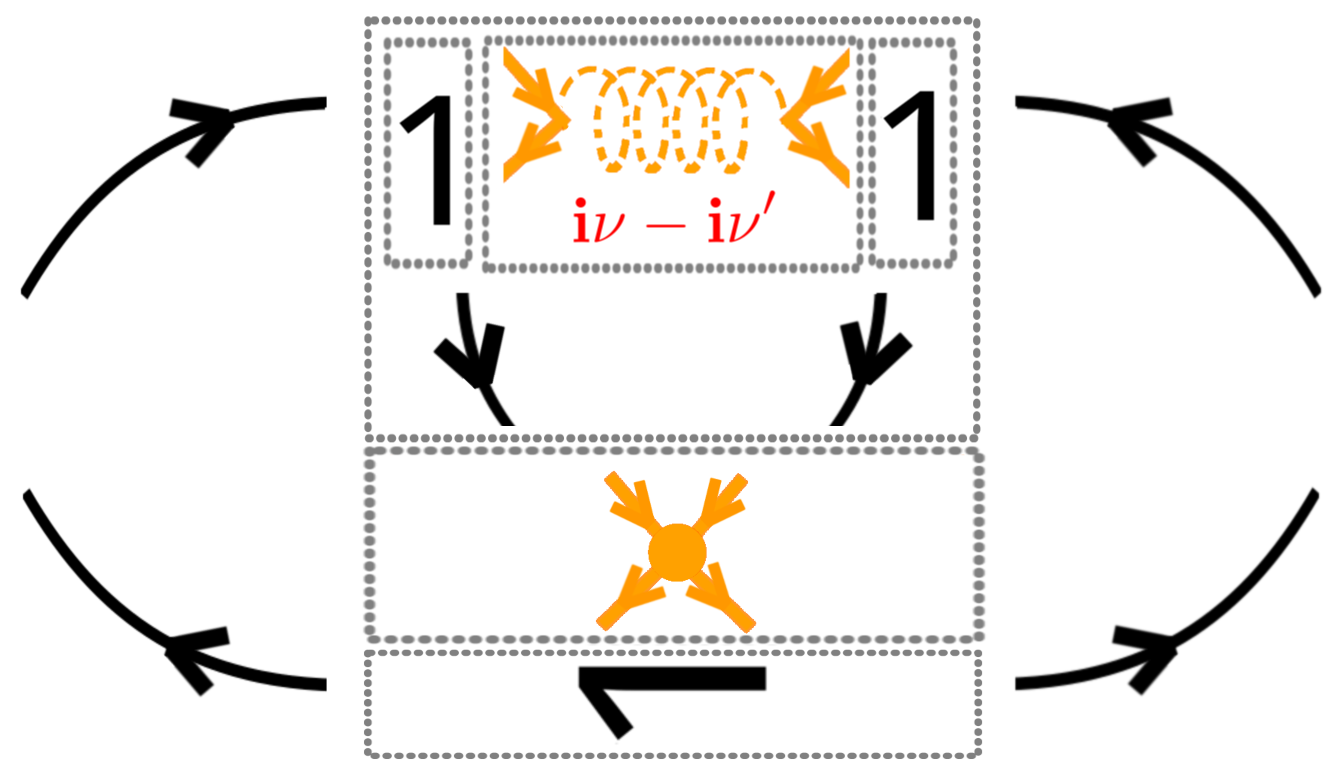}
    \caption{The leading-order diagram of $\mathcal{O}(U_\textrm{eff})$ in $\chi^\M_{\nabla^\SC}$, originating from the density contribution of the magnetic fermion-boson coupling. The leading-order diagrams in $\chi^\X_{\nabla^{\M}}$ have a similar form.}
    \label{fig:leading_sc_m_contributions_to_chi_X_omega_dependent}
\end{figure}

We now turn to the superconducting and magnetic contributions to $\chi^\X$. In this case, the leading order diagram of lowest order in $U_{\omega_0}\sim\mathcal{O}(U_\textrm{eff})$, see Fig.~\ref{fig:leading_sc_m_contributions_to_chi_X_omega_dependent}. In the contribution $\chi^\M_{\nabla^\SC}$, the $U_{\omega_0}$ interaction line originates from a $\nabla^\D_{\xph}$ contribution to $\lambda^{\SC}$, whereas the bare local interaction $U_\textrm{eff}^{\SC} = U_\textrm{eff}$ stems from the bare value of $w^\SC$. Since the $U_{\omega_0}$ interaction line is antiparallel in the $\SC$ channel, the integration over the frequencies in the loop samples all values of $U_{\omega_0}$ leading to an increase as $M_\textrm{ion}$ increases. 
We therefore have
\begin{align}
D_{\partial \chi^{\X}}^{\SC/\M} =&  - \text{Sgn}\,\Pi^\X \cdot \left(-D_{V^\X}^{\SC/\M}\right) \cdot \left(-D_{\lambda^{\SC/\M}}^\D\right) \cdot (+) \nonumber \\ & \cdot \text{Sgn}\, \Pi^\D \cdot \text{Sgn}\,U^{\SC/\M}_\textrm{eff} \cdot \text{Sgn}\,\Pi^\X\nonumber\\
=&-D_{V^\X}^{\SC/\M} \cdot D_{\lambda^{\SC/\M}}^\D\cdot\text{Sgn}\,U^{\SC/\M}_\textrm{eff} \,.
\end{align}
The first and the last bubbles come from Eq. \eqref{eq:susc_vertex_contribution}. In the expansion of the vertex, the sign of $(-D_{V^\X}^{\SC/\M})$ is acquired, the minus comes from the $\bar{\nabla}^{\SC/\M}$ component which in $D_{V}$ is taken to be negative. The remaining terms are the lowest order in $U_\textrm{eff}$ and $U_{\omega_0}$ coming from $\nabla^{\SC/\M}$. The lone plus sign comes from the $U_{\omega_0}$ term which increases as $M_\textrm{ion}$ increases. Evaluating, we find 
\begin{align}
    D_{\partial \chi^\X}^{\SC} = \mqty(+ \text{Sgn}\, U_\textrm{eff}\\ + \text{Sgn} \, U_\textrm{eff} \\ - \text{Sgn} \,U_\textrm{eff}), \qq{} D_{\partial \chi^\X}^{\M} = \mqty(- \text{Sgn} \,U_\textrm{eff}\\ - \text{Sgn} \,U_\textrm{eff} \\ + \text{Sgn} \,U_\textrm{eff}).
\end{align}

\section{Multiboson contributions to $\chi^\dSC$}
\label{app:multibosoncont}

In the main text, we saw how the single boson contribution to $d$-wave superconductivity is zero. Therefore, a $d$-wave superconducting instability can only reflect through a divergence of the multiboson part $M^\dSC$. Nonetheless, we claimed in the main text that this tendency to diverge is always preceded by a growth of the single-boson contribution ($\nabla^\M$) to $\chi^\dSC$. This was observed numerically in Ref.~\cite{Bonetti_2022}. We here take a closer look by analyzing the contributions of $M^\dSC$. We start from the definition~\cite{Krien2021b}
\begin{align}
M^\X_{n n'}(q, \nu, \nu') =& -\sum_{\mathclap{\nu'', m, m''}} \mathcal{I}_{n m}^\X(q, \nu, \nu'') \Pi^\X_{m m'}(q, \nu'') \nonumber \\
&\quad\;\; \times S^\X_{m' n'}(q, \nu'', \nu')
\end{align}
with $S^\X = \mathcal{I}^\X - M^\X$. Plugging in for $M^\dSC$ the definitions of $\mathcal{I}^\SC$ and $S^\SC$, and iterating the definition of $M^\X$, one generates  
a series of multiboson contributions  
\begin{align}
\label{eq:mboscontr}
M^\dSC =& M^\dSC_{\bar{\nabla}^\M, \bar{\nabla}^\M} + M^\dSC_{\bar{\nabla}^\D, \bar{\nabla}^\D} + M^\dSC_{\bar{\nabla}^\M, \bar{\nabla}^\D} + M^\dSC_{\bar{\nabla}^\D, \bar{\nabla}^\M}\nonumber \\&+\mathcal{O}(\bar{\nabla}^3),
\end{align}
where only the two-boson 
contributions are 
explicitly reported. These terms can be understood as the higher order terms to an expansion of the vertex $V$ in terms of the single boson vertices $\nabla^\X$ (and bubbles that connect them together).
The two-boson terms are given by
\begin{widetext}
\begin{subequations}
\label{eq:two_boson_terms}
\begin{align}
M^\dSC_{\bar{\nabla}^\M, \bar{\nabla}^\M}(q, \nu, \nu') &= -\left( -\frac{3}{2} \right)\left( -\frac{3}{2} \right) \;\sum_{\mathclap{\nu'', m, m''}} P^{\rightarrow \pp}\left[\bar{\nabla}^\M_{\xph,\ph}\right]_{\dwave, m}(q, \nu, \nu'') \Pi^\SC_{m m'}(q, \nu'') P^{\rightarrow \pp}\left[\bar{\nabla}^\M_{\xph,\ph}\right]_{m', \dwave}(q, \nu'', \nu'),\\
M^\dSC_{\bar{\nabla}^\D, \bar{\nabla}^\D} (q, \nu, \nu')&= -\left( \frac{1}{2} \right)\left( \frac{1}{2} \right) \;\sum_{\mathclap{\nu'', m, m''}} P^{\rightarrow \pp}\left[\bar{\nabla}^\D\right]_{\dwave, m}(q, \nu, \nu'') \Pi^\SC_{m m'}(q, \nu'') P^{\rightarrow \pp}\left[\bar{\nabla}^\D\right]_{m',\dwave}(q, \nu'', \nu'),\\
M^\dSC_{\bar{\nabla}^\M, \bar{\nabla}^\D}(q, \nu, \nu') &= -\left( -\frac{3}{2} \right)\left( \frac{1}{2} \right) \;\sum_{\mathclap{\nu'', m, m''}} P^{\rightarrow \pp}\left[\bar{\nabla}_{\xph,\ph}^\M\right]_{
\dwave,m}(q, \nu, \nu'') \Pi^\SC_{m m'}(q, \nu'') P^{\rightarrow \pp}\left[\bar{\nabla}^\D\right]_{m',\dwave}(q, \nu'', \nu')\,, 
\end{align}
\end{subequations} 
\end{widetext}
and similarly for $M^\SC_{\bar{\nabla}^\M, \bar{\nabla}^\D}$ 
with $\bar{\nabla}^\D$ and $\bar{\nabla}^\M$ exchanged, where we have denoted
\begin{align}
\bar{\nabla}^\M_{\xph,\ph} := \frac{2}{3}\left(\bar{\nabla}^\M + \frac{1}{2}P^{\rightarrow \ph}[\bar{\nabla}^\M] \right).
\end{align}
$P^{\rightarrow \pp}$ ($P^{\rightarrow \ph}$) is the TU projection onto the $\pp$ ($\ph$) channel \cite{Lichtenstein2017}. The prefactors appearing in Eq.~\eqref{eq:two_boson_terms} originate from the different components number 
of the associated order parameters (three for the magnetic and one for the density). 
Focusing  
on the ladder magnetic contributions in Eq.~\eqref{eq:mboscontr} which are the dominating ones for $U_\textrm{eff} > 0$, the $d$-wave projection of their resummation 
is 
\begin{align}
\left[\frac{-\frac{3}{2} P^{\rightarrow \pp}\left[\bar{\nabla}^\M\right]}{1 -  \Pi^\SC \circ \frac{3}{2}P^{\rightarrow \pp}\left[\bar{\nabla}^\M\right] } \right]_{\dwave, \dwave}(q, \nu, \nu')\,, \label{eq:dwave_ladder_magnetic_fluctuations}
\end{align}
where $\circ$ denotes a convolution as above and we have also included the single boson term. This (approximate) Stoner-like expression gives a condition for the onset of $d$-wave superconductivity
\begin{align}
\Pi^\dSC \circ \left[P^{\rightarrow \pp}\left[\bar{\nabla}^\M\right]\right]_{\dwave,\dwave}\simeq 1
\,
\end{align}
at $q = 0$, a statement nonetheless made in terms of the single-boson effective interaction. This is possible since $\Pi^\dSC$ is positive and the $d$-wave component of the $\pp$-projected $\nabla^\M$ peaked around $\bfq_* = (\pi,\pi)$ is also positive, see discussion in Section~\ref{sec:dwave_fluctuation_diagnostics}. Thus, we can see that the $d$-wave superconducting contributions are seeded by an antiferromagnetic $\nabla^\M$. We note furthermore that the reversal of the isotope effect through the dressing of the $d$-wave bubble by the self-energy has implications on the isotope effect of $T_c$.

\bibliography{citations}

\end{document}